\documentclass[aps,pre,longbibliography,notitlepage,floatfix,11pt,tightenlines,nofootinbib,superscriptaddress]{revtex4-1}

\usepackage[format=plain,labelfont={bf,small},textfont=small,justification=raggedright,singlelinecheck=false]{caption}

\usepackage{bm,amsmath,amssymb,graphicx,stmaryrd,xcolor}
\usepackage{subfig}
\usepackage[abs]{overpic}
\usepackage{float}
\usepackage{subcaption}
\usepackage{caption}
\usepackage{placeins}

\begin{document}

\title{
Elastic crumple pair potentials
}
\author{D. Gimeno}
    \email{D.Gimeno@sussex.ac.uk}
\affiliation{Department of Mechanical Engineering, University of Nevada, 1664  N.\  Virginia  St.\ (0312),  Reno,  NV  89557-0312,  U.S.A.}
\author{E. Hamm}
	\email{luis.hamm@usach.cl}
    \affiliation{Departamento de F\'isica, Facultad de Ciencia, Universidad de Santiago de Chile, Av.\ V\'ictor Jara 3493, Estaci\'{o}n Central, Santiago 9160000, Chile}
\author{J. A. Hanna}
    \email{jhanna@unr.edu}
    \affiliation{Department of Mechanical Engineering, University of Nevada, 1664  N.\  Virginia  St.\ (0312),  Reno,  NV  89557-0312,  U.S.A.}

\begin{abstract} 
	Thin structures across scales and materials, from graphene sheets to clothing to aircraft skins, respond to forcing and confinement by crumpling, with deformation localized into zones of characteristic shape. Much attention has been paid to isolated crescent vertex and ridge shapes, but there has been almost no investigation into the dynamics, creation/destruction, and interactions between multiple vertices (``crumples'') that are inherent in nearly all deformation processes. The next step beyond single crumples are pair interactions, mediated by background curvature.  Of these, we focus on two common types, using a quasi-1D setup allowing controlled boundary motion and force measurement coupled with tracking of crumple positions. We provide experimental evidence for non-monotonic pair potentials, with short-range attraction, long-range repulsion, and an intermediate forbidden region. We construct a phenomenological energy that reproduces the bifurcations and the interrelations between observed quantities. These results provide a new insight and input towards understanding and building reduced descriptions of dynamic deformation and pattern formation and evolution in thin elastic plate and shell structures. 
\end{abstract} 

\date{\today}

\maketitle

\section*{Crumples as particles?}

\begin{figure}[h]
	\includegraphics[width=\textwidth]{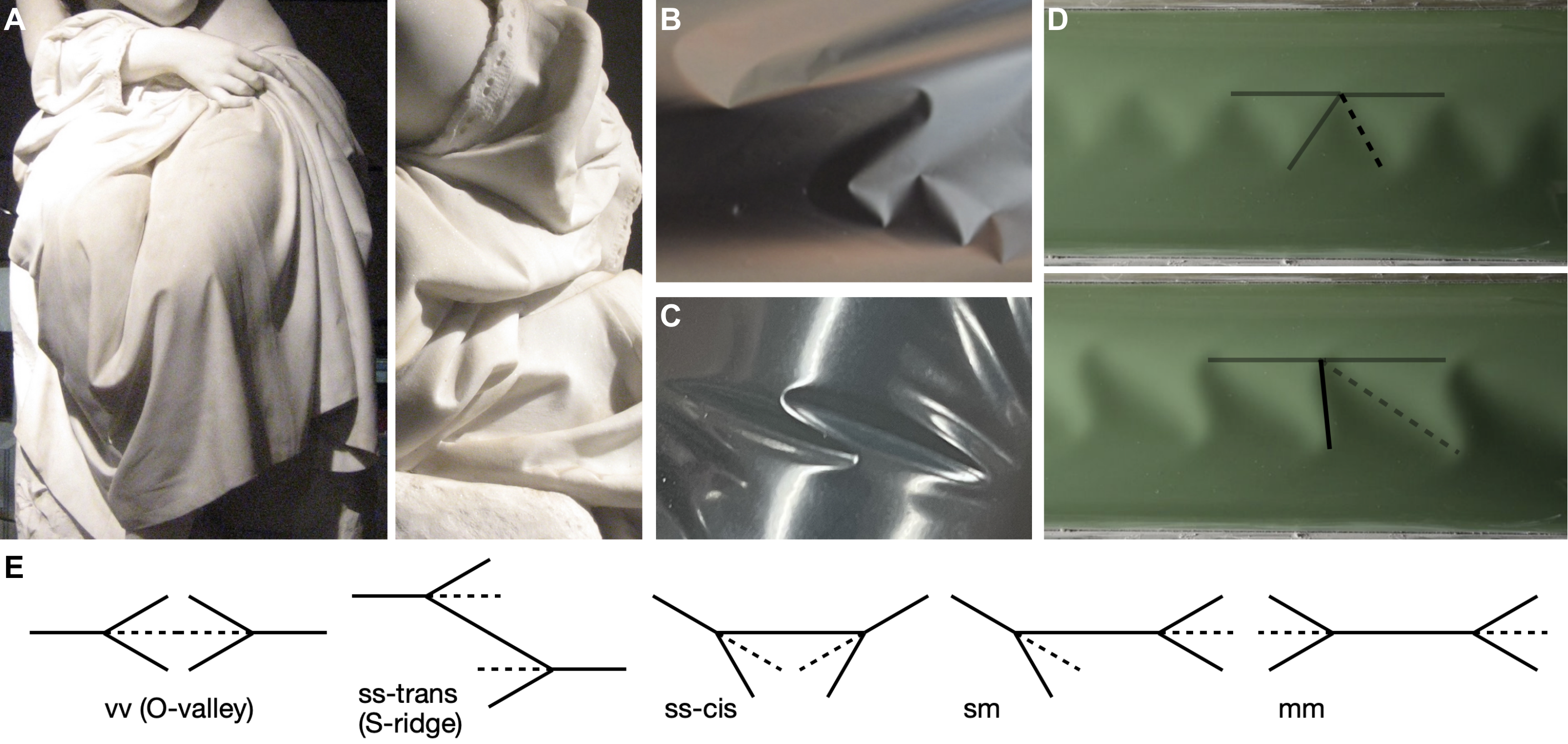}
%	\captionsetup{margin=0.5in}
	\caption{Crumples, ridges, and valleys in {\bf{A:}} a representation of fabric in a marble sculpture (Corbellini, \emph{Il pudore}, mid-19th century), {\bf{B:}} copper foil, {\bf{C:}} an inflated polymer bag.  {\bf{D:}} Single crumple motifs, consisting of three ridges (solid lines) and one valley (dashed line) superposed on arrays of O-valleys (upper, with valley shown darker) and S-ridges (lower, with relevant ridge shown darker) in a polyester sheet. {\bf{E:}} Schematic of the five possible pair interactions between crumples, involving links along a valley (v) or side (s) or middle (m) ridge. (copper and bag credit: C. Redman). 
%valley-valley  (Ovalley)
%side-side-trans (Sridge)
%side-side-cis 
%side-middle 
%middle-middle 
}\label{examples}
\end{figure}

Across materials and scales, the elastic deformation of thin membranes, plates, and shells is characterized by a pronounced focusing of curvature and strain energies into localized zones. %, such as the tip of a wrinkle.
Many such ``crumples''\footnote{We note different usage of the term crumple in \cite{King12, Timounay20, Croll19}; other common terms include d-cone and vertex or crescent singularity.} of both elastic and inelastic type can be seen in Figure \ref{examples}.  Much attention has been directed at static, isolated crumples (vertices) and inter-crumple ridges, viewed as elastic singularities of point and line type (early examples include \cite{Witten07,WittenLi93,Lobkovsky95,BenAmarPomeau97,Cerda99}), and regular lattices of crumples \cite{Yamaki, Timounay20}, including static pattern growth by snaking processes \cite{Hunt20, GrohHunt21}. %Groh21? 
These works offer fundamental insights into thin structure deformation, although many questions still remain around the energetic scaling and geometry of crumples. However, structures in use follow complex dynamic paths between states, featuring
ordered--- or more often disordered--- networks of mobile vertices, ridges, and valleys that form and develop through nucleation and rearrangement, and refine/coarsen through birth/death processes such as splitting/merging or entrance/escape from boundaries, with both external loads and (as yet unknown) internal inter-crumple interactions driving these motions and transitions. Examples include aircraft panels subject to fluid and thermal loading \cite{Spottswood19}, clothing affected by gravity and contact \cite{Goldenthal07}, and soft matter films under confinement \cite{Croll19}.  Videos V1-V3 show examples of these behaviors in two %polymeric
 materials at the desktop scale. 

Almost nothing is known about such dynamic, irregular, inter-crumple interactions which include--- as we suggest in the present work--- binding and unbinding, and exchange of bonds, between crumples. There are only a handful of experimental studies seeking to observe the \emph{real-time} evolution of such networks in the \emph{elastic} regime \cite{Hamm04, AharoniSharon10, RomanPocheau12, Hutton24}, and rare attempts to measure the induced motion of a crumple pair \cite{Boudaoud00, Hamm04, Walsh11}. Aside from the fascinating dynamics we observe in thin sheets, the current study is inspired by the great utility of reduced descriptions of linear field theories (among these, electromagnetics, creeping and ideal fluid flows, and linear elasticity) in terms of singular solutions, and by various phenomena in which particles disturb, and induce driving forces through, a deformable medium such as a fluid interface %or nematic liquid crystal %, or solid solution. %interactions through free-surface indentation/disturbance by objects 
\cite{Nicolson49, GiffordScriven71, Loudet05, Poulin97, Ardell66, VoorheesJohnson88}. As the governing nonlinear field theory of plate elasticity also displays localized solutions, perhaps it may also be amenable to a particle-like description. 

\section*{Observing pair interactions}

The first step towards understanding inter-crumple forces, and developing reduced physical descriptions of the deformation of thin bodies, is to consider pair interactions.  As shown in Figure \ref{examples}E, there are five distinct ways for two crumples to form a pair 
along a ridge or valley (solid or dashed line), having respectively the same or opposite sense of curvature as the local background. 
These are all observable during generic manipulations of thin materials. We focus on two of these motifs that commonly recur in our recent experiments (Figure \ref{examples}D, see also \cite{Hutton24}). The O-valley and S-ridge are, respectively, crumple pairs connected by a valley or by a side-side ridge with trans-valley arrangement; the O-valley resembles the common sight of a indentation in a soda can, while the S-ridge is easy to spot on shirt sleeves. 

These two pair interactions, and more, are displayed in a controlled experiment shown in Video V1 (0.003 inch polyester shim stock), and an uncontrolled demonstration with a different material in Video V2 (0.002 inch Tedlar$^{\circledR}$/polyvinyl fluoride), in which high-aspect-ratio, clamped, bent sheets are subject to lateral boundary displacement, resulting in a dynamic sequence of quasi-1D patterns of crumples. In the controlled experiment (Video V1), a zipper of O-valleys first forms in rapid sequence. This ``crystal'' then re-pairs into a similar, but more ``fluid'', array of S-ridges. After a sequence of mergers and edge escapes, a single S-ridge remains.  
Upon reversal of the boundary motion, the S-ridge's two constituent crumples split apart and briefly appear to weakly repel, before a new O-pair is nucleated in between (a process reminiscent of ridge instabilities observed elsewhere \cite{DiDonna02,Hutton24}), resulting in strong repulsion and splitting of the S-ridge into two widely spaced S-ridges.  A sequence of splittings continues until a full array of S-ridges returns, then re-pairs into O-valleys, which close through pairwise annihilation of their constituent crumples. Similar splitting and pair exchange behaviors can be seen in Video V3, and in irregular deformations. 

\begin{figure}[h]
	\includegraphics[width=0.5\textwidth]{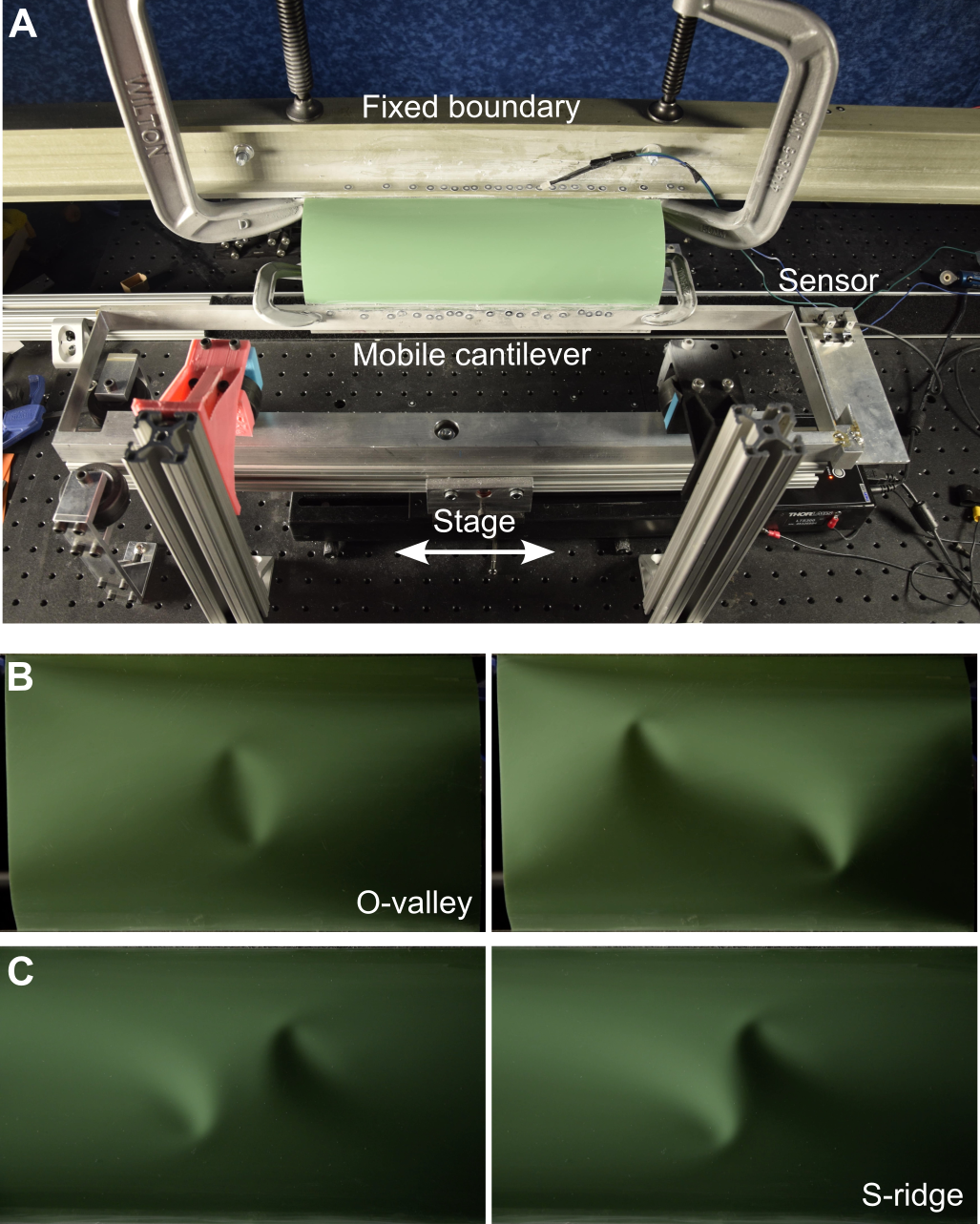}
%	\captionsetup{margin=0.5in}
	\caption{{\bf{A:}} Setup for pair interaction experiments using a laterally translating, instrumented boundary (details in text and Appendix~\ref{methods}). The sheet in the image is as used for the S-ridge pairs, 12 inches long and bent into a quarter-circle with 4 inches of free span; for O-valley pairs, a 6 inch long sheet is used. {\bf{B:}} Bound (left) and unbound (right) states of the O-valley.  {\bf{C:}} Unbound (left) and bound (right) states of the S-ridge.} \label{setup}
\end{figure}

The O and S pair interactions are of particular interest as they display certain characteristics of ``bonds''; both display bound and unbound states, as shown in Figure \ref{setup}B-C. Pairing leads to collective crumple motion; the S-ridge in particular translates as a rigid unit, which organized motion mediates much of the overall deformation, consisting primarily of inversion of the direction of curvature of the sheet. When we informally address the crumples through external poking, we observe different behavior in connected and separated states (Videos V4-V5).  The individual crumples in an unbound/broken S-ridge can be moved separately, but when bound, they are a tightly coordinated pair that moves together. The same can be said about an unbound O-valley, although in contrast to the S, the bound O pair appears less mobile. There are marked changes in velocity as individual crumples approach and recede from each other (Videos V6-V7), suggesting an instability at certain distances, as well as a transition from attraction to repulsion. 
Next, we seek to quantify these observations regarding pair interactions.

\section*{Measuring pair interactions}

This arrangement for lateral boundary displacement of a bent sheet is further exploited in order to study crumple pair interactions in relative isolation (Figure \ref{setup}A). The setup allows us to explore different behaviors than normal probing of the sheet \cite{HolstCalladine94, PauchardRica98, Boudaoud00, Hamm04, Walsh11, Das07}, while keeping track of the work done on, and therefore energy input to, the system.  Boundary loading is more akin to the setting experienced by a patch of material within a larger sheet. 
In our experiments, one boundary is instrumented, attached to a cantilever associated with an inductive position sensor.  We prepare the system, and displace this boundary to slowly separate and bring together O and S pairs (in 6 inch and 12 inch long sheets, respectively), and measure displacement and shear force around these loops, while simultaneously optically tracking the positions of crumples (further details in Appendix~\ref{methods}). 

Figure \ref{loops} shows two loops each of force $F$ versus boundary displacement $u$ (A-B) and the signed inter-crumple separation distance $D$ versus $u$ (C-D) for the O-valley (A,C) and S-ridge (B,D). Zeroes of $u$ and $D$ correspond to bound states near the inner stability limit for O or vertical orientation for S. Additional data is included in Appendix~\ref{data}, while information about repeatability within and across samples, and the force during unbinding of the S-ridge shown within the context of a wider loop of events, can be found in Appendix~\ref{methods}.  The boundary moves on the order of tens of microns, while crumples move on the order of millimeters.  Both force and crumple separation display jumps, corresponding to more rapid, unstable motion of crumples (the sheet edge is moved at 0.004 mm/s, while typical rates of crumple separation outside and during the jumps are, respectively, amplified to the order of 0.18 and 11 mm/s for the O-valley and 0.04 and 1 mm/s for the S-ridge... even the fast motion is not particularly so, as has been noted for other elastic bifurcations \cite{Gomez17}.) A range of displacements corresponds to two possible force/distance states, and some ranges of force and distance are inaccessible as equilibrium states, so only transiently observed. The O-valley data is the clearest to interpret, as measurements are bounded on either side by bifurcations or other changes in motion (see Appendix~\ref{methods} for details) and this range of observation is particularly limited for the S-ridge. Note that when closer together, the O-crumples will annihilate, but the S-crumples do not--- we are not probing this range of separations in either case. The loops do not close perfectly in some places, and hysteretic motion of crumple positions is also observed, particularly in the O-valley (see Appendix~\ref{data} for crumple trajectory data), reflecting the fact that this is not a simple 1D system, but has additional degrees of freedom \cite{XuVirgin19, Ehrhardt20}. 

From these observations, we infer that these pairs are jumping between two stable branches of solutions, perhaps corresponding to bound and unbound states.  Next, we further obtain and employ stability information from crumple velocities and, by measuring force and work-conjugate displacement, obtain changes in energy along each stable branch on either side of these bifurcations. By relating the displacement and the inter-crumple separation, we indirectly measure a ``pair potential'' between the crumples. We determine the %qualitative
shape of this potential, and present a minimal reduced model mimicing the behavior of these variables. 

\begin{figure}[h]
	\includegraphics[width=0.65\textwidth]{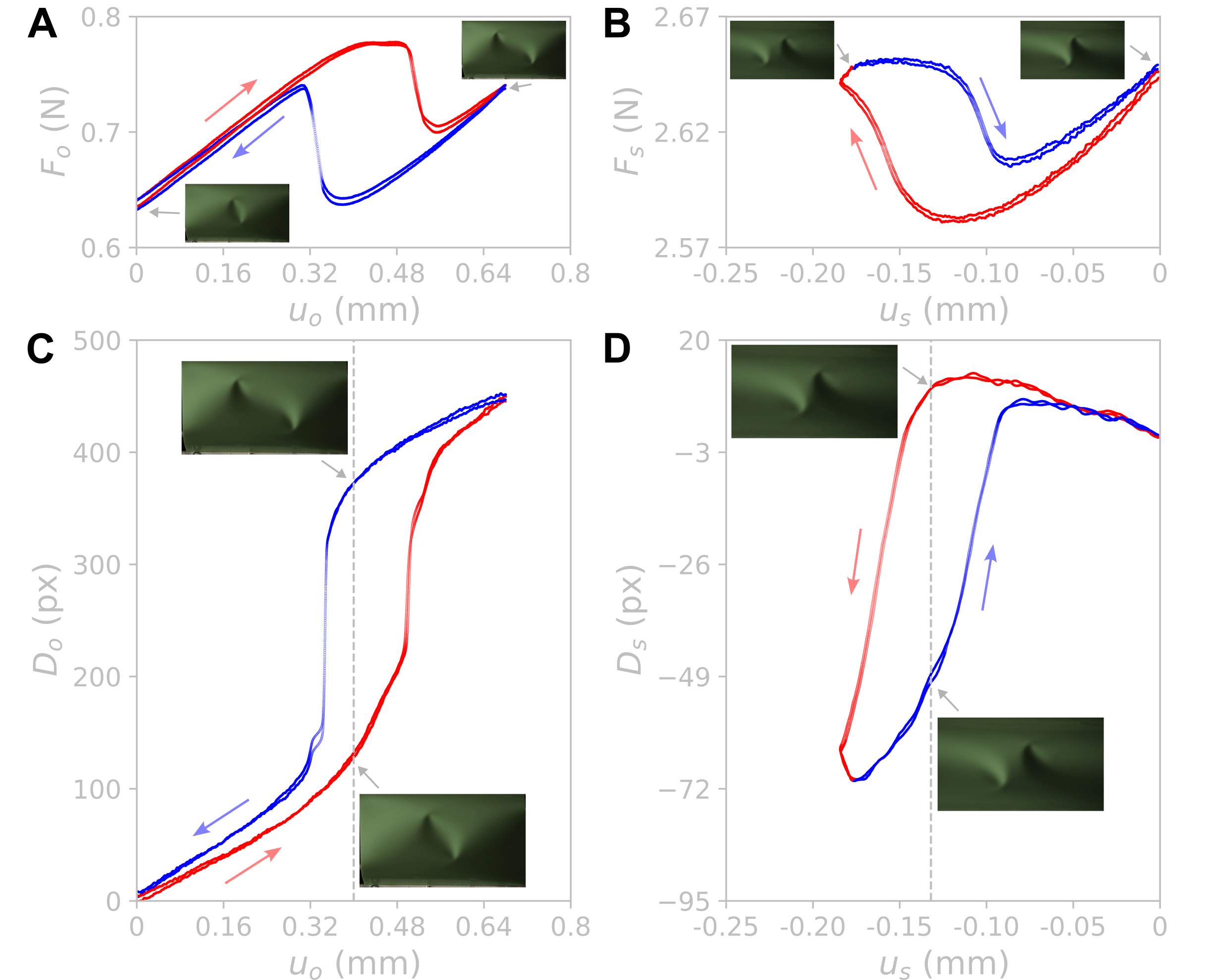}
%	\captionsetup{margin=0.5in}
	\caption{Examples of Force $F$ versus boundary displacements $u$ ({\bf{A-B}}), and signed inter-crumple separation distance $D$ versus $u$ ({\bf{C-D}}) for the O-valley ({\bf{A, C}}) and S-ridge ({\bf{B, D}}), with images of representative bound and unbound states. %, including coexisting states at the same $u$.  
	Large arrows indicate the direction of loop traversal. Red and blue indicate the state of the system, whether undergoing an unbinding or binding process, rather than the direction of motion of the boundary; brief retrograde motion occurs for the S-ridge (see Appendix~\ref{methods} for details). Zeroes of $u$ and $D$ correspond to bound states near the inner stability limit for O and vertical orientation for S. On the distance axes, about 9px corresponds to 1mm.  The two runs for the O-valley were taken about two hours apart, after running several loops of smaller width.  The two runs for the S-ridge were taken in immediate sequence.  
	 }
    \label{loops}
\end{figure}

\FloatBarrier

\section*{Swallowtail energy and pair potential}

The qualitative relationships between the measured crumple pair quantities are reproduced by a phenomenological model inspired by those arising in canonical structural examples such as a spring-loaded von Mises truss \cite{BazantCedolin} or tied arch \cite{Hunt20}. 
The model describes response under an applied force $P$ in terms of its work-conjugate variable $q_2$, here a proxy for the elastic response of the sheet background, and an internal variable $q_1$, here representing the inter-crumple distance (see Appendix~\ref{methods} for direct relationships and distinctions with abovementioned examples). The elastic energy of the sheet 
\begin{align}
	E = \tfrac{1}{4}\left(q_1^2-1\right)^2 + \tfrac{c}{2}\left(q_2-q_1\right)^2 \label{modelenergy}
\end{align}
is quartic in the internal variable, which is in turn negatively coupled to the work-conjugate variable with coupling constant $0 < c  < 1$.  
The work of the applied force through its work-conjugate quantity $\int\! P \mathrm{d}q_2$ is equal to the change in energy.  Along equilibrium paths, virtual work considerations dictate $P = \frac{\partial E}{\partial q_2}$ and $0 = \frac{\partial E}{\partial q_1}$ ; this provides two relationships between the force and variables, $	P = c\left(q_2-q_1\right)$ and $q_1\left(q_1^2-1\right) = c\left(q_2-q_1\right)$, and therefore also $P = q_1\left(q_1^2-1\right) = \left(q_2 - \tfrac{P}{c}\right)\left[ \left(q_2 - \tfrac{P}{c}\right)^2-1\right]$. Equilibrium force and energy can thus be expressed in terms of either variable. 

While the canonical examples mentioned above are bistable, the existence and location of energy minima are not relevant to the present discussion, nor is even the particular functional form of the energy.  The important features we require are a maximum in the internal contribution to the energy, and two fold (turning-point) bifurcations delimiting an unstable folded-back (regressive) branch in the force-conjugate displacement response. 

The force $P$ and internal variable $q_1$ as functions of the work-conjugate variable $q_2$ have two folds separating three equilibrium branches; for the force, the folds lie between two max/min $P$ values. Stability of these equilibria depends on the nature of loading and constraints \cite{BazantCedolin}. For load control, the system changes stability at the force max/min and jumps (horizontally) in $q_2$, and for displacement control it changes stability at the folds and jumps (vertically) in the force.  A system such as our cantilever would not be strictly displacement-controlled because of large but finite machine stiffness, but would have a slightly-sloped jump between branches, somewhere between max/min and folds. We illustrate the simple case of displacement control, which assumes $q_2$ is fixed and stability is given by the second derivative with respect to $q_1$. The middle folded-back branch is unstable (dotted lines).  

Integrating the double-overhanging implicit cubic $P$ along $q_2$, or inserting equilibrium results into $E$,  
gives us a multivalued swallowtail energy $E$ as a function of $q_2$, where cusps in energy correspond to folds in force.\footnote{What appears to be an erroneous attempt at such an integration may be seen in Salvador Dal{\'{\i}}'s last painting \cite{Dali} with its misaligned vertical lines; besides the cusp-fold correspondence, the energy minimum should align with a zero force at the cubic's inflection point. Whether these curves are meant to relate to each other, or whether they are simply a juxtaposition of two images from Thom's book \cite{ThomSSM} is an open question, but King \cite{King} has suggested that the cello holes are meant to represent integral signs. Note the entirely different meanings of ``structural stability'' in \cite{BazantCedolin} and \cite{ThomSSM}.}
Translating this into a function of inter-crumple distance, closely related to $q_1$, transforms the swallowtail into a mound-shaped pair potential with attractive and repulsive regions, separated by an only transiently accessible ``forbidden'' region of unstable solutions. 

\begin{figure}[h]
	\includegraphics[width=\textwidth]{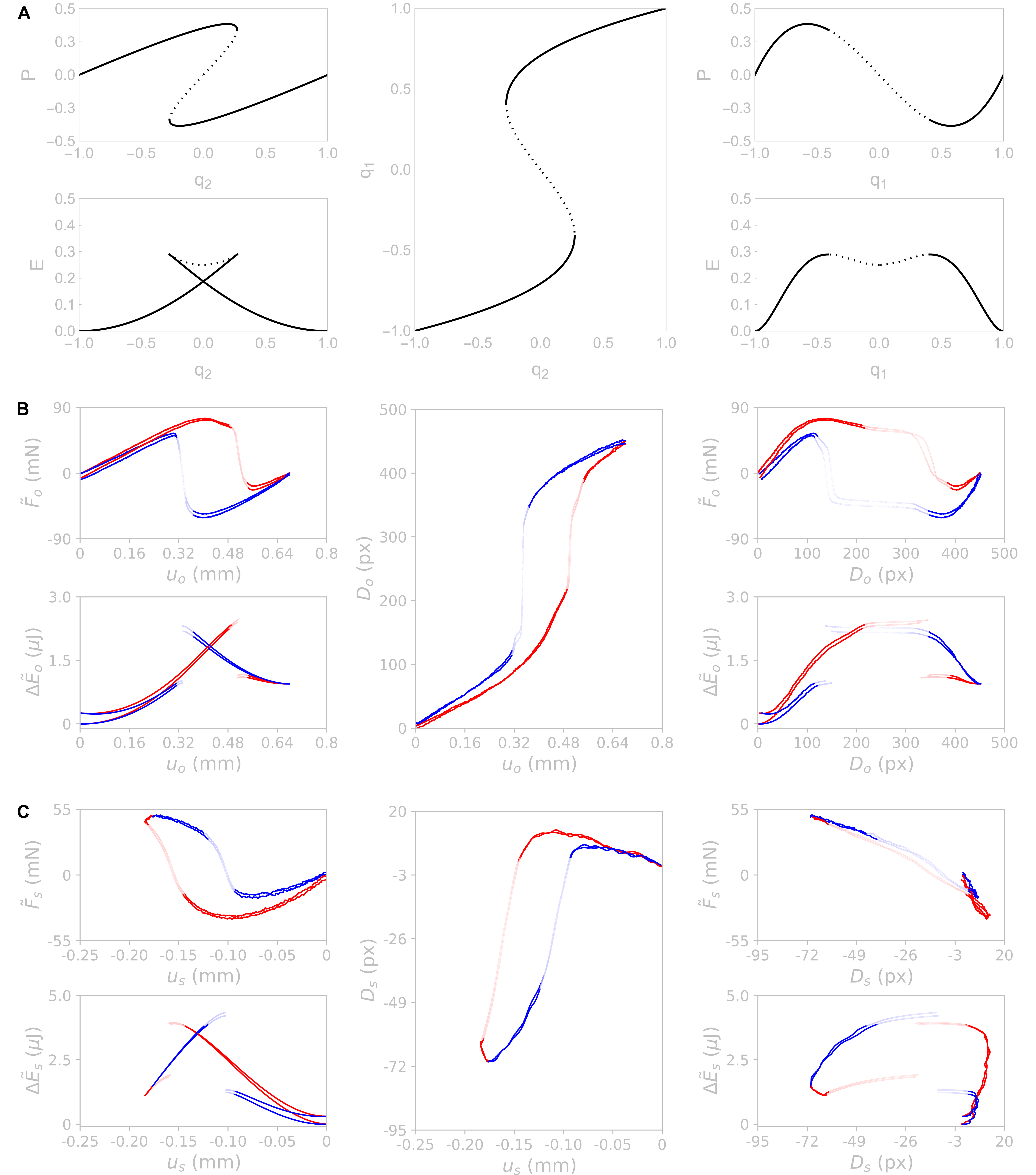}
%	\captionsetup{margin=0.5in}
	\caption{
	{\bf{A:}} Stable (solid) and unstable (dashed) equilibrium branches of the minimal model based on the energy \eqref{modelenergy}, with $c = 1/2$ and chosen for clarity without loss of generality, and irrelevant energy minima placed at $\pm1$. Force $P$ and energy $E$ as functions of work-conjugate variable $q_2$, relation between $q_2$ and internal variable $q_1$, and $P$ and $E$ as functions of $q_1$. 	
	Processed data from Figure \ref{loops} for {\bf{B:}} O-valley and {\bf{C:}} S-ridge, showing stable (dark) and unstable (light) regions. For explicit algebraic relations of internal and work-conjugate variables $q_1$ and $q_2$ to experimental boundary displacements $u$ and signed inter-crumple distances $D$, see Appendix~\ref{methods}. Zeroes of $u$ and $D$ correspond to bound states near the inner stability limit for O and vertical orientation for S. 
	On the distance axes, about 9px corresponds to 1mm. 
The same force subtraction was used for both runs in each case, with cuts made at the points of most rapid motion.  For the O-valley, energies were matched at the maximum value of $u_O \approx 0.68$ mm. For the S-ridge, energies were matched at a nominally stable point $u_S \approx -0.18$ mm on the retrograde branch.
}
    \label{compare}
\end{figure}

This model response is qualitatively compared with data obtained from representative curves of each type in Figure \ref{compare}. In the experiments, the double-fold system is embedded in a background sheet response significantly larger than the force differences of interest, and also contributing a slope to the force curves.  The system is pre-loaded; we are effectively pulling the O-valley apart and pushing the S-ridge together, so the zero/bound state is on the left for O and on the right for S. The background has been implicitly removed in the model \eqref{modelenergy}, but could be accounted for with an additional term of the form $\tfrac{a}{2}(q_2-b)^2$ with constants $a$ and $b$. To facilitate comparison, we subtract out some of the experimentally measured trend to obtain a force $\tilde F$ and associated energy $\tilde E$; this process is interrelated with the existence of irrelevant energy minima. Unstable branches are not observable. Rather than follow them, the system will jump and release an unknown amount of energy, during which the work integral of $P$ along $q_2$ no longer reflects the energy absorbed in elastic response. Those portions of the curves in which motions are too fast to be considered quasistatic are shown in lighter colors.  We cut the curves at the point of most rapid change, and displace the energy curves in a symmetric manner, although the real response need not be symmetric. Details of subtraction, cutting, and displacement approaches, including alternate approaches, can be found in Appendix~\ref{methods}. 

The O-valley pair interaction clearly displays the same qualitative behavior as the minimal model, namely its bifurcation structure and interrelations between the three quantities.  There is short-range attraction near zero, long-range repulsion, and an unstable range in between. The data for the S-ridge is consistent with similar behavior, but is obtainable only over a range insufficient to be conclusive.  Very little data can be obtained from the unbound side of the jump due to the proximity of the splitting bifurcation, while on the bound side near zero, the stable attractive region is very small, after which the ridge pair pass around each other and adjust their opening angles, rather than continue to approach each other (see crumple trajectories in Appendix~\ref{data}).  This corresponds to a post-peak decline in the displacement-distance response, and a nearly vertical, apparently multivalued, portion of the energy plot, in which this effectively one-dimensional description fails to capture the crumple kinematics.

\FloatBarrier

Interpretation of the data in the light of the minimal model reveals a pair potential with attractive, forbidden, and repulsive regions, arising from a negative coupling between the inter-crumple distance and the overall sheet elasticity. Given this new observation, it is perhaps notable that non-monotonicity of applied torque versus valley-like separations on an indented, twisted cylinder can also be inferred from discrete values of indentation depth in vertical slices through figure 17 of \cite{Hamm04}.

\section*{Questions raised} 

As crumples often induce other crumples, and because high aspect ratio experiments are alignment sensitive, isolating single crumple pairs requires us to work with moderately-sized sheets. This raises the question of the significance of finite size effects. Crumples are not isolated, but connected with edges through ridges and valleys, just as they would be connected to other crumples if present. These additional connections have energies associated with them, suggestive of a spring network, although deformations involve not just lengthening but changes in angles (see Figure \ref{examples}D for an example). 

A full description of a pair potential for anisotropic objects will depend on orientation. There will further be some dependence on the anisotropic background curvature field. To what extent does this field mediate the interaction, and is this asymmetric with respect to ridge and valley bond behaviors, given the bias in extrinsic curvature?  Before addressing this question, the background's effects on the energy, depth, and other geometry of single crumple should be systematically examined. The untested scaling predictions of \cite{Hutton24} are a possible starting point.   
Furthermore, we did not induce crumple motion simply along the bond direction, and in particular experienced a breakdown of the one-dimensional description for the S-ridge bond. We have effectively been measuring imperfect cross sections of some anisotropic potential. It is also possible that the qualitative form of the $u$-$D$ relation for the S-ridge differs from that of the O-valley and minimal model. 

A negative coupling suggests that background elastic energy is relieved by the elongation of valleys, regions that oppose the background curvature, which extend from crumples to crumples in the O-valley and from crumples to edges in the S-ridge. In either case, rigid motion of a pair should be a relatively soft mode of deformation, costing little or no energy.  All this suggests a possible form of a reduced model in which valley length lowers the overall energy, a more specific state variable than that proposed for elastoplastic sheets in \cite{Gottesman18}. But what is the physical origin of the repulsive interaction inherent in the energetic dependence on the separation, and what happens to curvature fields as valley or ridge bonds are formed or broken? Currently available techniques are not sufficient to measure curvature energy changes, as the mean curvature values relevant to the energy have very large spreads (see for example \cite[supplementary document figure S2]{ChopinKudrolli16} for a result similar to what we have been able to obtain).  A more precise method for obtaining mean curvature fields from point cloud or similar data would be of great value in this and other investigations. 

To complete the picture of crumple interactions, additional measurements are needed, and some will require different experimental arrangements. Single crumple-boundary interactions (image forces) have not been examined; observations suggest an energy barrier upon approach, as well as induction of new crumples entering from the edge. The next step up from pair interactions, pair-pair interactions, are prominent during the coarsening/refining process when S-ridges adjust their positions. This involves crumples with the same, rather than opposite, orientation, and it's not clear whether the interaction is primarily through a valley or through a side-middle ridge, which we have not yet observed as a pure pair interaction. From observation, the behavior may also be of non-monotonic form with a forbidden region; S-ridges attract and merge at short distances, but appear to strongly repel at medium distances. There are two other possible pairs.  The middle-middle ridge interaction, explored in \cite[Figures 3 and 5]{Walsh11}, is unlikely to display attraction, only annihilation upon colliding. The side-side cis ridge (in contrast to the trans S) interaction seems related to the instability of a single crumple to splitting \cite{PauchardRica98, Das07}; this ridge appears when bending a long fold \cite{SatouFuruya14}.  The five possible interaction types (Figure \ref{examples}E) can be seen \emph{in situ} in the various examples in Figure \ref{examples}A-D.  It is unknown whether other ``bonding'' type interactions may exist beyond the two explored here. In Video V1, we observe a transition from an ordered lattice of O-valleys to an apparently more fluid array of S-ridges, and a reverse transition upon return; it seems the relative strength of ridge versus valley bonds changes, or some exchange of bonds occurs, in conjunction with spreading or contraction of the pattern of pairs, although this can be masked by additional nucleation, entrances, or escapes \cite{Gimeno1}.  Future work will attempt to measure signatures of this collective transition. We have been unable to set up a single bond exchange, as seen also in more irregular dynamics. The contrast between the highly mobile S-ridge and the more, but not entirely, sessile O-valley is intriguing. 

Finally, a forbidden region in the pair potential should leave some signature in statistics of ridge lengths or crumple pair correlations for \emph{elastically} crumpled sheets, such as a gap in the distribution of lengths.

\section*{acknowledgments}
We thank R. Groh for much helpful discussion, B. Nagy for machining cantilevers, and C. Redman for help with videos and images. 
JAH acknowledges support from the U.S. National Science Foundation under grant CMMI-2210797. 
EH is grateful for the support of DICYT (USACH) under grant 042431HH. 

\section*{additional files}
\noindent Videos: \url{https://doi.org/10.5281/zenodo.18673731}  \\
\noindent Data: \url{https://doi.org/10.5281/zenodo.18673738} 

\section*{author contributions}
EH and JAH conceived and guided the study. DG and EH designed the apparatus. DG performed the experiments. DG and JAH analyzed data and developed the model.  DG and JAH made figures.  JAH wrote the first draft, and all authors contributed to the final draft.

\bibliographystyle{unsrt}
\bibliography{refs_pairs}

@article{Gimeno1,
  title        = {A sequence of elastic patterns in a sheared bent sheet},
  author       = {D. Gimeno and B. K. Meghwar and G. Fisher and R. S. Hutton and E. Hamm and J. A. Hanna},
JOURNAL="International Journal of Solids and Structures",
VOLUME="339",
PAGES="114152",
YEAR="2026"
}

@misc{Dali,
  title        = {{L}a queue d'aronde / {T}he swallow's tail},
  author       = {S. Dal{\'{\i}}},
  year         = 1983
}

@incollection{King,
  title        = {Dal{\'{\i}} After 1940},
  author       = {E. H. King},
  year         = 2010,
  booktitle    = {Salvador Dal{\'{\i}}: The Late Work},
editor={E. H. King and D. A. Brenneman},
publisher    = {Yale University Press},
  pages        = {12--55}
}

@Book{BazantCedolin,
	author="Z. P. Ba{\v{z}}ant and L. Cedolin",
	title="Stability of Structures",
	publisher="World Scientific",
	address="Singapore",
	year="2010"}

@Book{ThomSSM,
	author="R. Thom",
	title="Structural Stability and Morphogenesis",
	publisher="W. A. Benjamin",
	address="Reading, MA",
	year="1975"}

@Book{Yamaki,
	author="N. Yamaki",
	title="Elastic Stability of Circular Cylindrical Shells",
	publisher="North-Holland",
	address="Amsterdam",
	year="1984"}

@Article(AharoniSharon10,
AUTHOR="H. Aharoni and E. Sharon",
TITLE="Direct observation of the temporal and spatial dynamics during crumpling",
JOURNAL="Nature Materials",
VOLUME="9",
PAGES="993-997",
YEAR="2010")

@Article(Ardell66,
AUTHOR="A. J. Ardell and R. B. Nicholson and J. D. Eshelby",
TITLE="On the Modulated Structure of Aged {N}i-{A}l Alloys, with an {A}ppendix, {O}n the Elastic Interation Between Inclusions",
JOURNAL="Acta Metallurgica",
VOLUME="14",
PAGES="1295-1309",
YEAR="1966")

@Article(BenAmarPomeau97,
AUTHOR="Ben Amar, M. and Pomeau, Y.",
TITLE="Crumpled Paper",
JOURNAL="Proceedings of the Royal Society of London. Series A",
VOLUME="453",
PAGES="729-755",
YEAR="1997")

@Article(Boudaoud00,
AUTHOR="A. Boudaoud and P. Patricio and Y. Couder and M. Ben Amar",
TITLE="Dynamics of singularities in a constrained elastic plate",
JOURNAL="Nature",
VOLUME="407",
PAGES="718-720",
YEAR="2000")

@article{Cerda99,
  title={Conical dislocations in crumpling},
  author={Cerda, E. and Chaieb, S. and Melo, F. and Mahadevan, L.},
  journal={Nature},
  volume={401},
  number={6748},
  pages={46--49},
  year={1999}
 }

@article{ChopinKudrolli16,
	title={Disclinations, e-cones, and their interactions in extensible sheets},
	author={Chopin, J. and Kudrolli, A.},
	journal={Soft matter},
	volume={12},
	number={19},
	pages={4457-4462},
	year={2016}
}

@Article(Croll19,
AUTHOR="A. B. Croll and T. Twohig and T. Elder",
TITLE="The compressive strength of crumpled matter",
JOURNAL="Nature Communications",
VOLUME="10",
PAGES="1502",
YEAR="2019")

@Article(Das07,
AUTHOR="Das, M. and Vaziri, A. and Kudrolli, A. and Mahadevan, L.",
TITLE="Curvature Condensation and Bifurcation in an Elastic Shell",
JOURNAL="Physical Review Letters",
VOLUME="98",
PAGES="014301",
YEAR="2007")

@Article(DiDonna02,
AUTHOR="B. A. DiDonna",
TITLE="Scaling of the buckling transition of ridges in thin sheets",
JOURNAL="Physical Review E",
VOLUME="66",
PAGES="016601",
YEAR="2002")

@Article(Ehrhardt20,
AUTHOR="D. Ehrhardt and L. N. Virgin and S. M. Spottswood",
TITLE="Experiments on Probing the Configuration Space of Post-Buckled Panels",
JOURNAL="Journal of Applied Mechanics",
VOLUME="87",
PAGES="121005",
YEAR="2020")

@Article(GiffordScriven71,
AUTHOR="W. A. Gifford and L. E. Scriven",
TITLE="On the attraction of floating particles",
JOURNAL="Chemical Engineering Science",
VOLUME="26",
PAGES="287-297",
YEAR="1971")

@Article(Goldenthal07,
AUTHOR="Goldenthal, R. and Harmon, D. and Fattal, R. and Bercovier, M. and Grinspun, E.",
TITLE="Efficient Simulation of Inextensible Cloth",
journal="ACM Transactions on Graphics",
volume="26",
number="3",
pages="Article 49",
YEAR="2007")

@Article(Gomez17,
AUTHOR="M. Gomez and D. E. Moulton and D. Vella",
TITLE="Critical slowing down in purely elastic `snap-through' instabilities",
JOURNAL="Nature Physics",
VOLUME="13",
PAGES="142-145",
YEAR="2017")

@Article(Gottesman18,
AUTHOR="O. Gottesman and J. Andrejevic and C. H. Rycroft and S. M. Rubinstein",
TITLE="A state variable for crumpled thin sheets",
JOURNAL="Communications Physics",
VOLUME="1",
PAGES="70",
YEAR="2018")

@Article(GrohHunt21,
AUTHOR="R. M. J. Groh and G. W. Hunt",
TITLE="Localization and snaking in axially compressed and internally pressurized thin cylindrical shells",
JOURNAL="IMA Journal of Applied Mathematics",
VOLUME="86",
PAGES="1010-1030",
YEAR="2021")

@Article(Hamm04,
AUTHOR="Hamm, E. and Roman, B. and Melo, F.",
TITLE="Dynamics of developable cones under shear",
JOURNAL="Physical Review E",
VOLUME="70",
PAGES="026607",
YEAR="2004")

@Article(HolstCalladine94,
AUTHOR="J. M. F. G. Holst and C. R. Calladine",
TITLE="Inversion problems in elastic thin shells",
JOURNAL="European Journal of Mechanics, A/Solids",
VOLUME="13",
number="4 suppl.",
PAGES="3-18",
YEAR="1994")

@Article(Hunt20,
AUTHOR="G. W. Hunt and R. M. J. Groh and T. J. Dodwell",
TITLE="Maxwell tipping points: the hidden mechanics of an axially compressed cylindrical shell",
JOURNAL="Proceedings of the Royal Society A",
VOLUME="476",
PAGES="20200273",
YEAR="2020")

@Article(Hutton24,
AUTHOR="R. S. Hutton and E. Vitral and E. Hamm and J. Hanna",
TITLE="Buckling mediated by mobile localized elastic excitations",
JOURNAL="PNAS Nexus",
VOLUME="3",
PAGES="pgae083",
YEAR="2024")

@Article(King12,
AUTHOR="H. King and R. D. Schroll and B. Davidovitch and N. Menon",
TITLE="Elastic sheet on a liquid drop reveals wrinkling and crumpling as distinct symmetry-breaking instabilities",
JOURNAL="Proceedings of the National Academy of Sciences",
VOLUME="109",
number="25",
PAGES="9716-9720",
YEAR="2012")

@Article(Lobkovsky95,
AUTHOR="Lobkovsky, A. and Gentges, S. and Li, H. and Morse, D. and Witten, T. A.",
TITLE="Scaling Properties of Stretching Ridges in a Crumpled Elastic Sheet",
JOURNAL="Science",
VOLUME="270",
NUMBER="5241",
PAGES="1482-1485",
YEAR="1995")

@Article(Loudet05,
AUTHOR="J. C. Loudet and A. M. Alsayed and J. Zhang and A. G. Yodh",
TITLE="Capillary Interactions Between Anisotropic Colloidal Particles",
JOURNAL="Physical Review Letters",
VOLUME="94",
PAGES="018301",
YEAR="2005")

@Article(Nicolson49,
AUTHOR="M. M. Nicolson",
TITLE="The interaction between floating particles",
JOURNAL="Mathematical Proceedings of the Cambridge Philosophical Society",
VOLUME="45",
NUMBER="2",
PAGES="288-295",
YEAR="1949")

@Article(PauchardRica98,
AUTHOR="Pauchard, L. and Rica, S.",
TITLE="Contact and compression of elastic spherical shells: the physics of a `ping--pong' ball",
JOURNAL="Philosophical Magazine B",
VOLUME="78",
NUMBER="2",
PAGES="225-233",
YEAR="1998")

@Article(Poulin97,
AUTHOR="P. Poulin and H. Stark and T. C. Lubensky and D. A. Weitz",
TITLE="Novel Colloidal Interactions in Anisotropic Fluids",
JOURNAL="Science",
VOLUME="275",
PAGES="1770-1773",
YEAR="1997")

@article{RomanPocheau12,
  title={Stress defocusing in anisotropic compaction of thin sheets},
  author={Roman, B. and Pocheau, A.},
  journal={Physical Review Letters},
  volume={108},
  pages={074301},
  year={2012}
}

@Article(SatouFuruya14,
AUTHOR="Y. Satou and H. Furuya",
TITLE="Local Buckling in Crease Induced by Wrapping Fold of Space Membrane",
JOURNAL="Journal of Spacecraft and Rockets",
VOLUME="51",
number="2",
PAGES="595-603",
YEAR="2014")

@Article(Spottswood19,
AUTHOR="S. M. Spottswood and T. J. Beberniss and T. G. Eason and R. A. Perez and J. M. Donbar and D. A. Ehrhardt and Z. B. Riley",
TITLE="Exploring the response of a thin, flexible panel to shock-turbulent boundary-layer interactions",
JOURNAL="Journal of Sound and Vibration",
VOLUME="443",
PAGES="74-89",
YEAR="2019")

@Article(Timounay20,
AUTHOR="Y. Timounay and R. De and J. L. Stelzel and Z. S. Schrecengost and M. M. Ripp and J. D. Paulsen",
TITLE="Crumples as a Generic Stress-Focusing Instability in Confined Sheets",
JOURNAL="Physical Review X",
VOLUME="10",
PAGES="021008",
YEAR="2020")

@Article(VoorheesJohnson88,
AUTHOR="P. W. Voorhees and W. C. Johnson",
TITLE="Development of Spatial Correlations during Diffusional Late-Stage Phase Transformations in Stressed Solids",
JOURNAL="Physical Review Letters",
VOLUME="61",
number="19",
PAGES="2225-2228",
YEAR="1988")

@Article(Walsh11,
AUTHOR="Walsh, L. and Meza, R. and Hamm, E.",
TITLE="Weakening of a thin shell structure by annihilating singularities",
JOURNAL="Journal of Physics D: Applied Physics",
VOLUME="44",
PAGES="232002",
YEAR="2011")

@Article(Witten07,
AUTHOR="Witten, T. A.",
TITLE="Stress focusing in elastic sheets",
JOURNAL="Reviews of Modern Physics",
VOLUME="79",
PAGES="643-675",
YEAR="2007")

@Article(WittenLi93,
AUTHOR="Witten, T. A. and Li., H.",
TITLE="Asymptotic Shape of a Fullerene Ball",
JOURNAL="Europhysics Letters",
VOLUME="23",
NUMBER="1",
PAGES="51-55",
YEAR="1993")

@Article(XuVirgin19,
AUTHOR="Y. Xu and L. N. Virgin",
TITLE="Probing the Force Field to Identify Potential Energy",
JOURNAL="Journal of Applied Mechanics",
VOLUME="86",
PAGES="101008",
YEAR="2019")

\newpage

\appendix

\section{Additional data}\label{data}

\subsection*{Crumple trajectories corresponding to the example data in the main text}

\begin{figure}[h]
    \centering
    \includegraphics[width=0.49\textwidth]{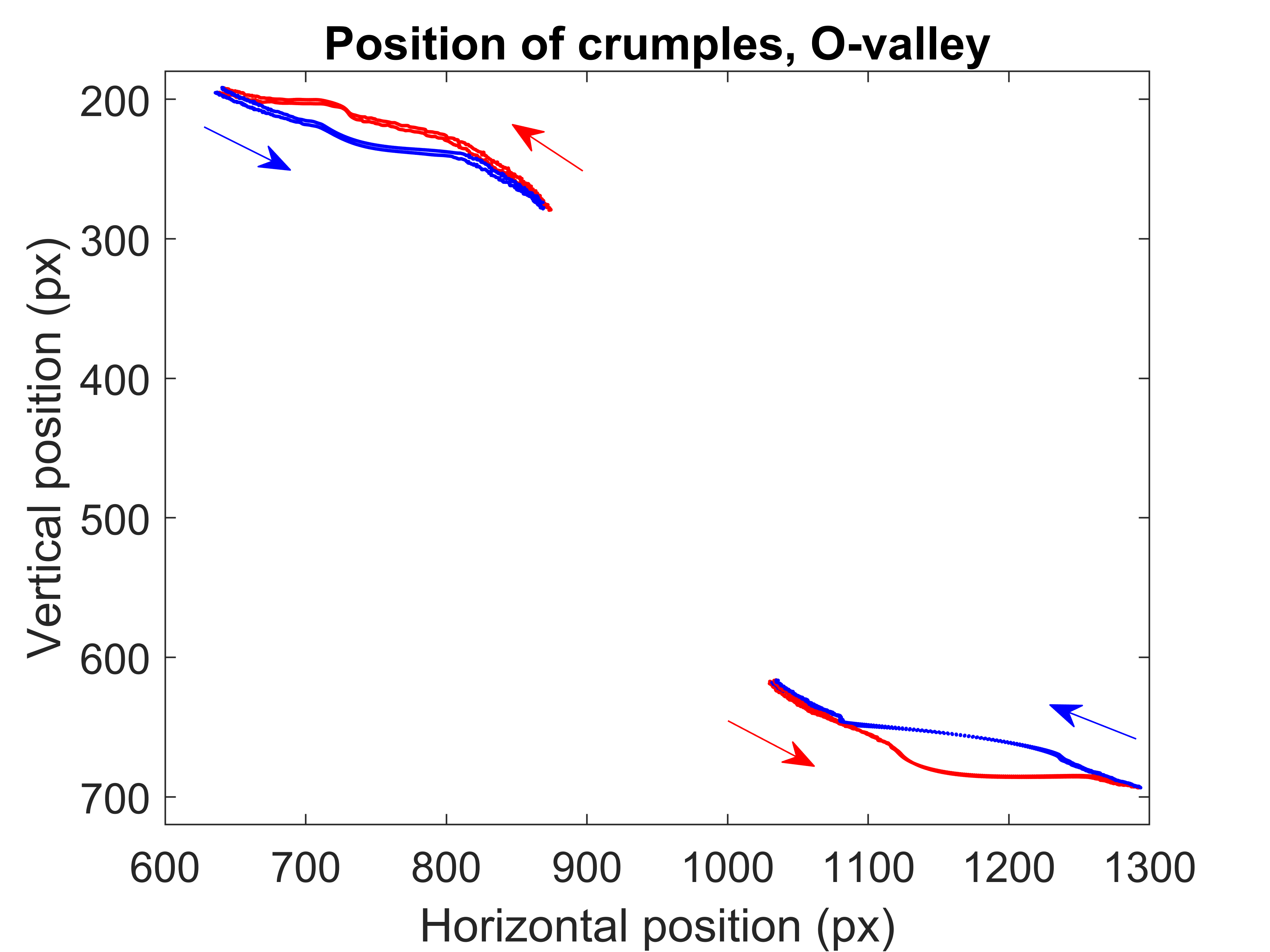}
    \includegraphics[width=0.49\textwidth]{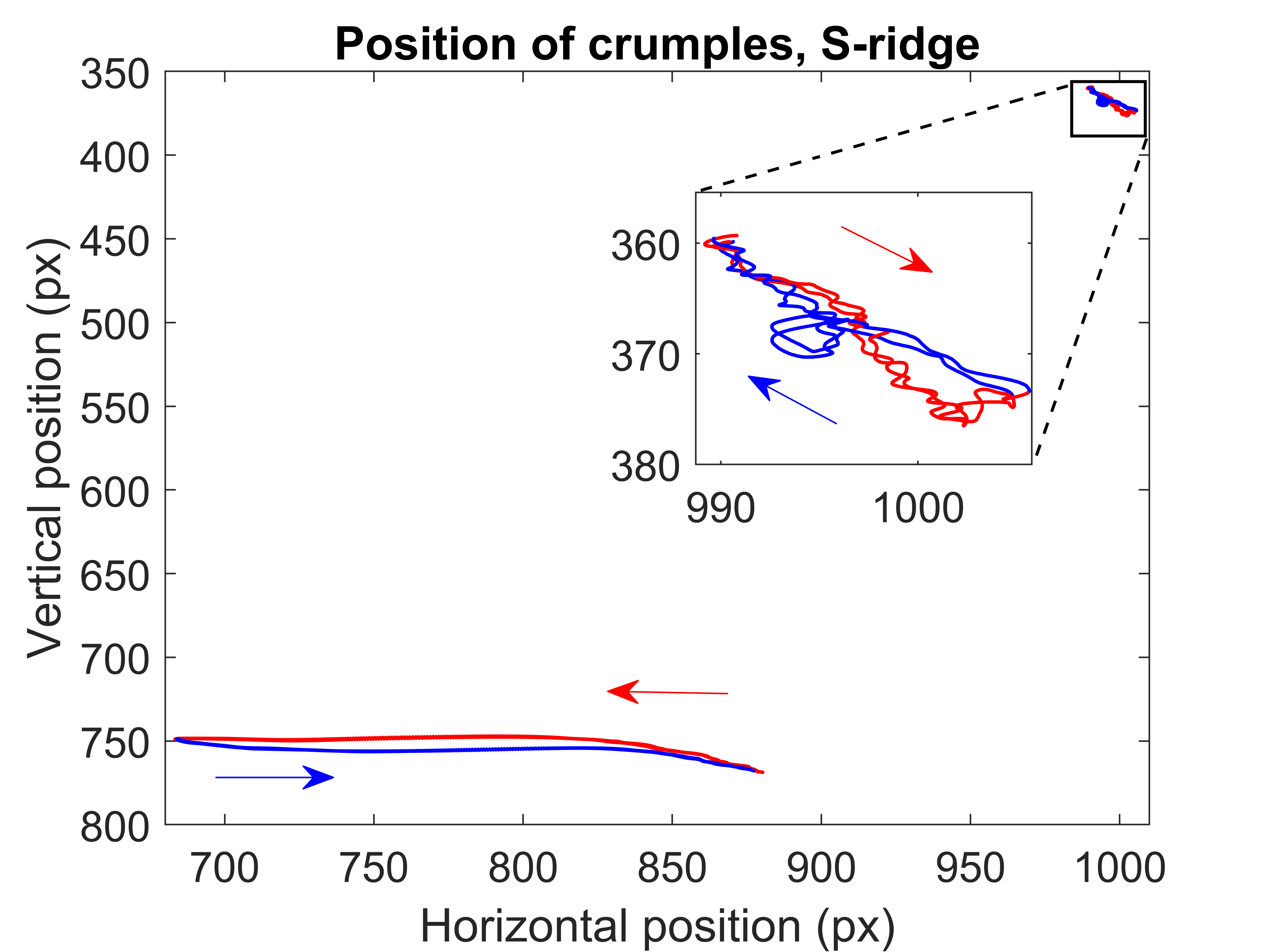}
%	\captionsetup{margin=0.5in}
    \caption{Crumple positions for O-valley and S-ridge corresponding to the data shown in Figure \ref{loops}. These are sheets O7 trials 1 and 2, and S12 trials 8 and 9.} 
    \label{fig:positions}
\end{figure}

Figure \ref{fig:positions} shows the spatial positions of crumples during the experiments (two runs each) corresponding to Figure \ref{loops} of the main text.  As before, about 9px corresponds to 1mm. Note hysteresis in the O-valley trajectories. In the S-ridge example, the motion is primarily of only one crumple, but this behavior varies across different experiments.  The bend around 850 px is related to the transition to crumples passing around each other rather than approaching each other. 

\subsection*{Remaining measurements}

See Appendix~\ref{methods} for further information. 

\begin{figure}[H]
	\includegraphics[width=\textwidth]{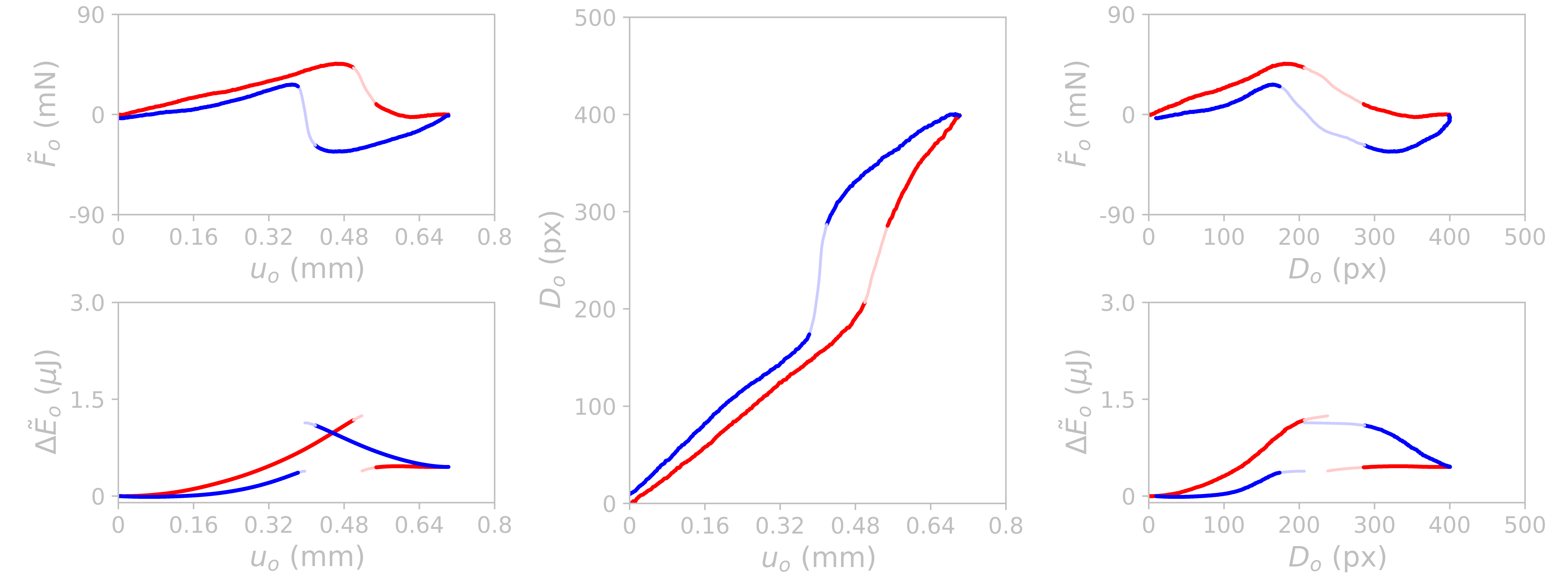}
	\includegraphics[width=0.49\textwidth]{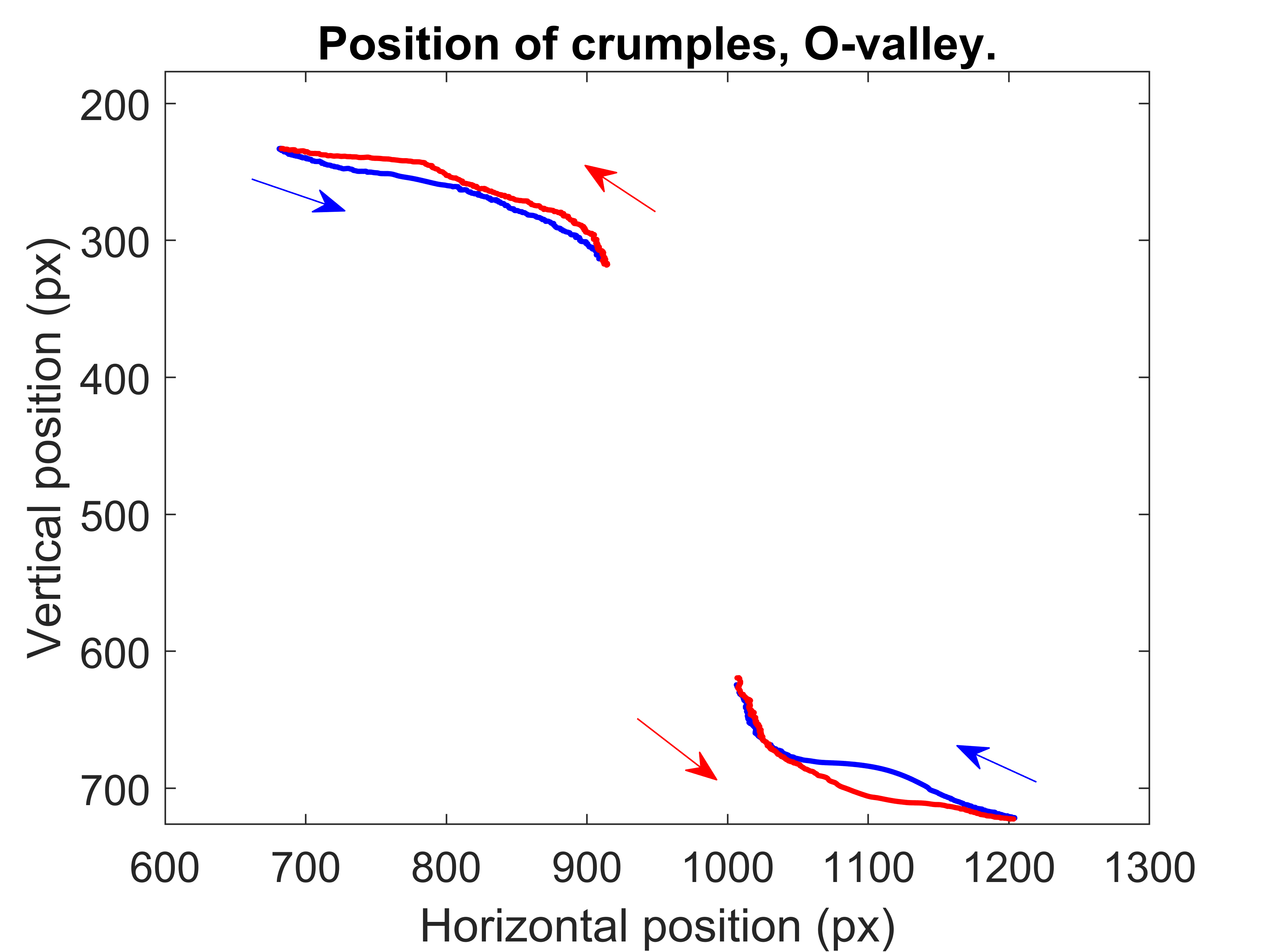}
%	\captionsetup{margin=0.5in}
	\caption{O-valley sheet O6 trial 1.} %60fps
    \label{Osheet6-1}
\end{figure}

\begin{figure}[H]
	\includegraphics[width=\textwidth]{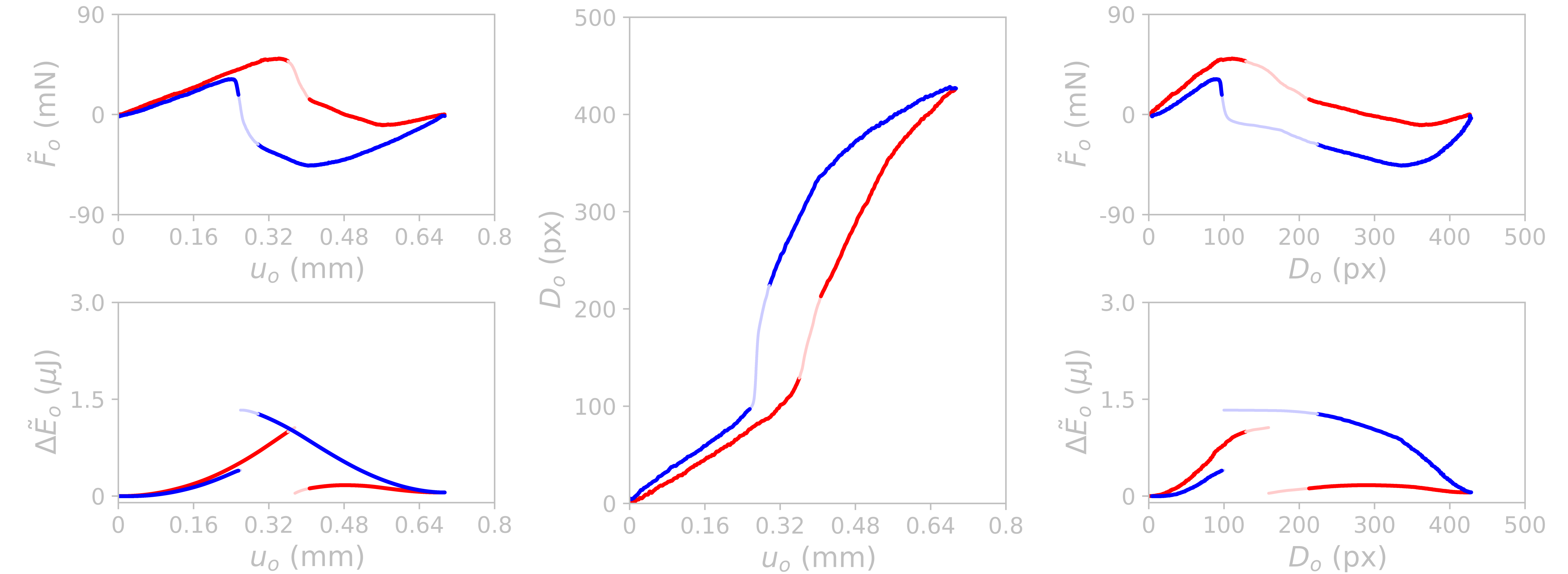}
	\includegraphics[width=0.49\textwidth]{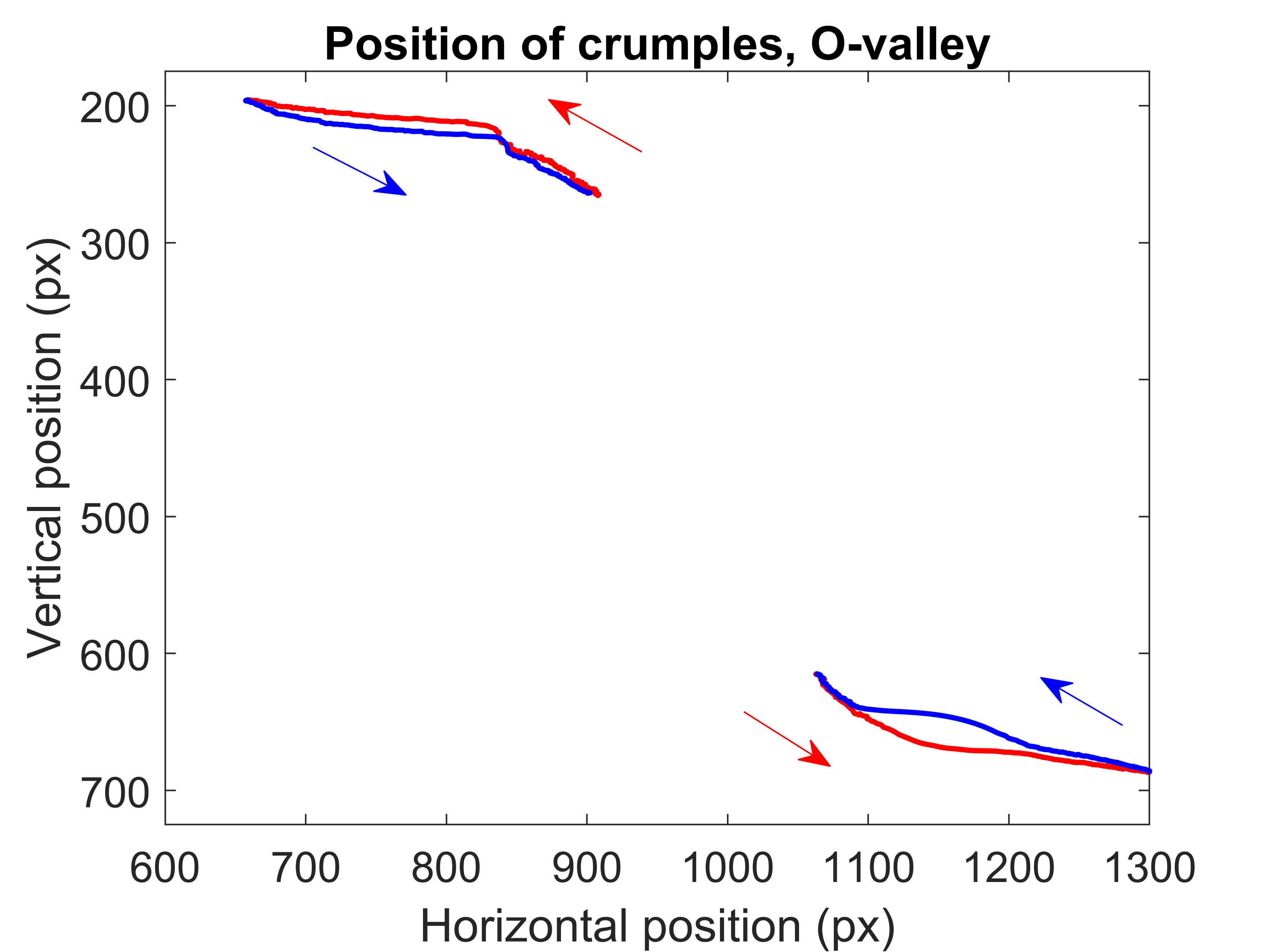}
%	\captionsetup{margin=0.5in}
	\caption{O-valley sheet O6 trial 2.} %60fps
    \label{Osheet6-2}
\end{figure}

\begin{figure}[H]
	\includegraphics[width=\textwidth]{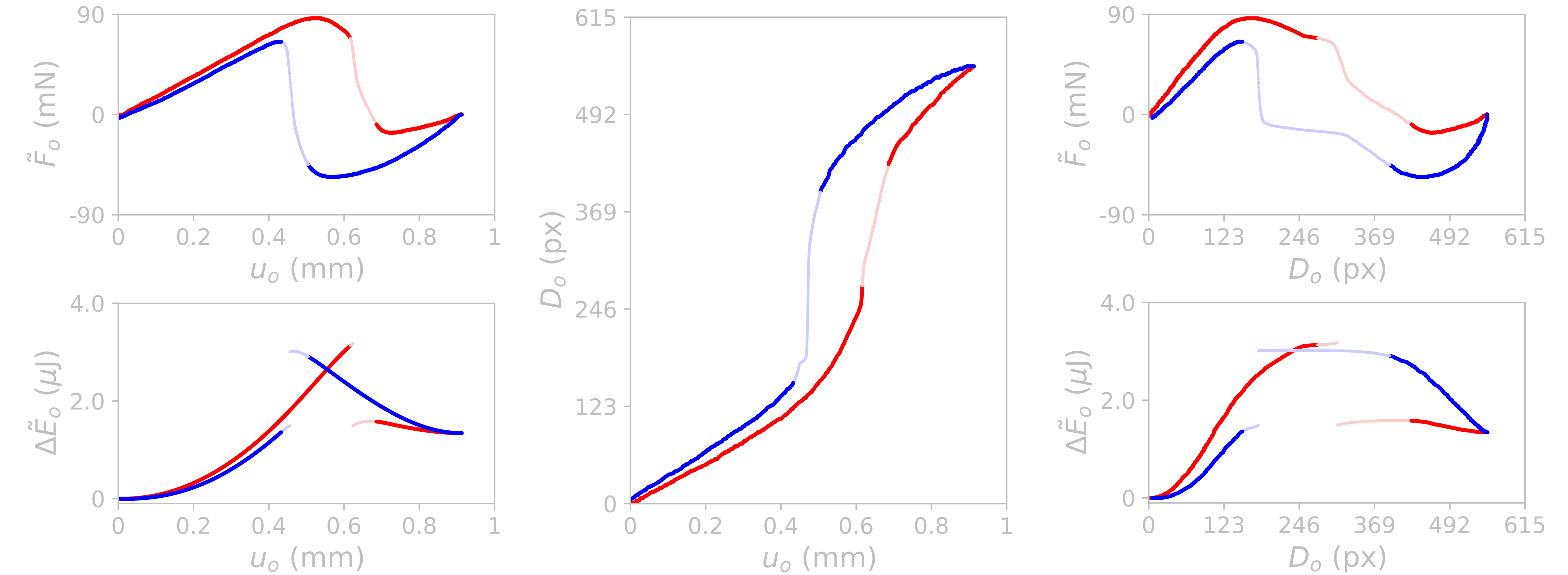}
	\includegraphics[width=0.49\textwidth]{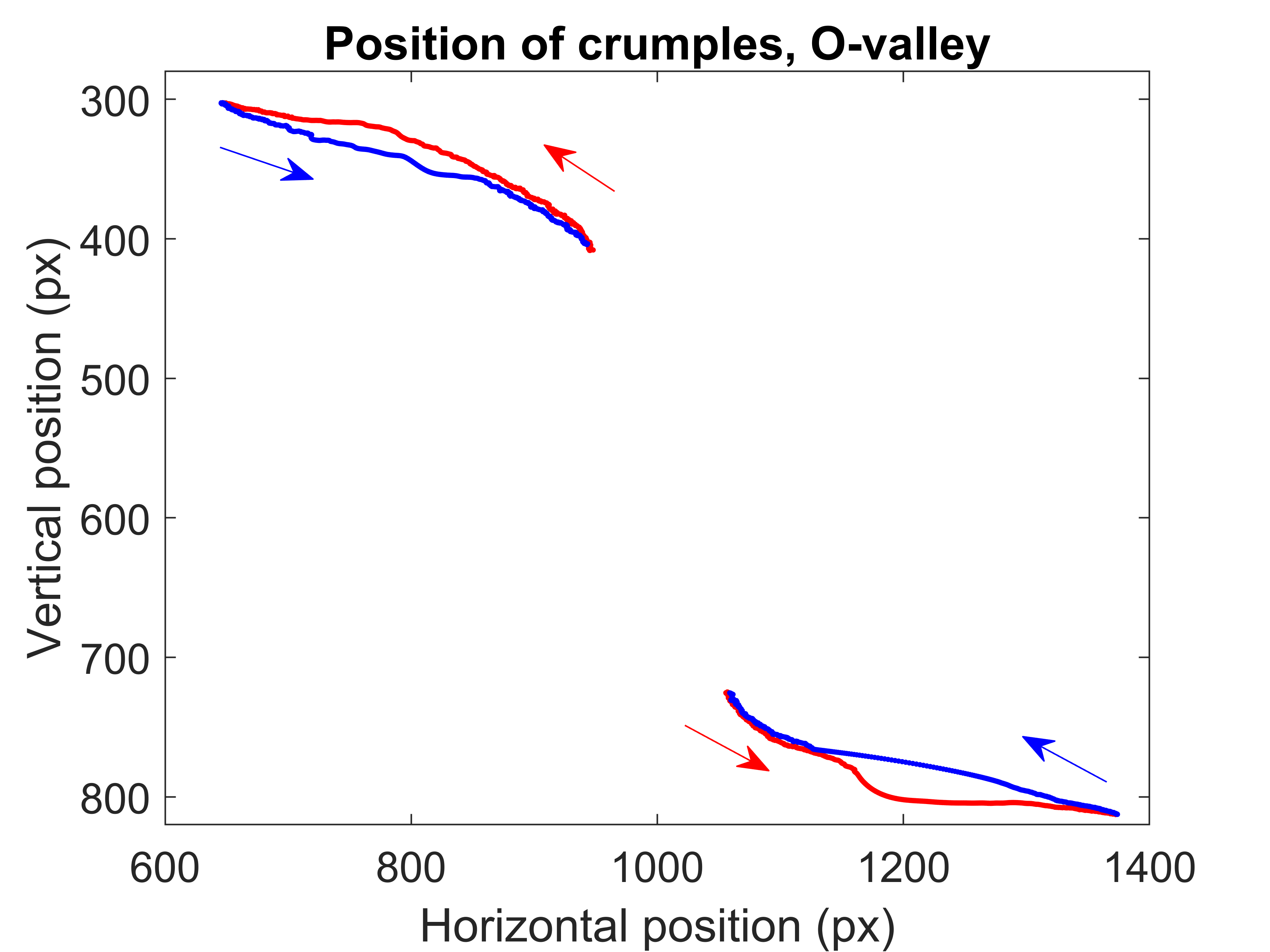}
%	\captionsetup{margin=0.5in}
	\caption{O-valley sheet O7 trial 3. Note Different scale from other O data. Corresponds to Video V6.} %60fps. 
    \label{Osheet7-3}
\end{figure}

\begin{figure}[H]
	\includegraphics[width=\textwidth]{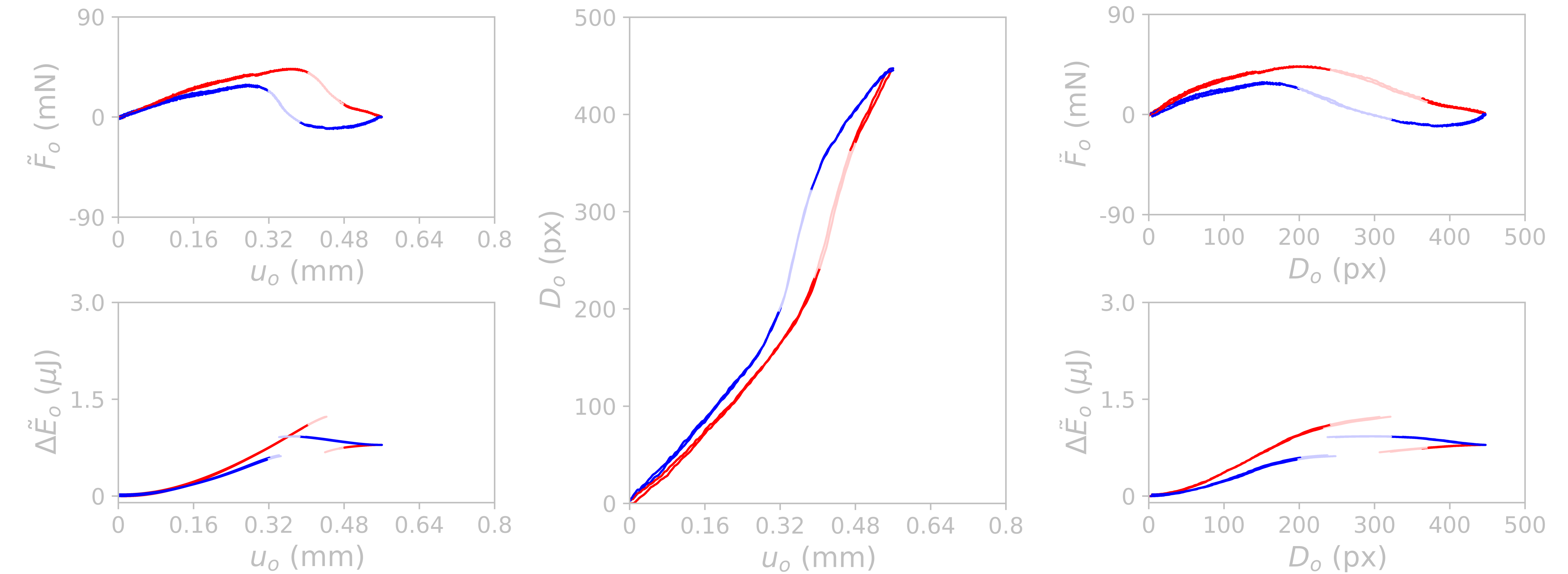}
	\includegraphics[width=0.49\textwidth]{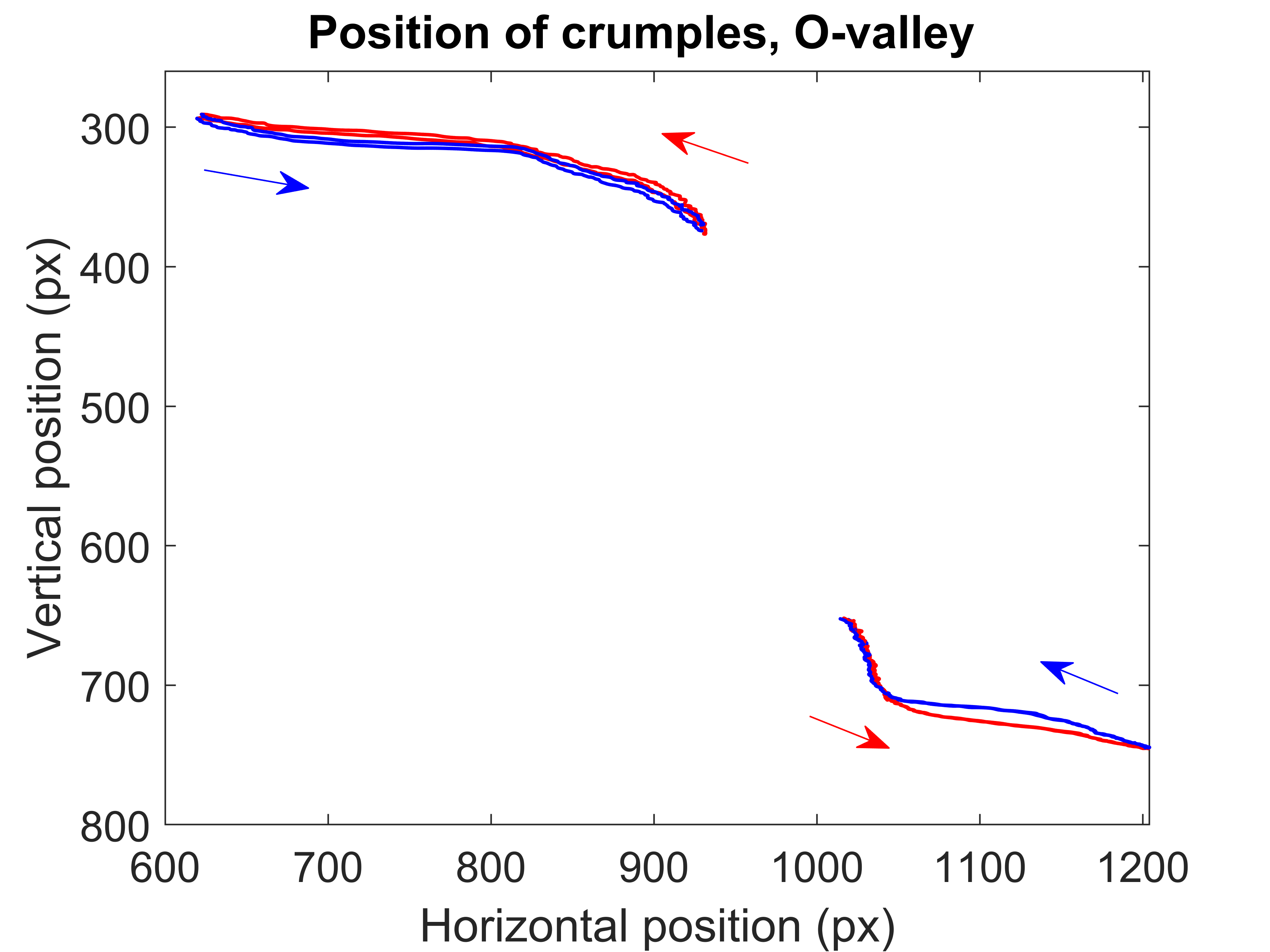}
%	\captionsetup{margin=0.5in}
	\caption{O-valley sheet O28, trials 1 and 2. Energies matched at $u_O=0.56$ mm. A different framerate was used here (see Appendix~\ref{methods}).} %30 fps
    \label{Osheet28-1}
\end{figure}

\begin{figure}[H]
	\includegraphics[width=\textwidth]{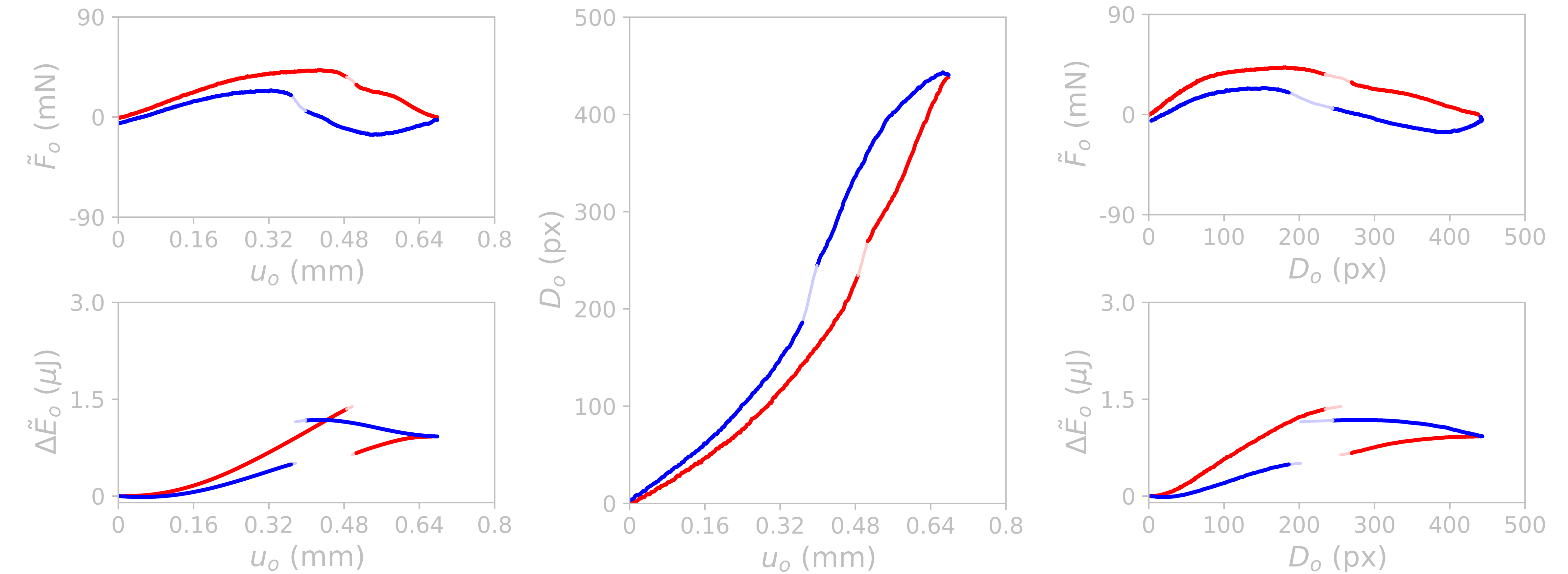}
	\includegraphics[width=0.49\textwidth]{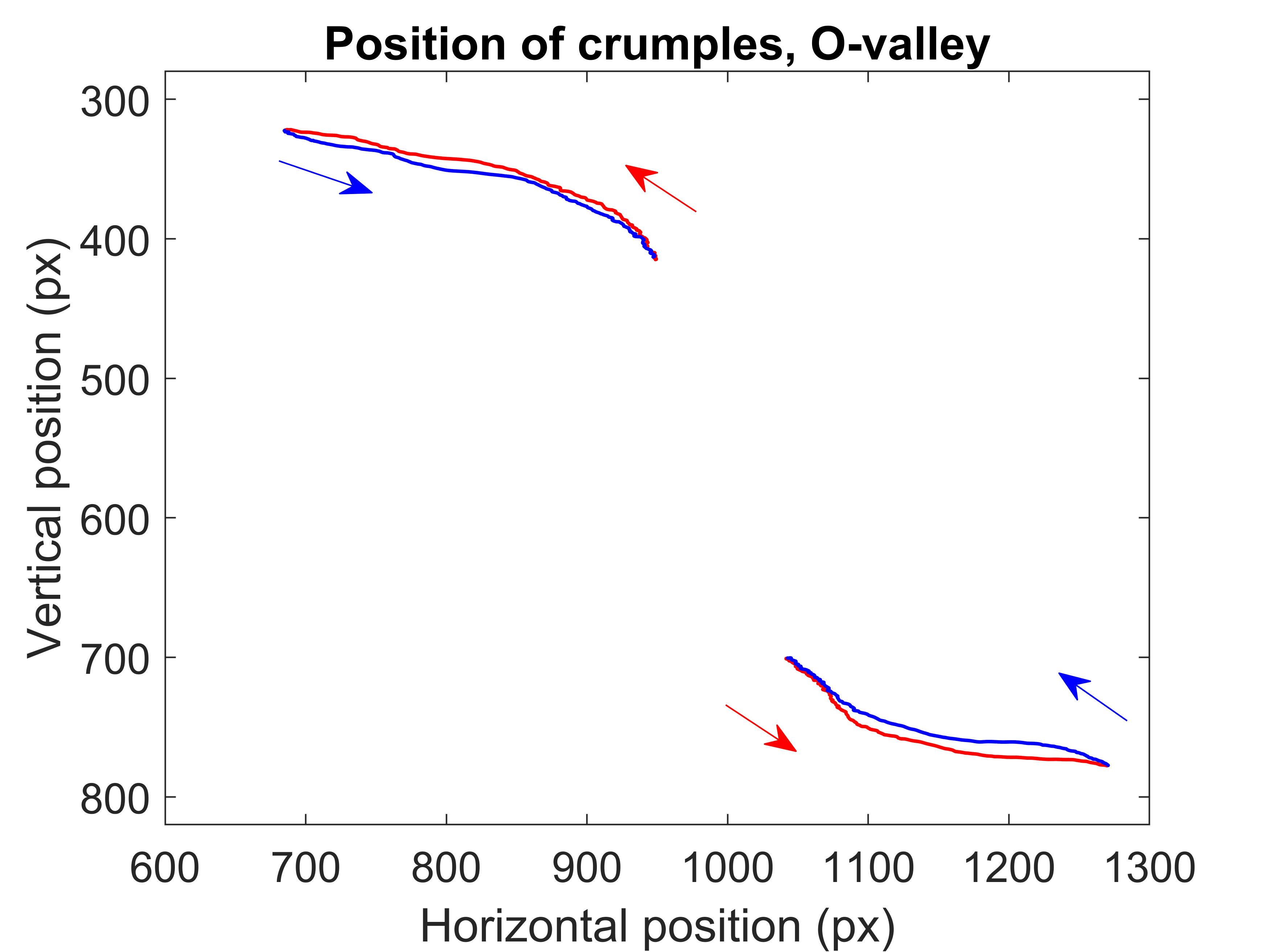}
%	\captionsetup{margin=0.5in}
	\caption{O-valley, sheet O18, trial 1.} %, 60 fps.
    \label{Osheet18-1}
\end{figure}

\begin{figure}[H]
	\includegraphics[width=\textwidth]{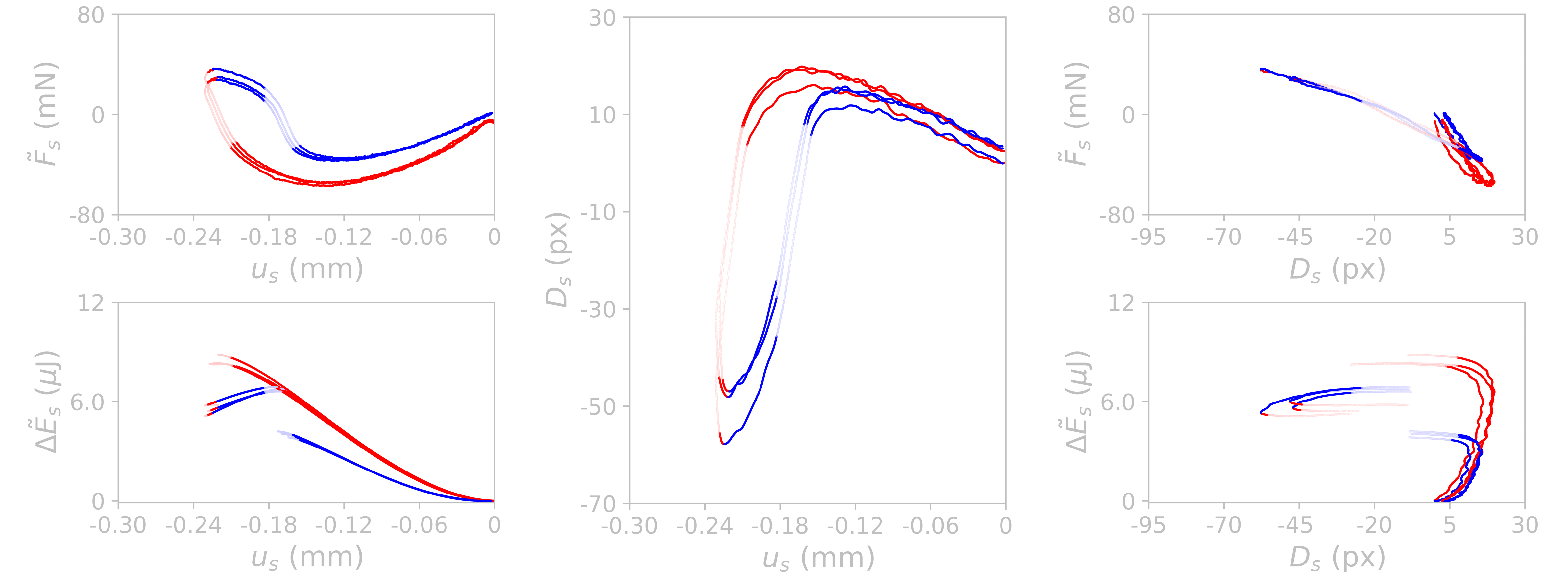}
	\includegraphics[width=0.49\textwidth]{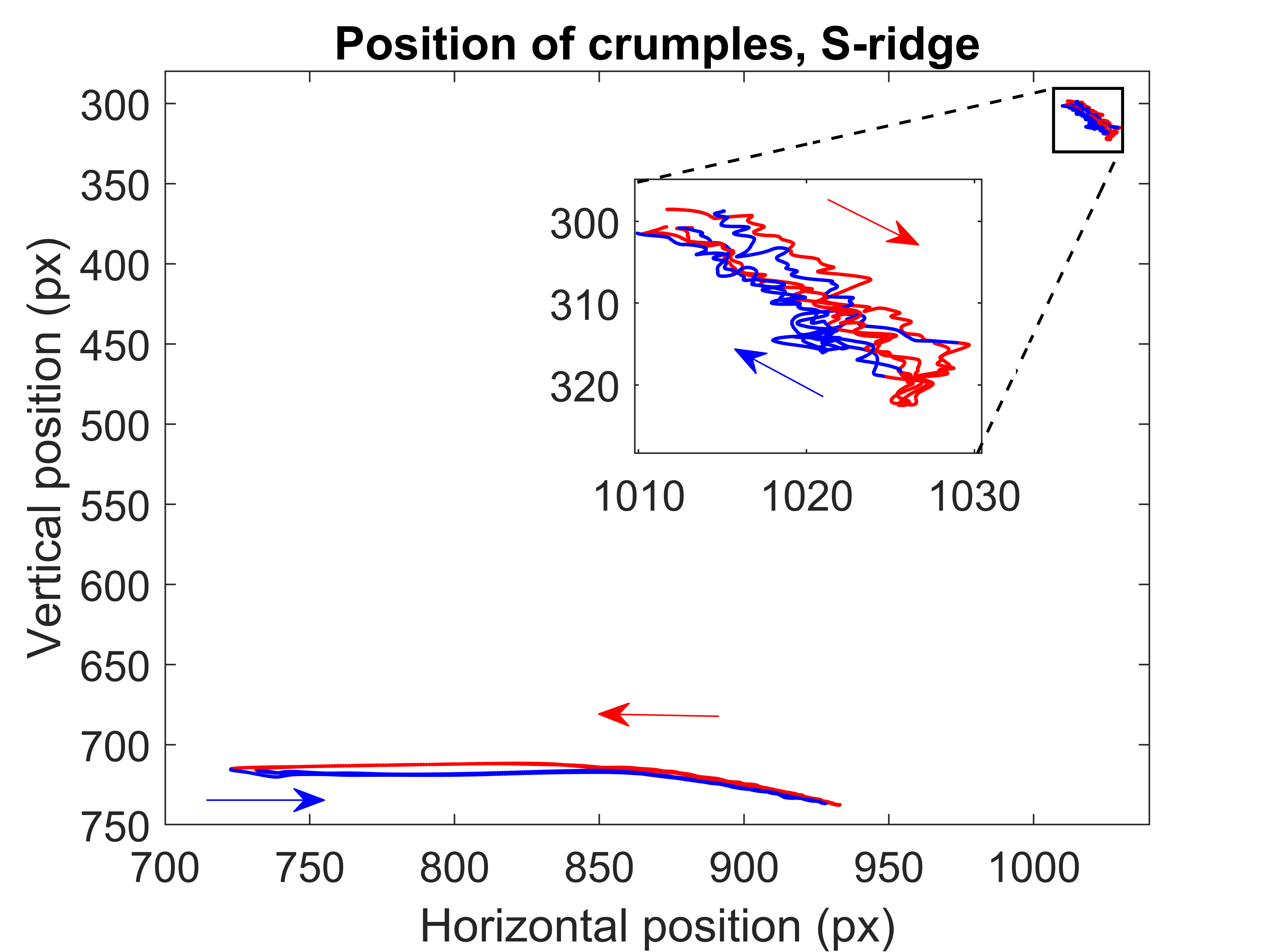}
%	\captionsetup{margin=0.5in}
	\caption{S-ridge sheet S12, trials 5,6 and 7. Energies matched at $u=0$ on the blue curves. Note different scale from other S data. Trial 7 corresponds to Video V7.}%30 fps
    \label{Ssheet12-5}
\end{figure}

\begin{figure}[H]
	\includegraphics[width=\textwidth]{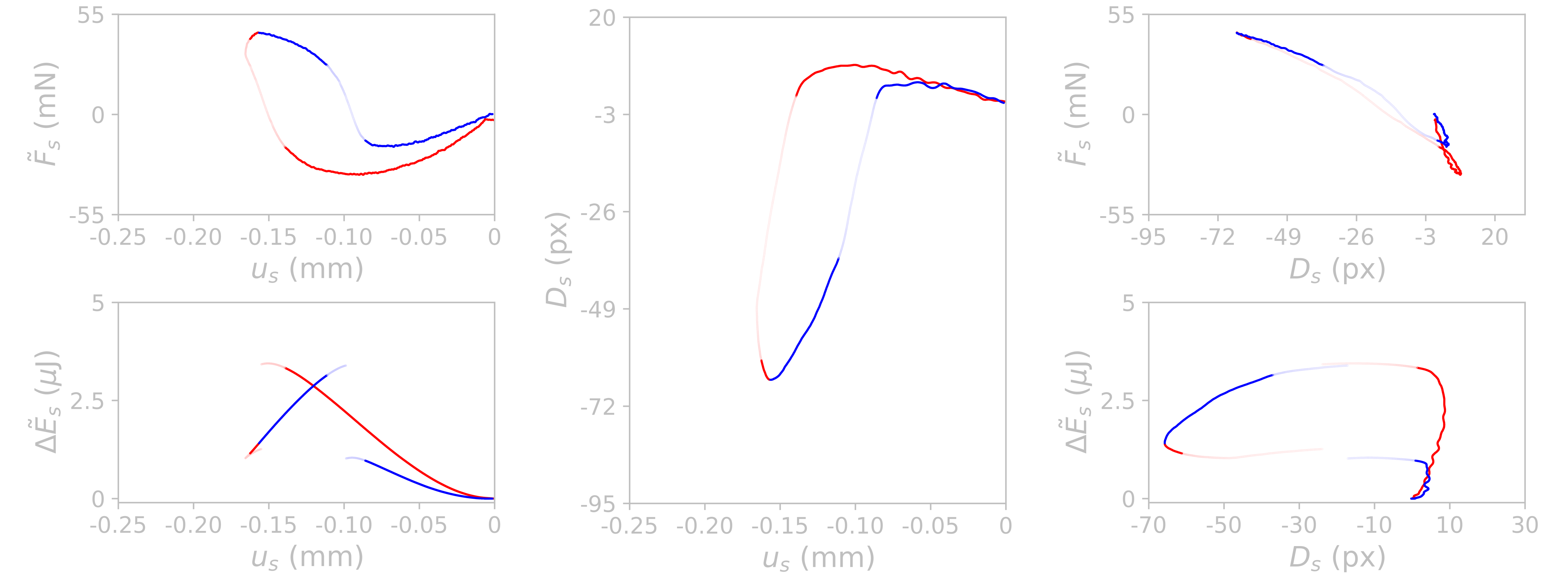}
	\includegraphics[width=0.49\textwidth]{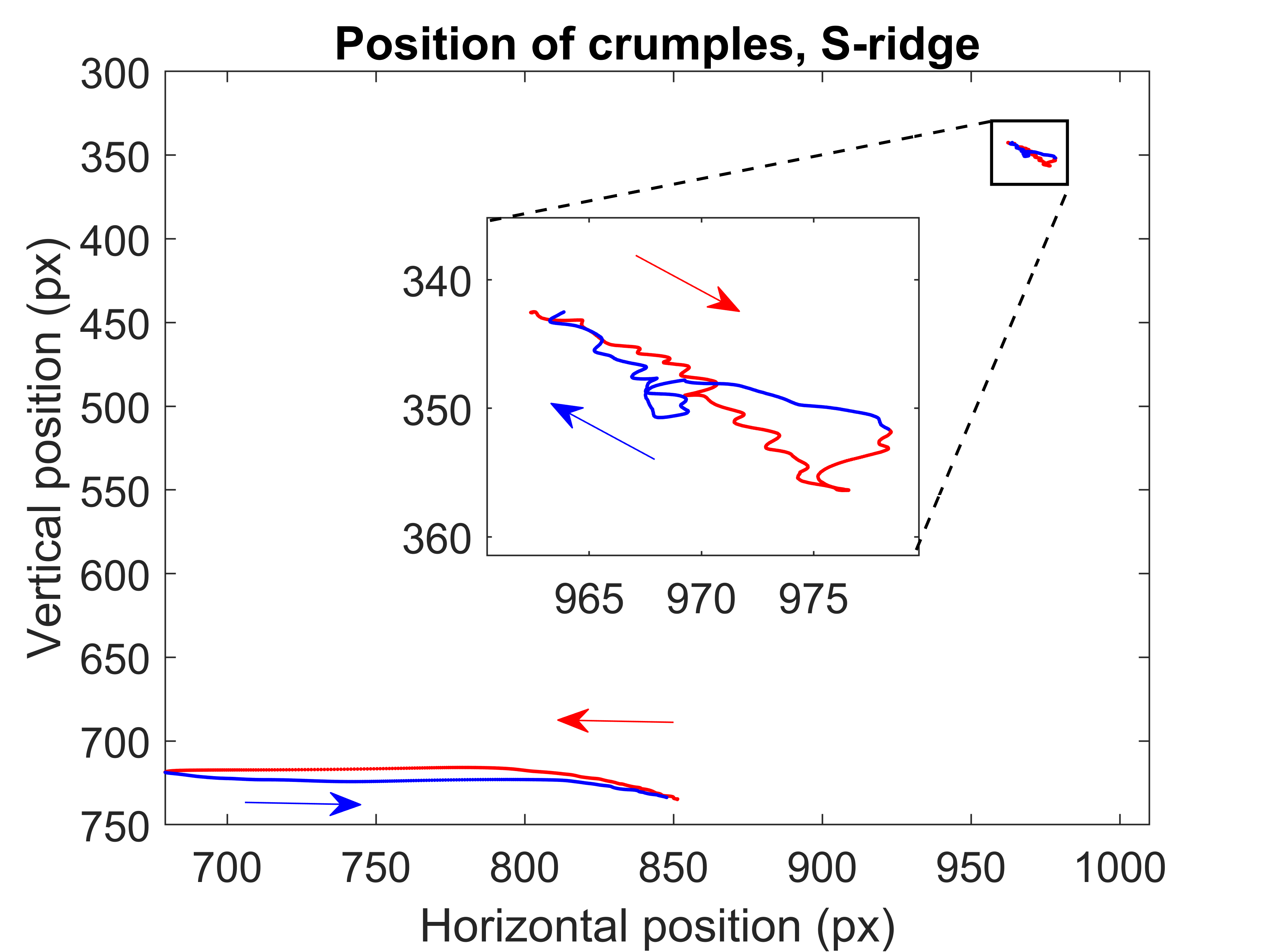}
%	\captionsetup{margin=0.5in}
	\caption{S-ridge sheet S12, trial 13.} %30 fps
    \label{Ssheet12-13}
\end{figure}

\begin{figure}[H]
	\includegraphics[width=\textwidth]{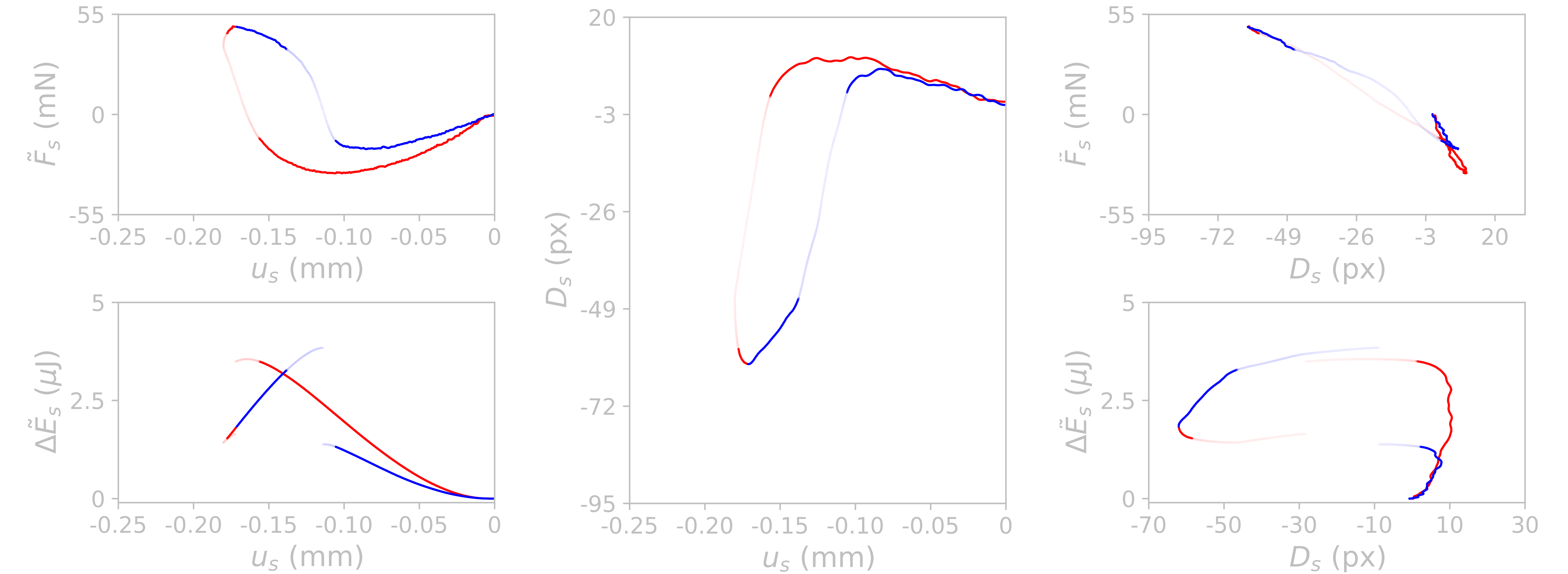}
	\includegraphics[width=0.49\textwidth]{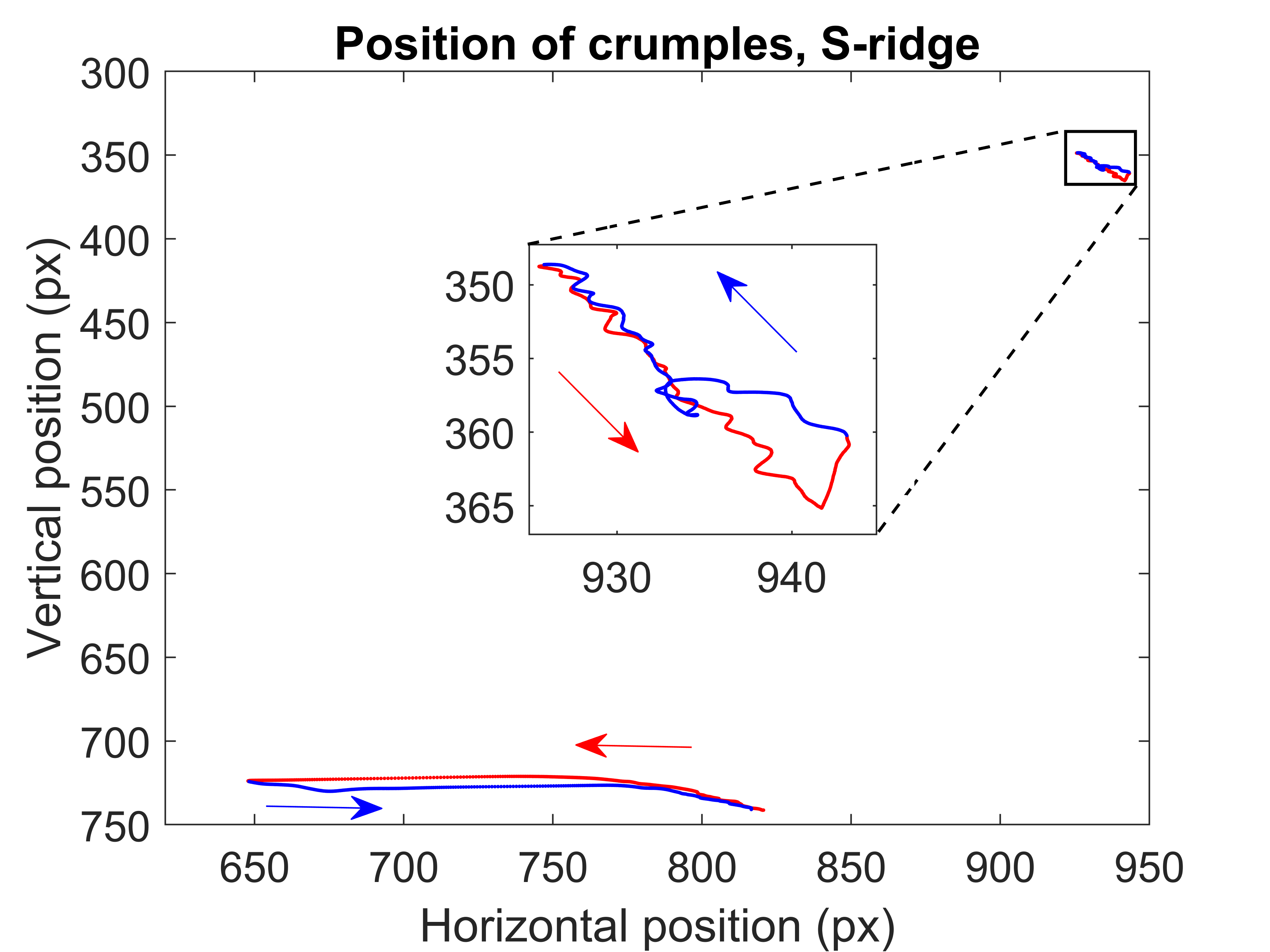}
%	\captionsetup{margin=0.5in}
	\caption{S-ridge sheet S12, trial 14.} %30 fps
    \label{Ssheet12-14}
\end{figure}

\begin{figure}[H]
	\includegraphics[width=\textwidth]{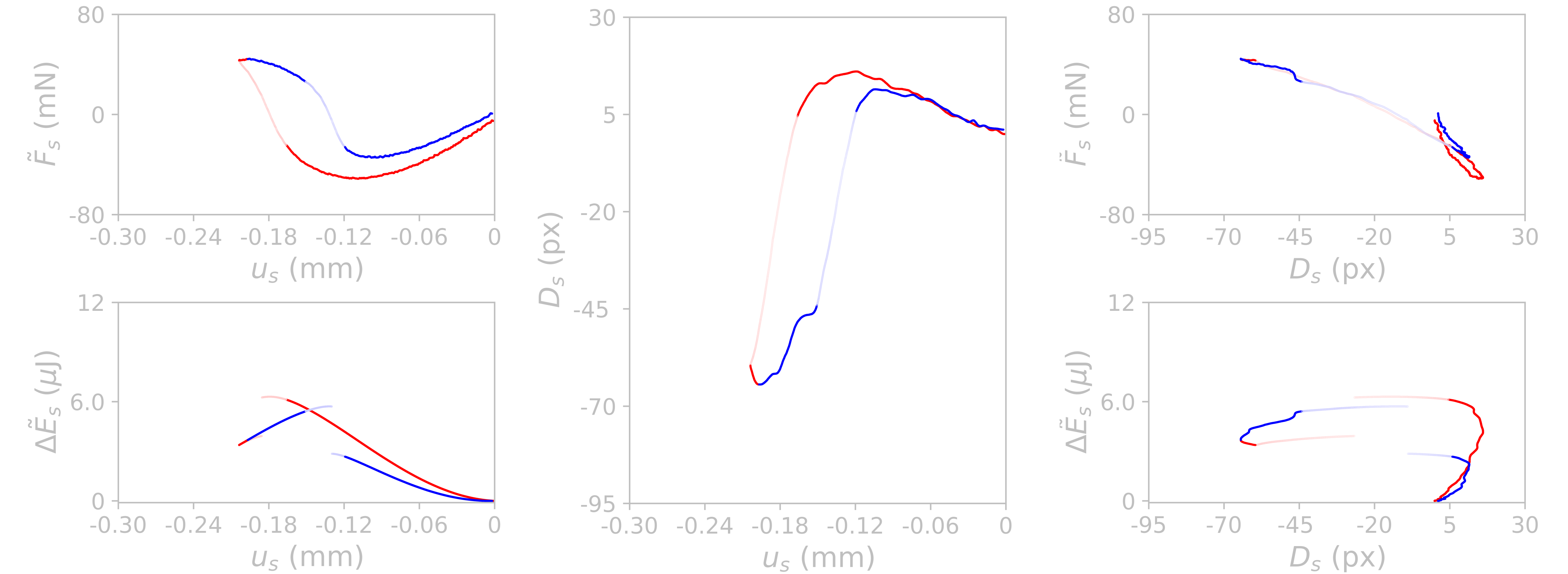}
	\includegraphics[width=0.49\textwidth]{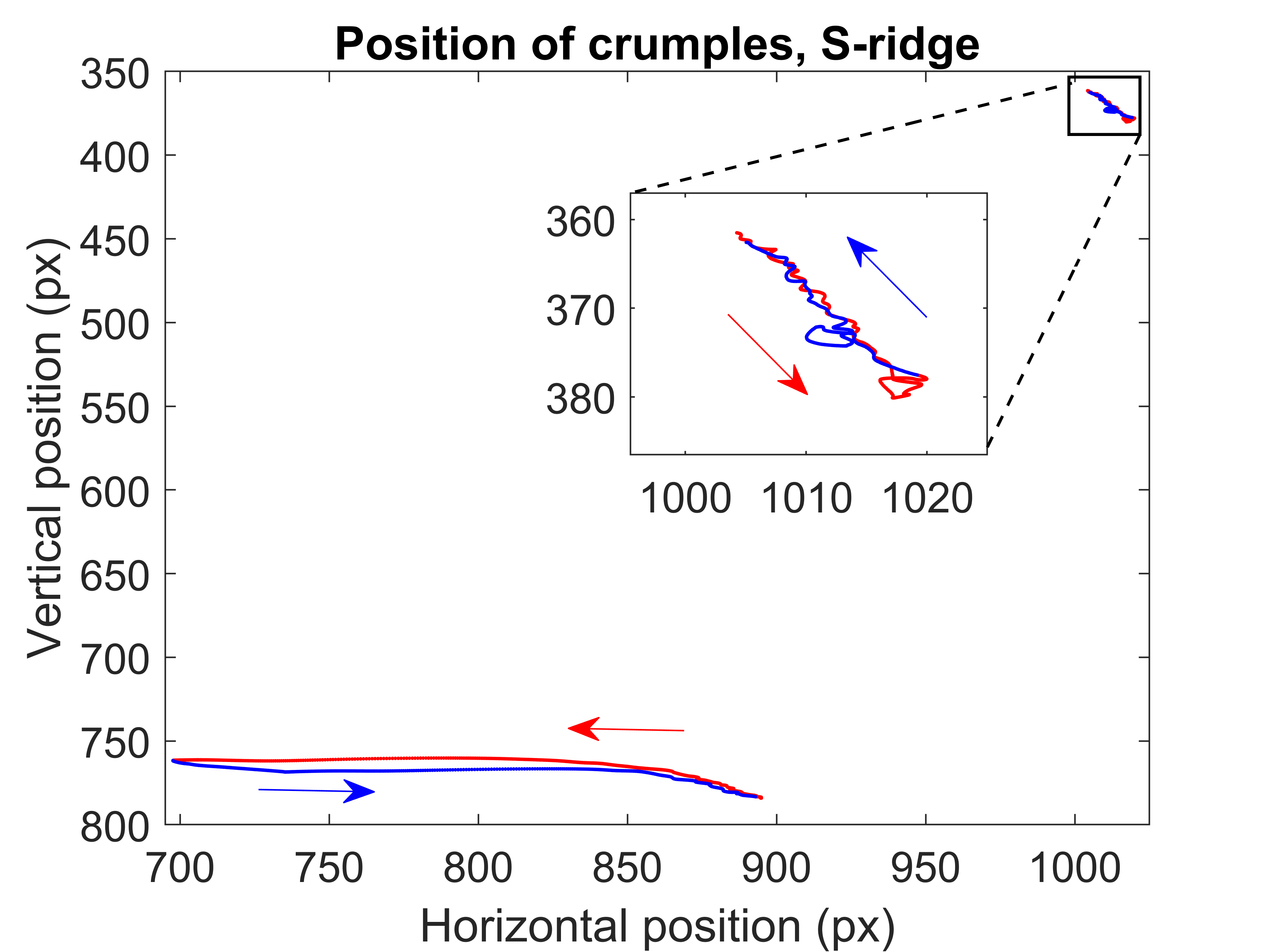}
%	\captionsetup{margin=0.5in}
	\caption{S-ridge sheet S12, trial 15. Note different scale from other S data.} % 30 fps
    \label{Ssheet12-15}
\end{figure}

\begin{figure}[H]
	\includegraphics[width=\textwidth]{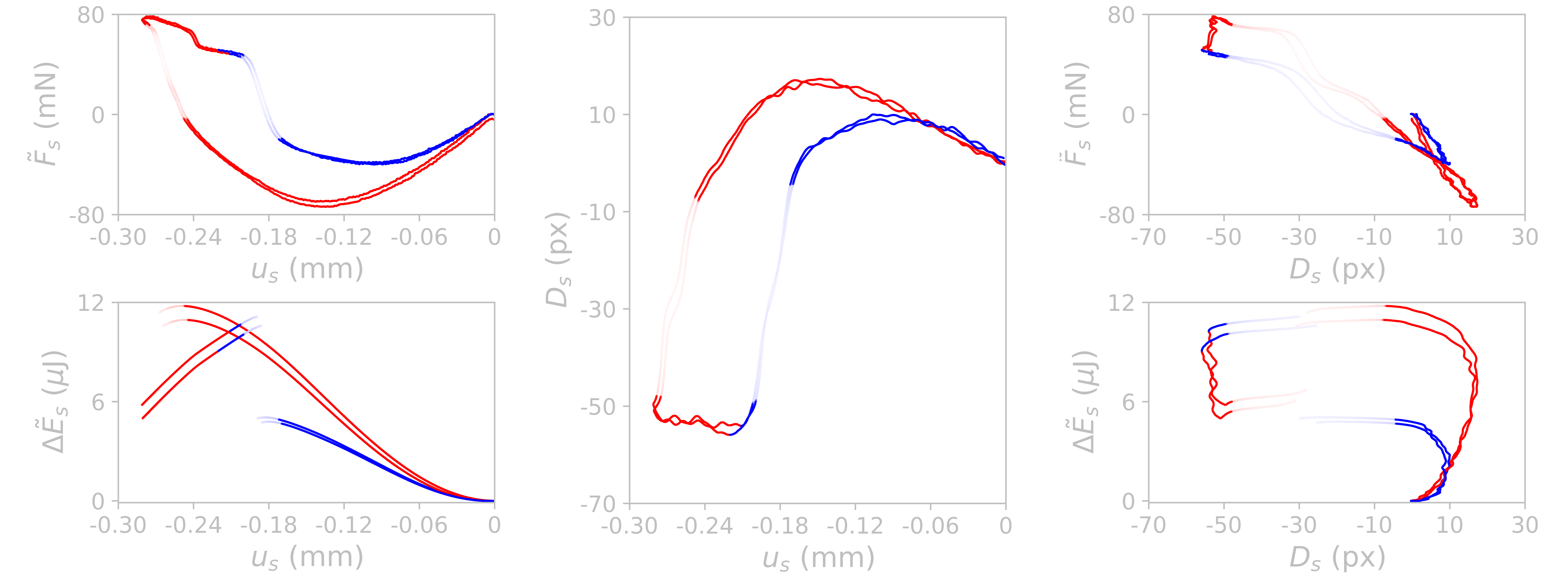}
	\includegraphics[width=0.49\textwidth]{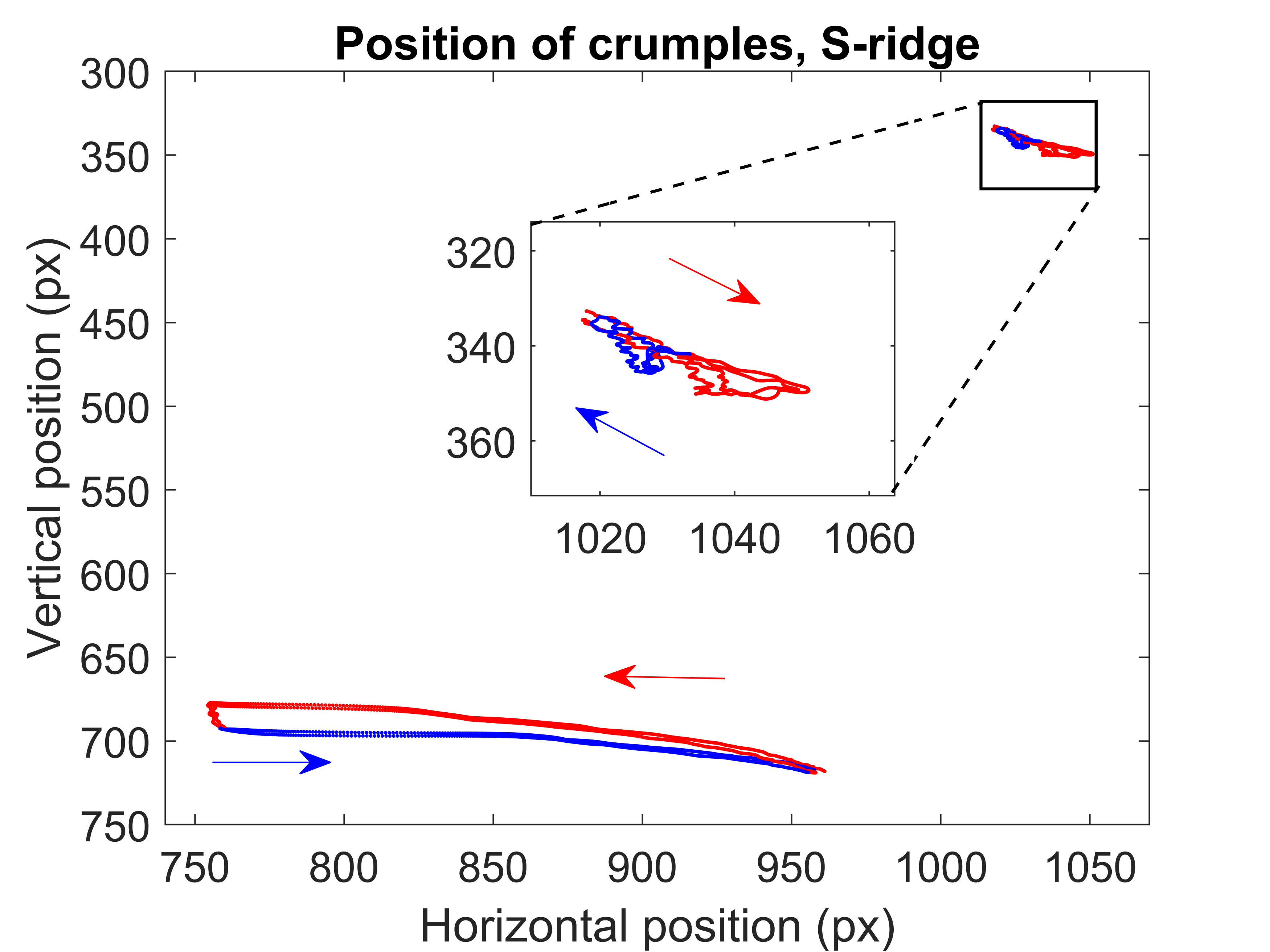}
%	\captionsetup{margin=0.5in}
	\caption{S-ridge sheet S14, trials 1 and 2. Energies matched at $u=0$ on the blue curves. Note different scale from other S data. Behavior differs from other sheets.} %30 fps
    \label{Ssheet14-1}
\end{figure}

\begin{figure}[H]
	\includegraphics[width=\textwidth]{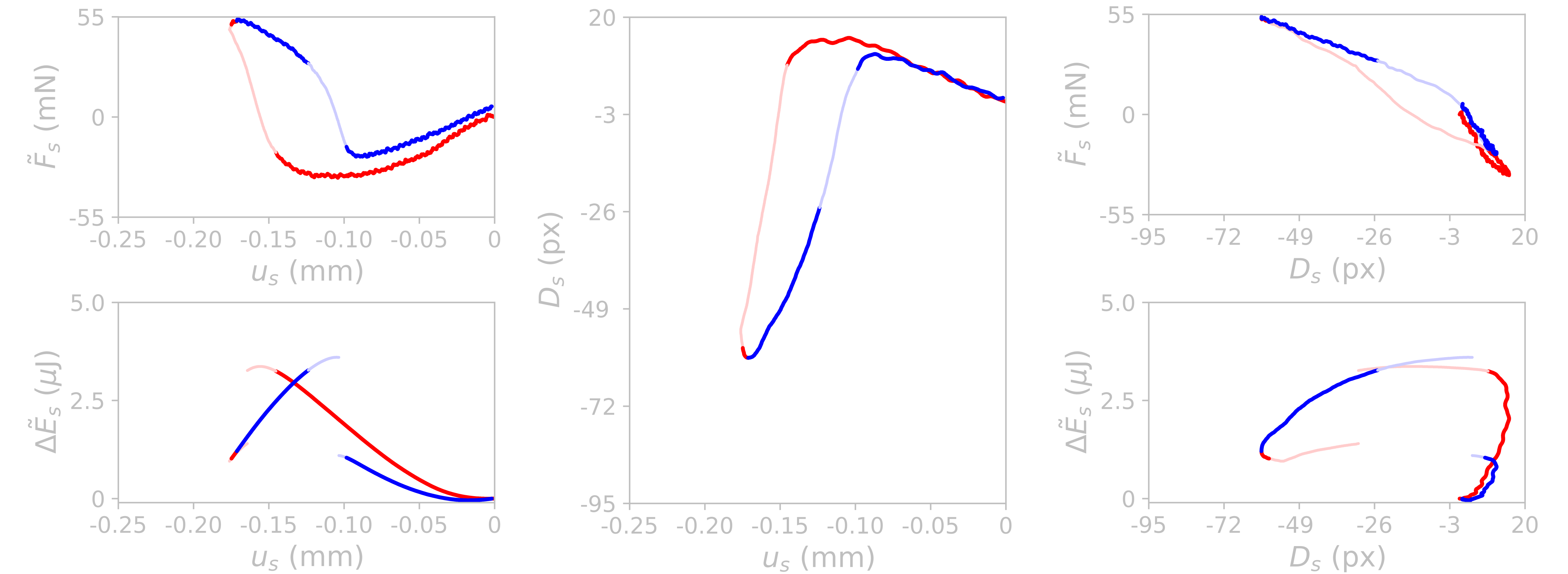}
	\includegraphics[width=0.49\textwidth]{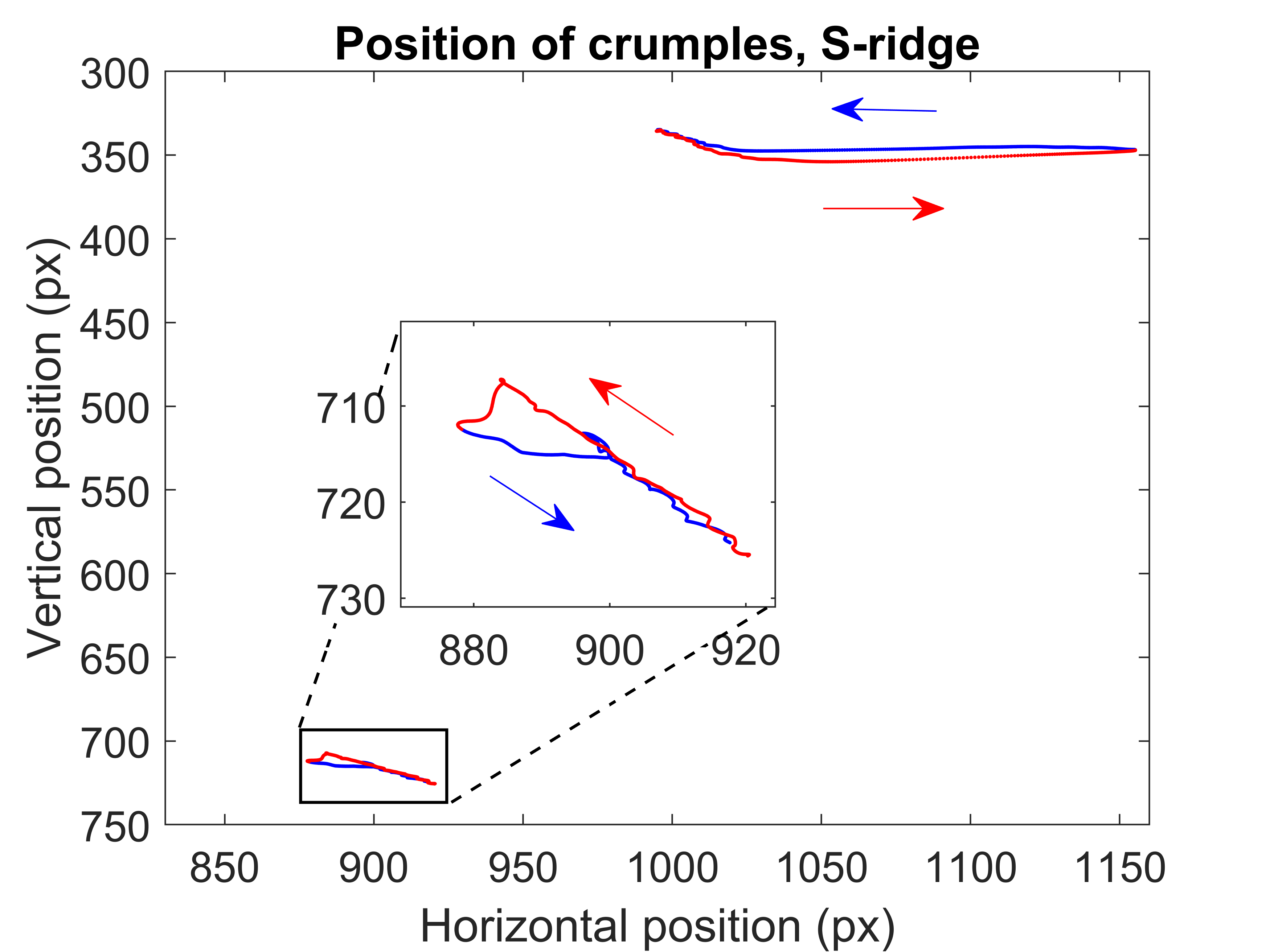}
%	\captionsetup{margin=0.5in}
	\caption{S-ridge sheet S18, trial 1.}%30 fps
    \label{Ssheet18-1}
\end{figure}

\FloatBarrier
\newpage

\section{Materials and Methods}\label{methods}

\subsection*{Experiments: materials and measurements}

Thin polyester sheets (0.003 $\pm$0.0002 inch thick shim stock, Artus Corp., Englewood, NJ, USA) were mounted in a quarter-cylinder configuration with 4 inch free span (2.55 inch radius), their flat sides attached to parallel beams (one rigid and one flexible cantilever). O-valley pairs were measured in a 6 inch long sheet, as additional O-valleys would form in longer sheets. S-ridge pairs were measured in a 12 inch long sheet.  As the material is mildly anisotropic, sheets were oriented with their machine direction parallel to the beams.  Sheets with marks (plastic indentations that might pin a crumple) were avoided. Sheets were attached to the beams using rubber adhesive polypropylene cloth mounting tape, then secured with an additional clamping plate and corner clamps to prevent edge peeling. 
The cantilever was mounted on a translation stage (Thorlabs, Inc., Newton, NJ, USA). 
Alignment was achieved with additional positioning stages, and lateral displacement was guided by four stabilizers to prevent twisting or deflection that could occur due to the inclusion of a soft pad between the translation stage and cantilever to damp external vibrations.  Still, errors on the order of 1-2 mm (half a percent of the sheet dimensions) are expected. Small differences in clamping and alignment from sheet to sheet had noticeable effects on the results. 

The samples were quasistatically sheared by laterally displacing the cantilever boundary. The cantilever is machined from a 24 inch long, 1 inch thick aluminum plate, leaving two thin beams approximately 5.31 inch long and 0.05 inch thick. A mild iron target was attached to one beam to enable measurement of deflection using an inductive sensor (Baumer Ltd., Southington, CT, USA), attached to a soldered external piece bolted to the cantilever. 

The stage-cantilever-sensor arrangement was calibrated by pushing against a rigid object, and by pushing against a load cell (Sentran LLC, Ontario, CA, USA). This allowed us to relate the position of the stage, the deflection of the cantilever, and the force applied by the cantilever.  Displacements reported are those of the sheet edge mounted on the deflected end of the cantilever, not the base of the cantilever mounted on the stage. The cantilever is operated in a linear force-deflection regime, with sub-millimeter deflections. From the deadload measurements, we also determined that the stage backlash can be empirically corrected throughout the relevant load regime by applying a 5 micron shift between forward and backward load curves.  This is consistent with manufacturer-supplied accuracy and backlash values. Force data from the sensor had a noisy width estimated as 0.008 Newton, and was smoothed using a Gaussian filter (Matlab smoothdata function which has an SD of 1/5 of window, window 50, acquired at 100 Hz). The filtering introduces an artifact that the fast-moving (unstable) data for the O-valley are spread over an interval comparable to the error in position. Sensor drift of the same order as the noise was occasionally observed over periods spanning tens of minutes. We estimate the associated drift width as 0.008 Newton. The noise and drift error estimates are consistent with conservative bounds on root mean square deviation of unfiltered and filtered signals over tens to hundreds of minutes over the range of voltages used. We did not attempt to determine lower estimates for \emph{e.g.} low voltages which display lower drift.

\begin{figure}[ht!]
    \centering
    \includegraphics[width=0.4\linewidth]{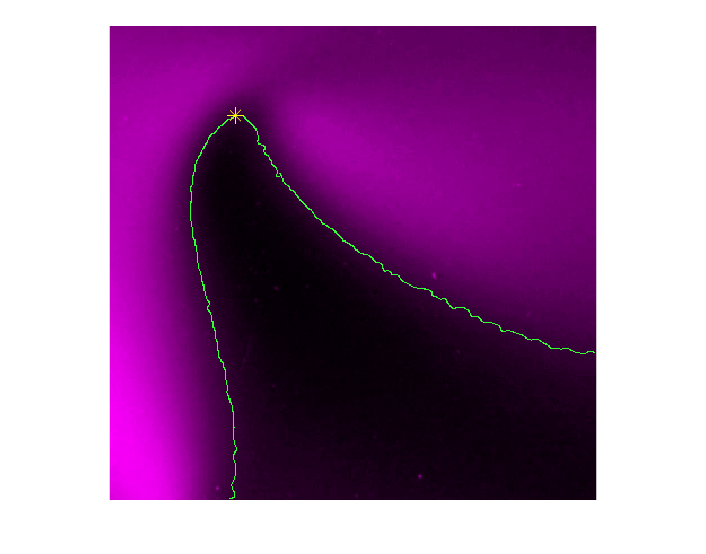}
%	\captionsetup{margin=0.5in}
    \caption{Example showing crumple tip identification (star).}
    \label{fig:tracking}
\end{figure}

Crumples are filmed with a Nikon D5300 camera (with AF-S DX NIKKOR 18-105mm f/3.5-5.6 G ED VR lens) positioned about two feet above the sheet when in a cylindrical configuration, leading to a correspondence of approximately 9 pixels per millimeter.  Note that the region of deformed sheet being filmed is generally flatter than a cylinder. Sheets are illuminated from the side at a consistent angle to enable crumple identification. Crumple positions are tracked using a binary filter and Canny edge detection, followed by identification of the edge pixel closest to the top of the frame, which in our experimental geometries always corresponds to the sharp tip of the crumple (see Figure \ref{fig:tracking}). Position data had a noisy width of 10 pixels, and was smoothed using a Gaussian filter (Matlab smoothdata function which has an SD of 1/5 of window, window 100, S-ridge data acquired at 30 Hz and O-valley data usually acquired at 60 Hz with the exception of Sheet O28 which was acquired at 30 Hz). The filtering may introduce artifacts in the fast-moving (unstable) regions. Distance data were then upsampled to match the force data rate. Films are synchronized to force data to within one frame.

\subsection*{Experiments: procedures for pair interactions}

To manipulate the crumples, the stage is moved at a speed of 0.004 mm/s, with accelerations of 1 mm/s$^2$ at the beginning and end of the loops.  Executing a single S-ridge loop takes about two minutes, an O-valley loop about 8 minutes. 
We zero the sensor (corresponding to the calibration zero) on the cantilever without any sample attached, then attach and clamp a sheet and move the stage until the sensor is zeroed again. This serves as a reference symmetric unsheared ``U'' state of the sheet.  Data is acquired after ``warming up'' the system by cycling it around the relevant measurement loop approximately ten times at a speed of the same order of magnitude as that used for data collection. 

Observations of single O-valley pairs were performed in a 6 inch long sheet, which was a compromise between the tendency for induction of additional valley pairs in long sheets and edge-entering crumples in shorter sheets. Crumple approach was limited by an instability to merging and annihilation. Crumple separation was limited by the induction of crumples from the edges. The O-valley is created by poking a mildly sheared configuration. Increasing/decreasing shear will cause crumples to separate/approach (unbind/bind).  The loops are bounded by the collapse of the valley on the approach side ($u_O=0$), and the induction of new edge crumples on the separation side. 

Observations of single S-ridge pairs were performed in a 12 inch long sheet.  Crumple separation was limited by the nucleation of a new valley pair and concomitant repulsive splitting into two S-ridges.  Crumple approach leads eventually to the crumples passing around each other and adjusting their opening angles.  Only a small range of stable approach and separation of S-ridge pairs can be observed. The S-ridge is created by shearing the structure through several bifurcations until a single ridge remains.  Decreasing/increasing shear will cause crumples to separate/approach (unbind/bind).  On the approach side, we terminate the loop when the S-ridge is in an approximate vertical orientation ($u_S=0$). The loops are limited on the separation side by the splitting that occurs shortly after ridge breaking and repulsion is observed. To prevent the splitting, we reverse the stage motion just as the system is slowing down while unstably separating. This results in brief retrograde motion of the crumples with respect to the boundary, in the red unbinding portion of the plotted curves, as the crumples continue to separate as the boundary reverses direction. We are able to confirm stability of the unbound state by pausing motion shortly after stage reversal; both intercrumple distance and force remain at a constant value for up to 70 seconds.  Similar observations were made in a few other stable S and O configurations.  However, our determination of stability or instability along the measured loops was based on the observed speed of crumples, as explained in the following subsection. 

\begin{figure}[H]
	\includegraphics[width=0.49\textwidth]{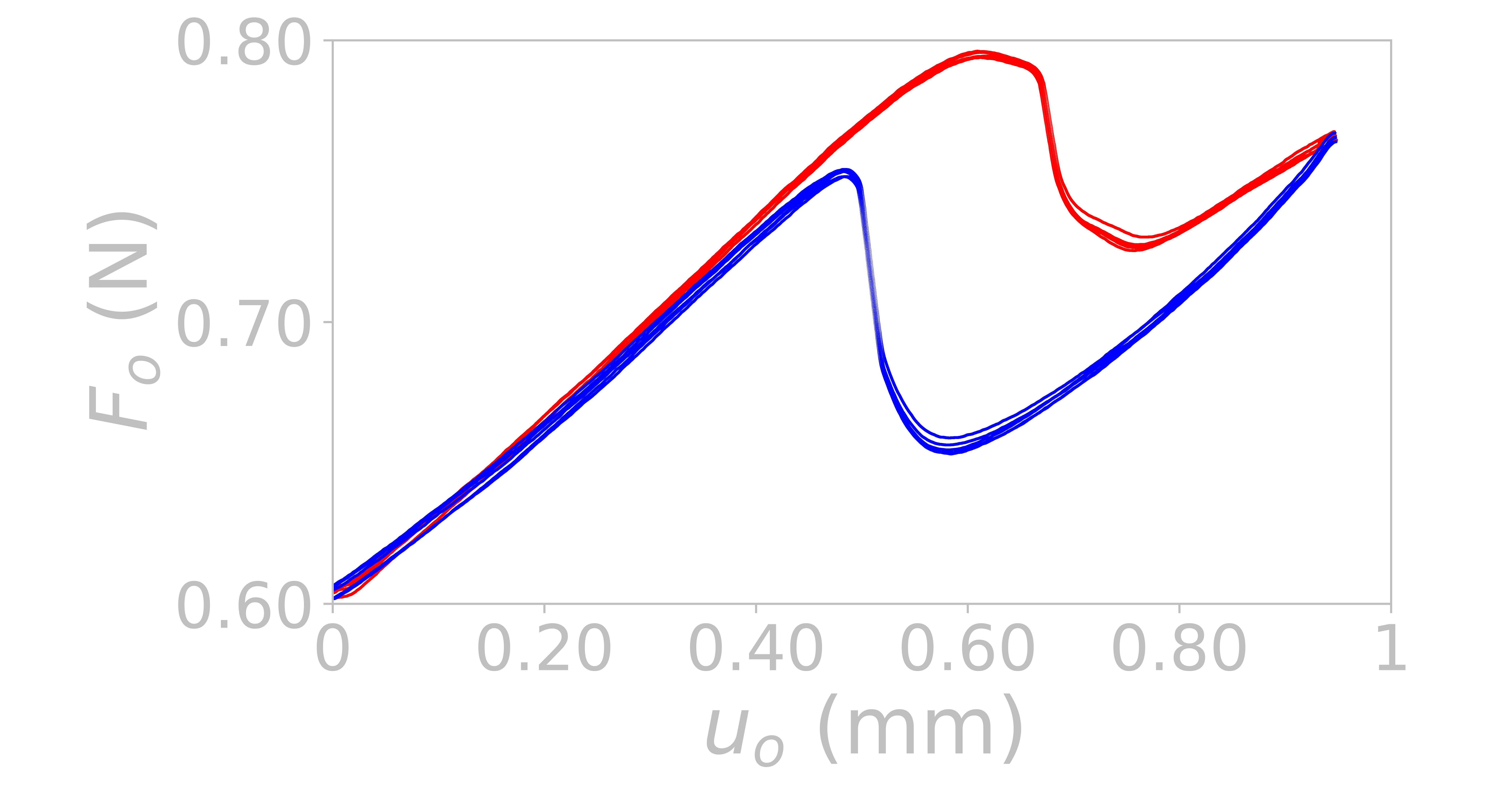}
	\includegraphics[width=0.49\textwidth]{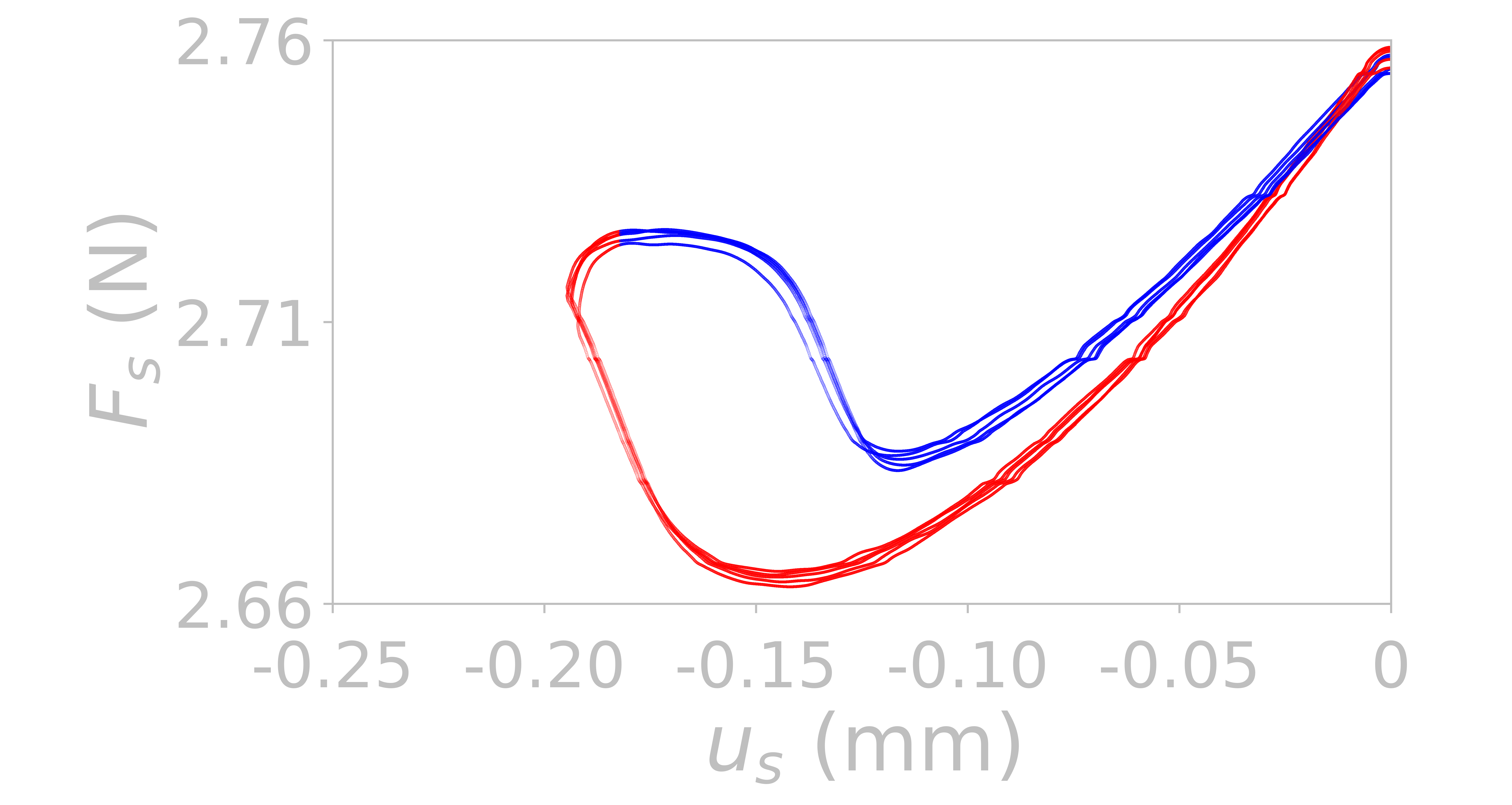}
%	\captionsetup{margin=0.5in}
	\caption{Five force loops collected in sequence. Using O-valley sheet O7 and S-ridge sheet S12.}
    \label{repeatable}
\end{figure}

Quantitative results were highly sensitive to the alignment and boundary conditions, which differed between sheets. 
Some qualitative results, such as the measurable range of stable S-ridge separation before splitting, also showed variation with alignment. 
In the data shown in the paper and Supplement, those trials (loops) performed without re-creation of crumples are plotted on top of each other in a single figure, using a single background subtraction (see next subsection) corresponding to one of the loops.  If crumples are expelled/destroyed and re-created, crumple interactions will not occur in the same locations with respect to the bent sheet background, so we analyze these in separate figures. Repeatability of force measurements between trials is illustrated in Figure \ref{repeatable}.

\begin{figure}[H]
	\includegraphics[width=0.49\textwidth]{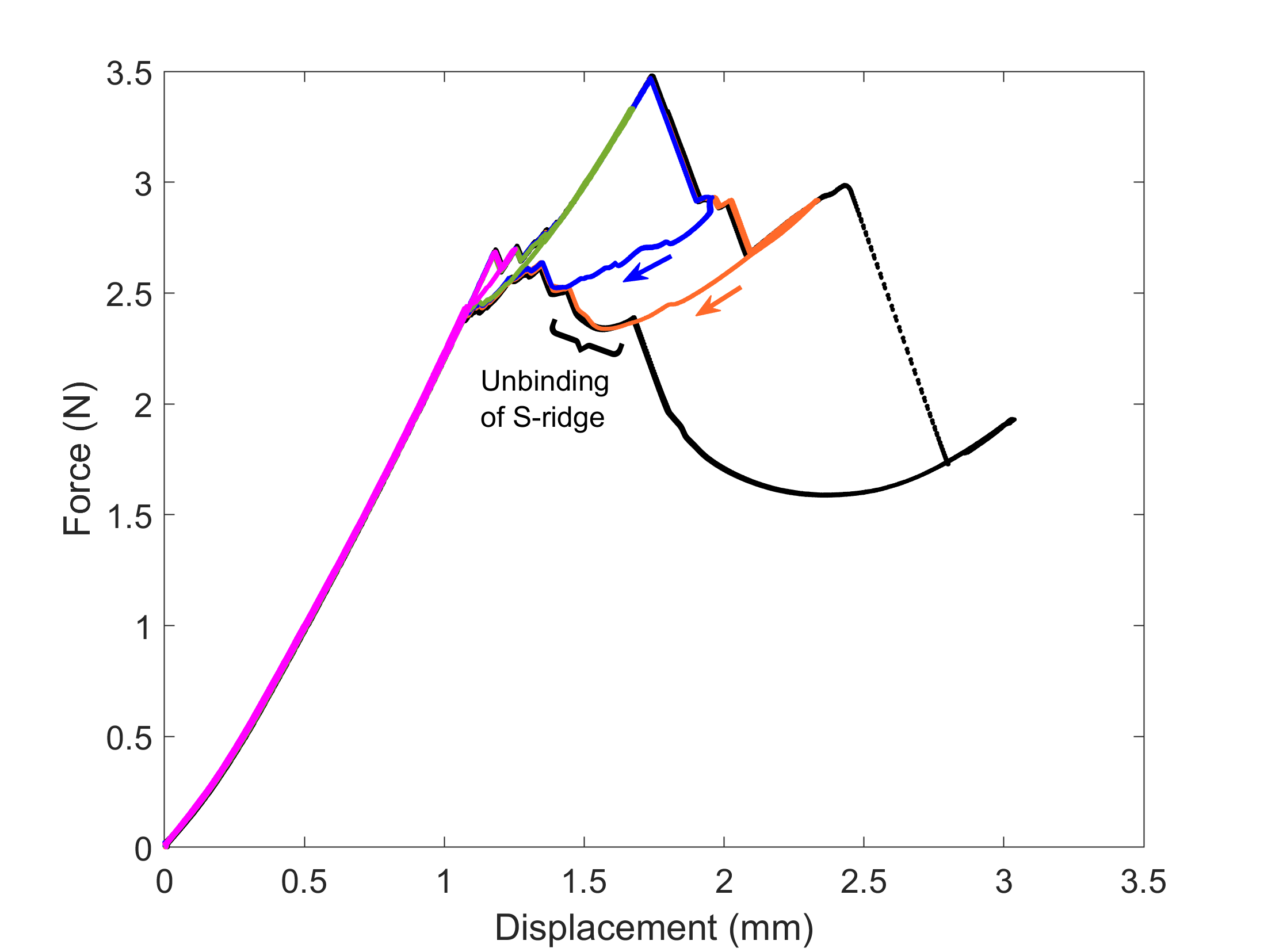}
%	\captionsetup{margin=0.5in}
	\caption{Multiple loops (decreasing length: black, orange, blue, green, purple) of partial or complete snap-through of a 12 inch sheet, with S-ridge unbinding indicated.  More information is provided in the paper \cite{Gimeno1}.}
    \label{UMU}
\end{figure}

Figure \ref{UMU} is included to emphasize the relatively small scales involved in the processes measured in this paper.  The figure shows multiple loops involving processes like those of Video V1, the largest of which goes even farther by expelling the final S-ridge into a fully snapped through state.  The unbinding of the S-ridge is indicated. Repeatability can also be inferred from the overlap of loops. 

\FloatBarrier

\subsection*{Data analysis}

In order to integrate force loops to obtain energies, several steps must be performed. 
First, the background force should be subtracted for clarity. Then we must determine which data are taken from stable and unstable regions, cut the data in the unstable regions, and displace the energy to reflect jumps in energy at bifurcations.  All of these steps can be done in different ways, requiring us to make consistent, reasonable choices. 

The subtraction process can create zeroes of force, and thus energy minima, at locations such as the endpoints of the loops (in the primary method used). While the endpoints of the loops do at times have physical significance as stability limits, there is no reason for them to correspond to energy minima.  Nor was there any need for our energy to have minima at $\pm 1$, rather than off at infinity, for example, or to have minima at all. These aspects of the subtraction method, and the model, are present purely for simplicity and approximate symmetry, and can be adjusted without loss of generality. 

We demonstrate our methods using the example data shown in the main text Figures \ref{loops} and \ref{compare}, and also show alternate approaches to force subtraction that change certain arbitrary but irrelevant aspects of the analysis.

\subsubsection*{Primary methods used} 

\begin{figure}[h]
    \centering
        \includegraphics[width=0.485\linewidth]{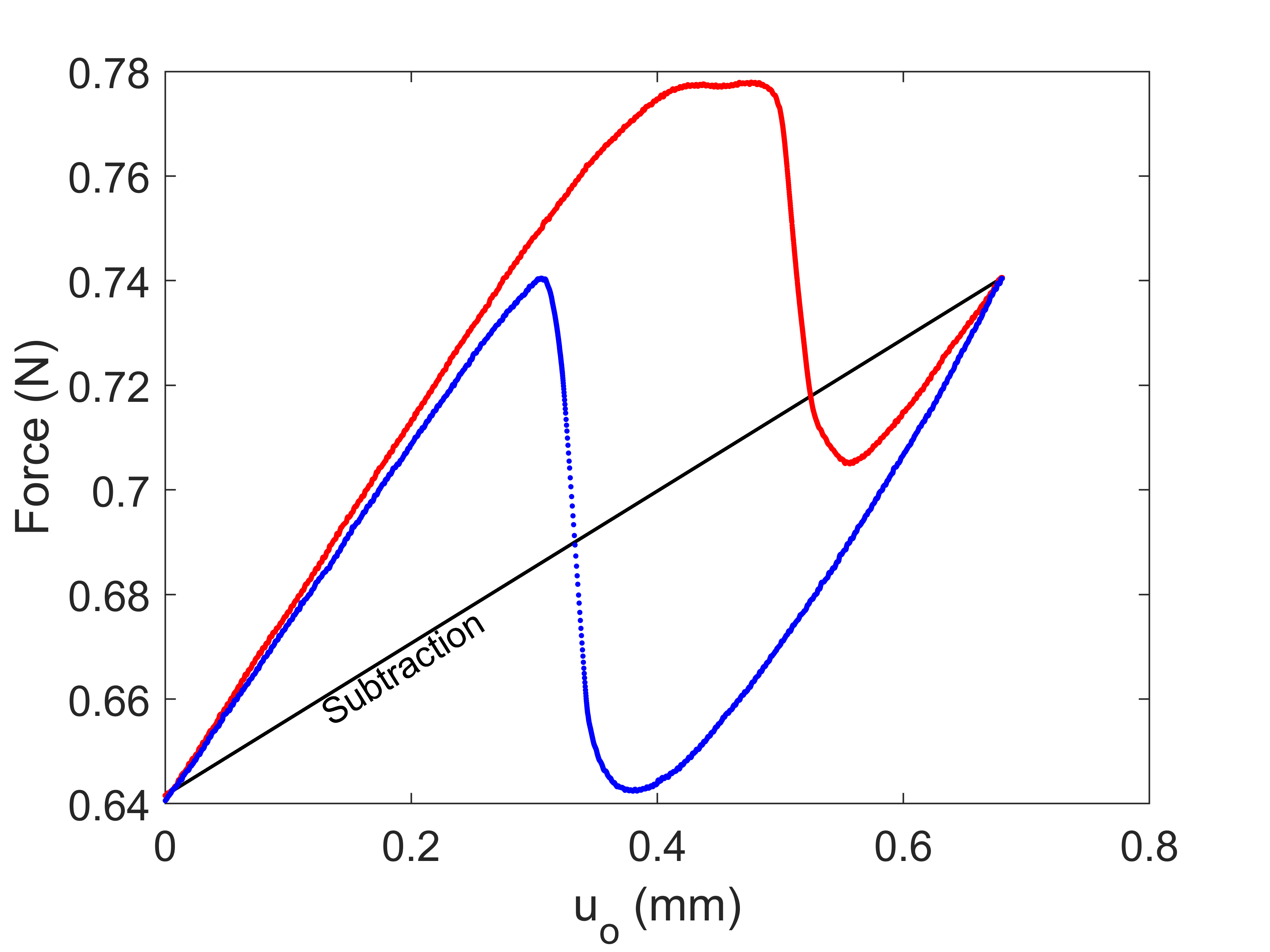}
        \includegraphics[width=0.485\linewidth]{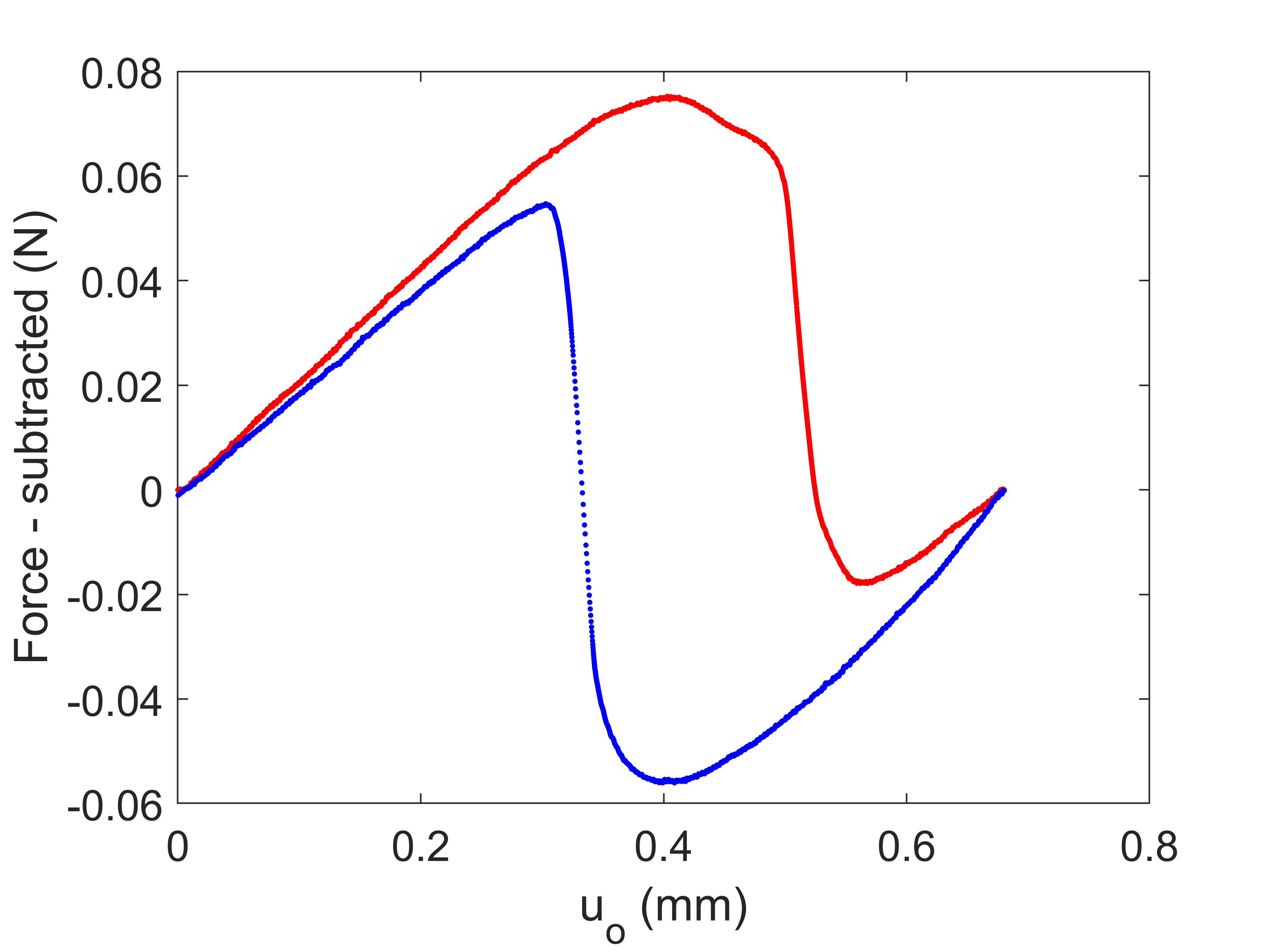}
    \caption[]{Force subtraction for the O-valley. Subtracted line and resulting curve.}
    \label{fig:subtractionO}
\end{figure}

For the O-valley, we subtract a straight line between the beginning and end of the unbinding (red) curve, as shown in Figure \ref{fig:subtractionO}.  This creates irrelevant energy minima at the endpoints. 

\begin{figure}[h]
    \centering
        \includegraphics[width=0.485\linewidth]{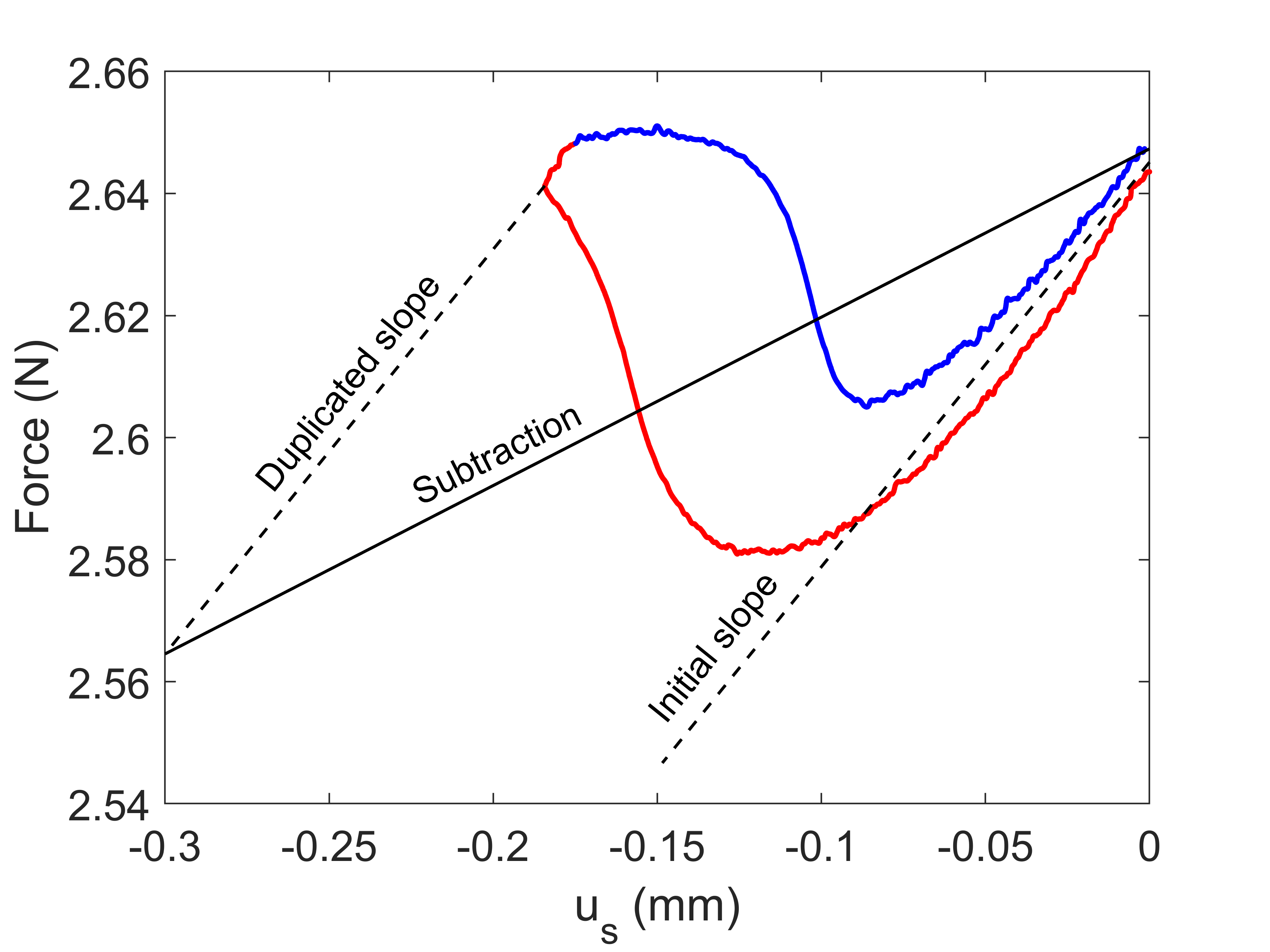}
        \includegraphics[width=0.485\linewidth]{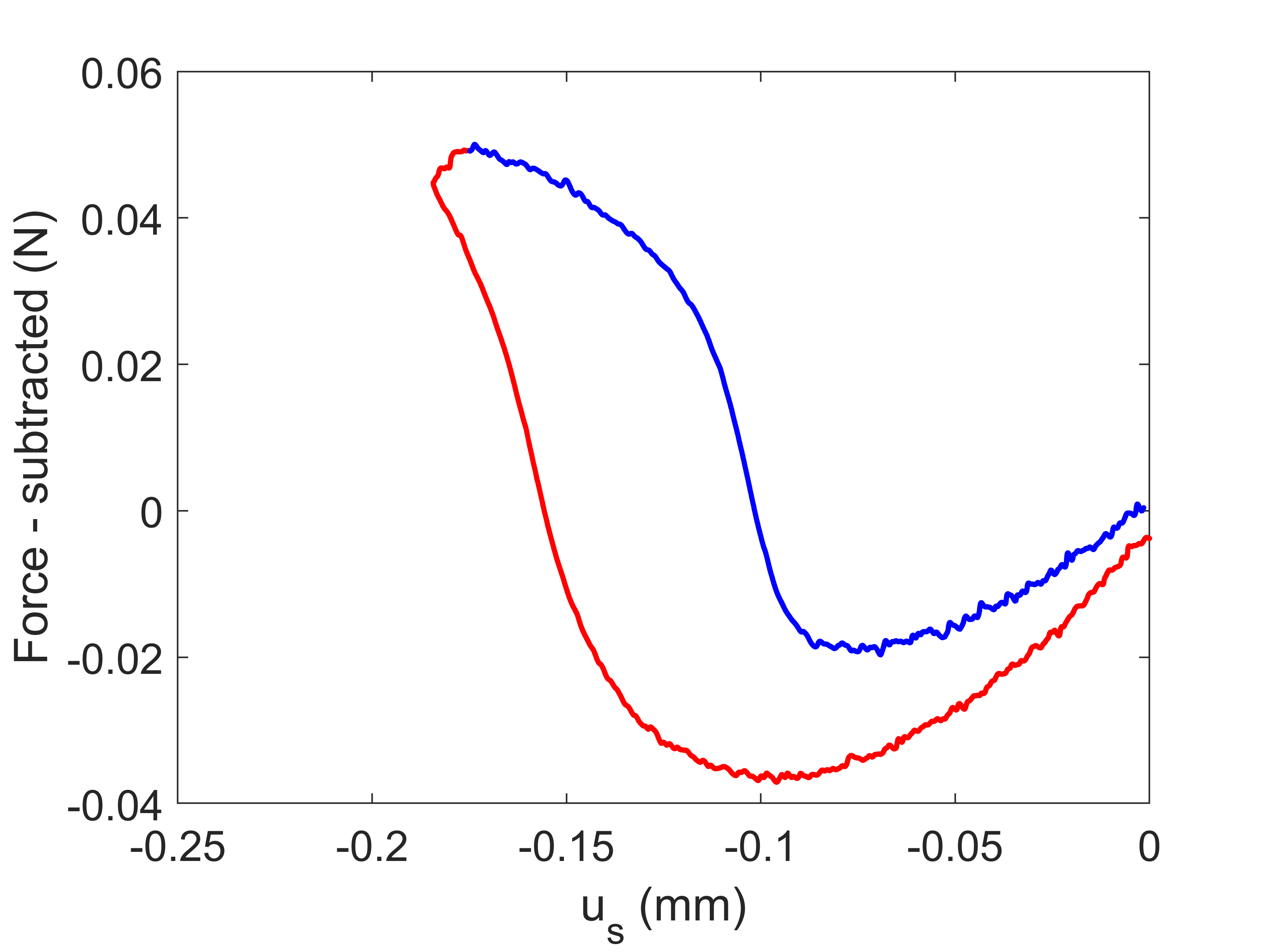}
    \caption[]{Force subtraction for the S-ridge. Subtracted line and resulting curve (note different horizontal range).}
    \label{fig:subtractionS}
\end{figure}

For the S-ridge, we draw a line starting at the beginning of the binding (blue) curve, parallel to that made by close crumples during unbinding, and shifted so that it ends at $u_s=0.3$, as shown in Figure \ref{fig:subtractionS}. 
This creates an irrelevant energy minimum at one endpoint. 

\FloatBarrier

\begin{figure}[h]
    \centering
        \includegraphics[width=0.485\linewidth]{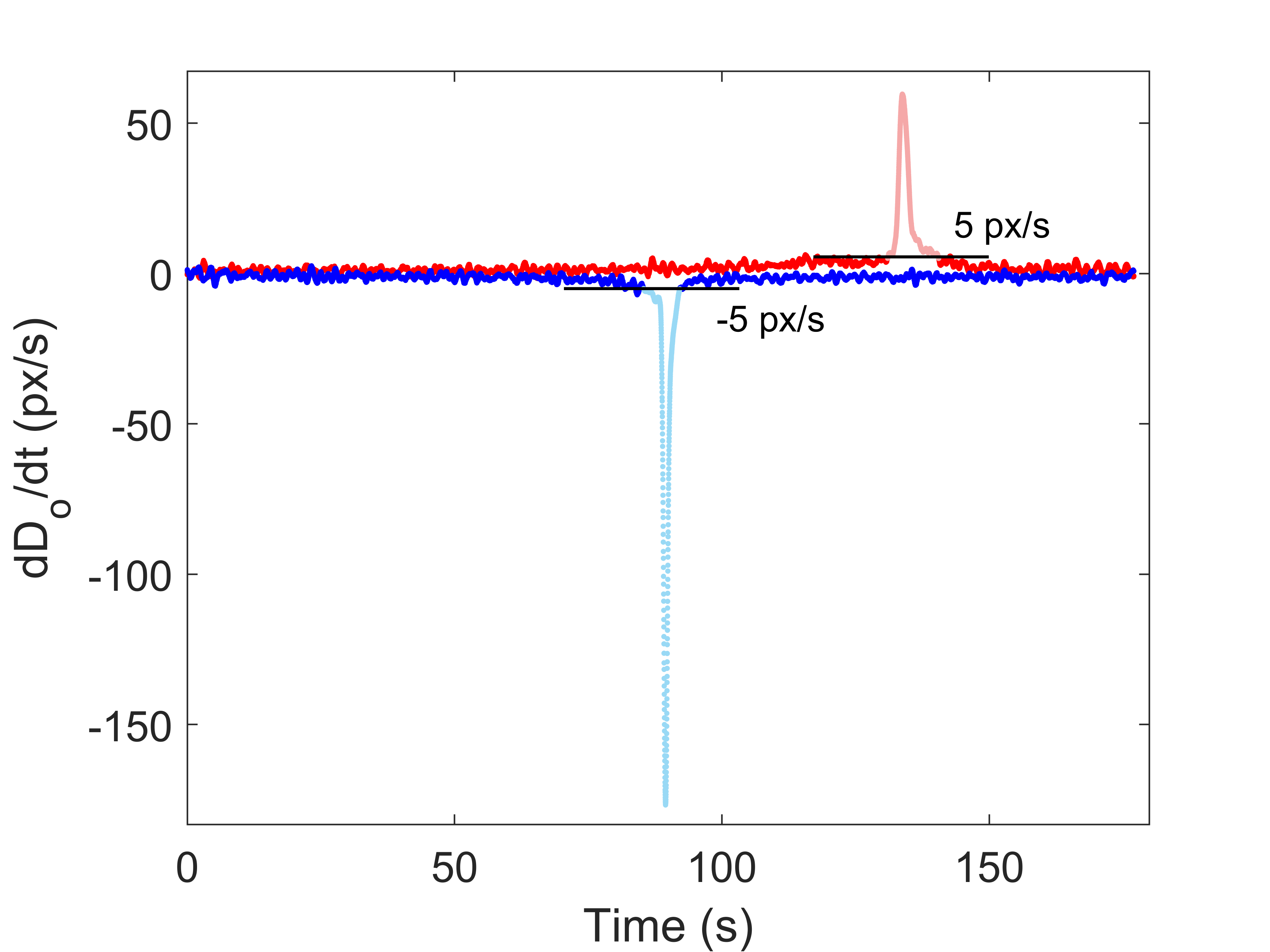}
        \includegraphics[width=0.485\linewidth]{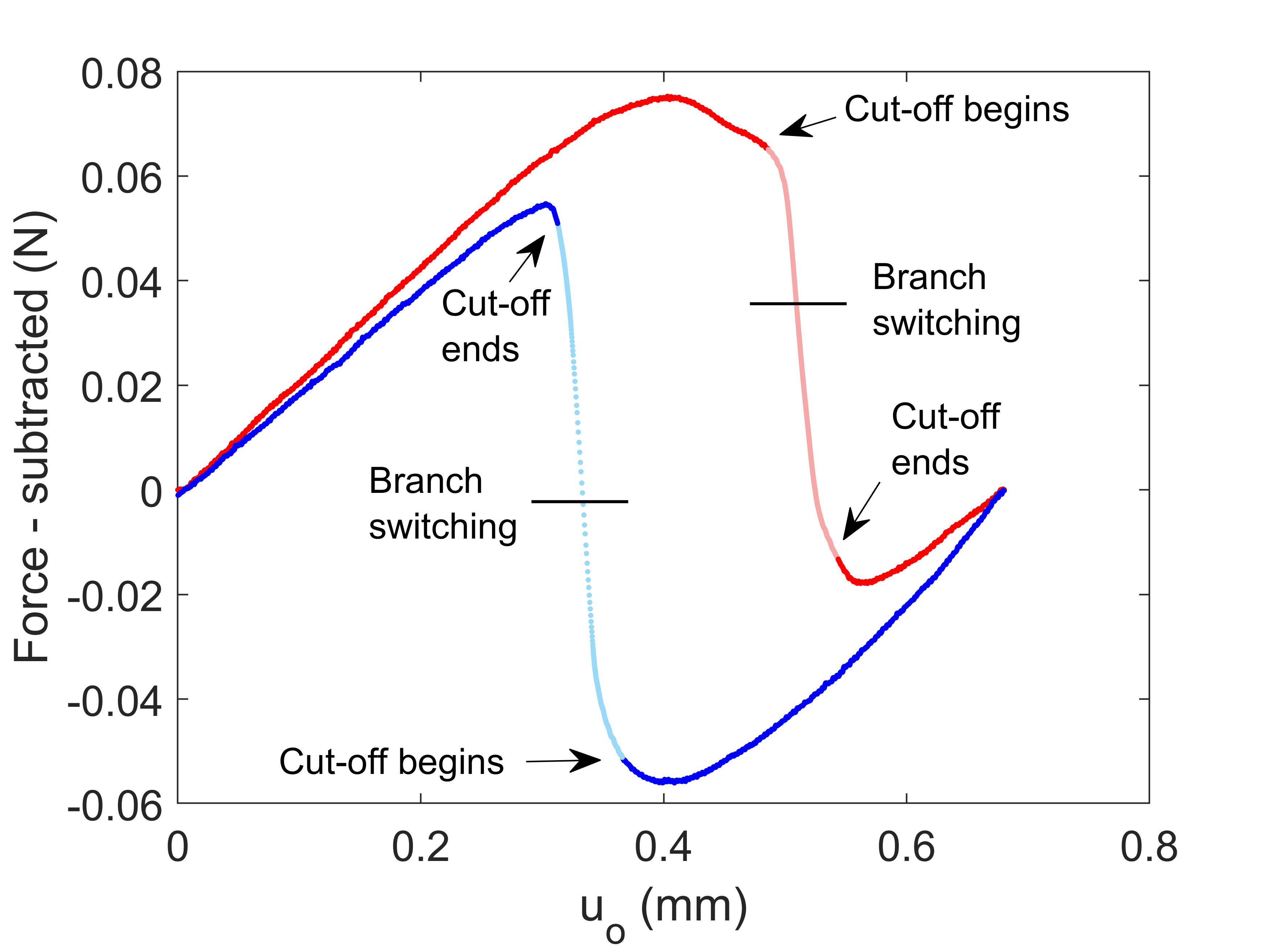}
    \caption[]{Stability and branch switching for the O-valley. Rate of change of crumple separation with cutoff thresholds at $\pm 5$ px/s separating stable (dark) and unstable (light) regions. Subtracted force curve showing stability and the locations where we cut the data to switch between energy branches.}
    \label{fig:cutoffO}
\end{figure}

\begin{figure}[h]
        \includegraphics[width=0.485\linewidth]{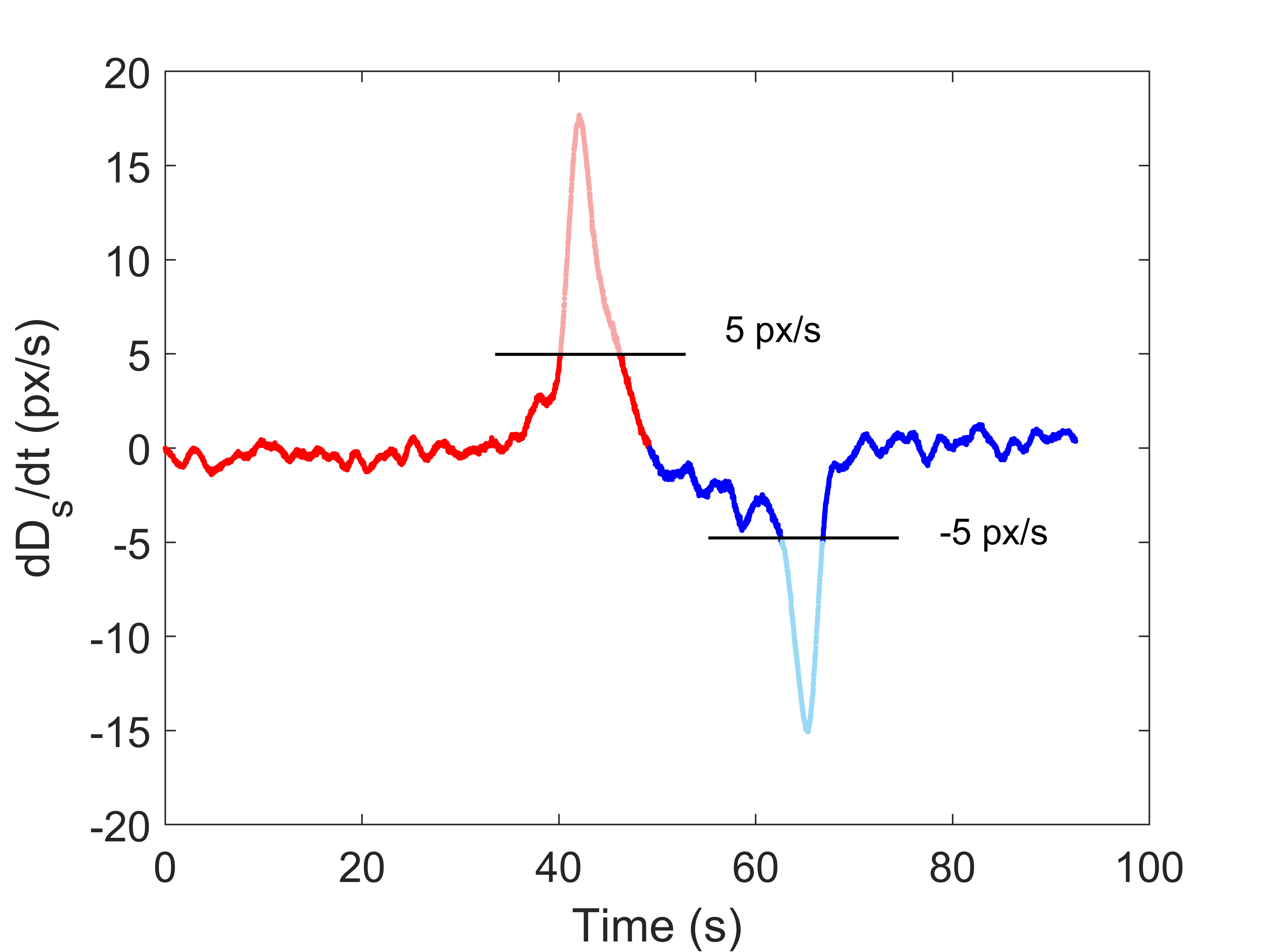}
        \includegraphics[width=0.485\linewidth]{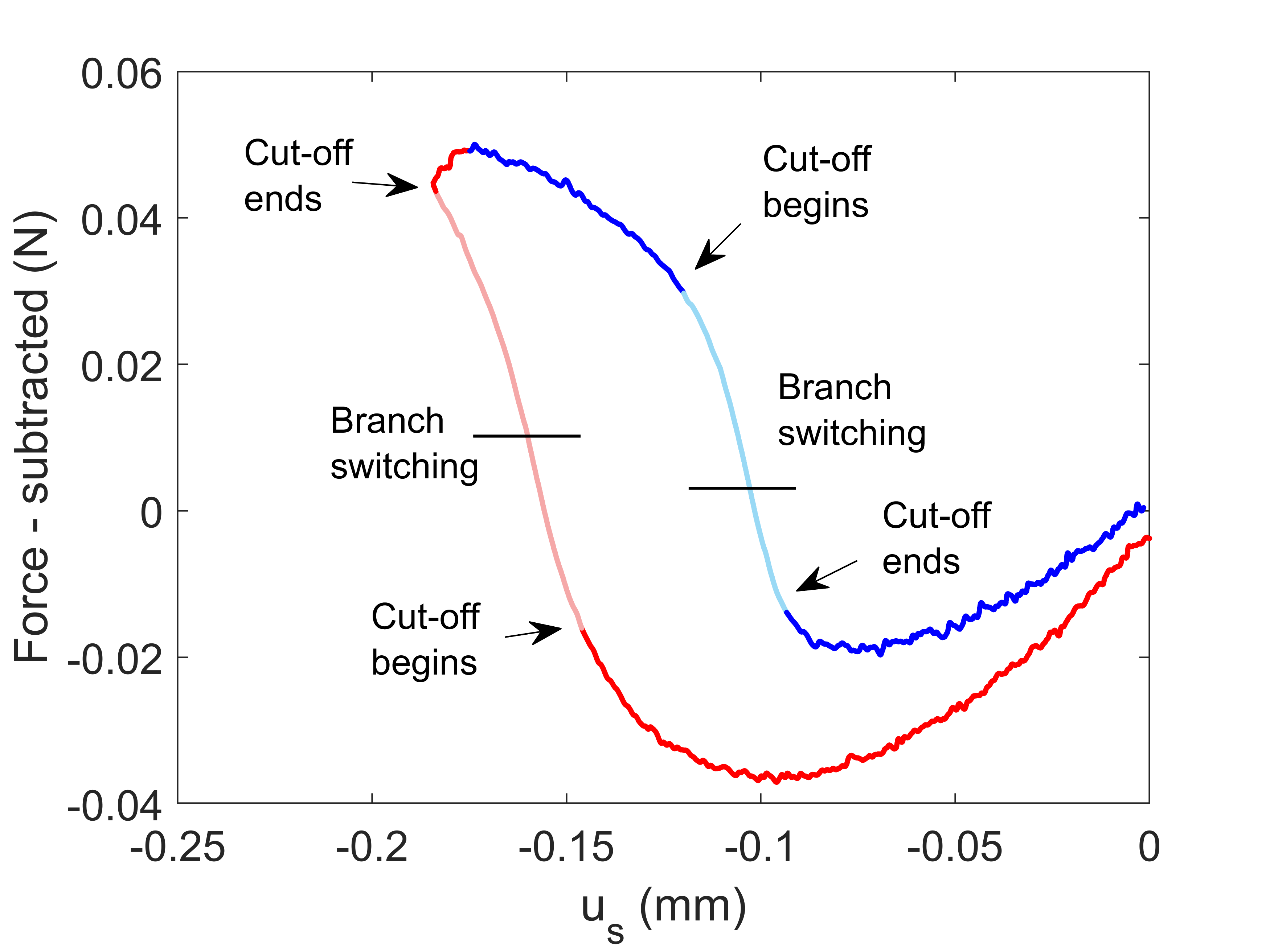}
   \caption[]{Stability and branch switching for the S-ridge. Rate of change of crumple separation with cutoff thresholds at $\pm 5$ px/s separating stable (dark) and unstable (light) regions. Subtracted force curve showing stability and the locations where we cut the data to switch between energy branches.}
    \label{fig:cutoffS}
\end{figure}

A threshold value for rate of crumple separation $| dD/dt |$ of 5 px/s (approximately half a millimeter per second) was chosen as a cutoff between stable and unstable motion.  These points indicate where we presume the system to be leaving stable equilibrium branches of the energy.  Rather than cut in these two places, we plot all the data and cut to switch between energy branches at points where the rate of change of force $| dF/dt |$ reached a maximum. The two types of rate have their maxima coincide in the case of the O-valley, and are slightly different for the S-ridge. The cut-offs and branch-switching cut points are shown in Figures \ref{fig:cutoffO} and \ref{fig:cutoffS}. 

Changes in energy were obtained by cumulative integration along subtracted force-displacement curves. We cannot determine the energy losses during the bifurcations, so jumps in energy between branches were chosen to provide equal drops across the unstable regions (that is, between the cut-off points), although it is unknown whether the real system behaves so symmetrically. 

All this leads to the energy and other plots in Figure \ref{compare}.

\FloatBarrier

\subsubsection*{An alternate subtraction method}

\begin{figure}[p]
    \centering
        \includegraphics[width=0.485\linewidth]{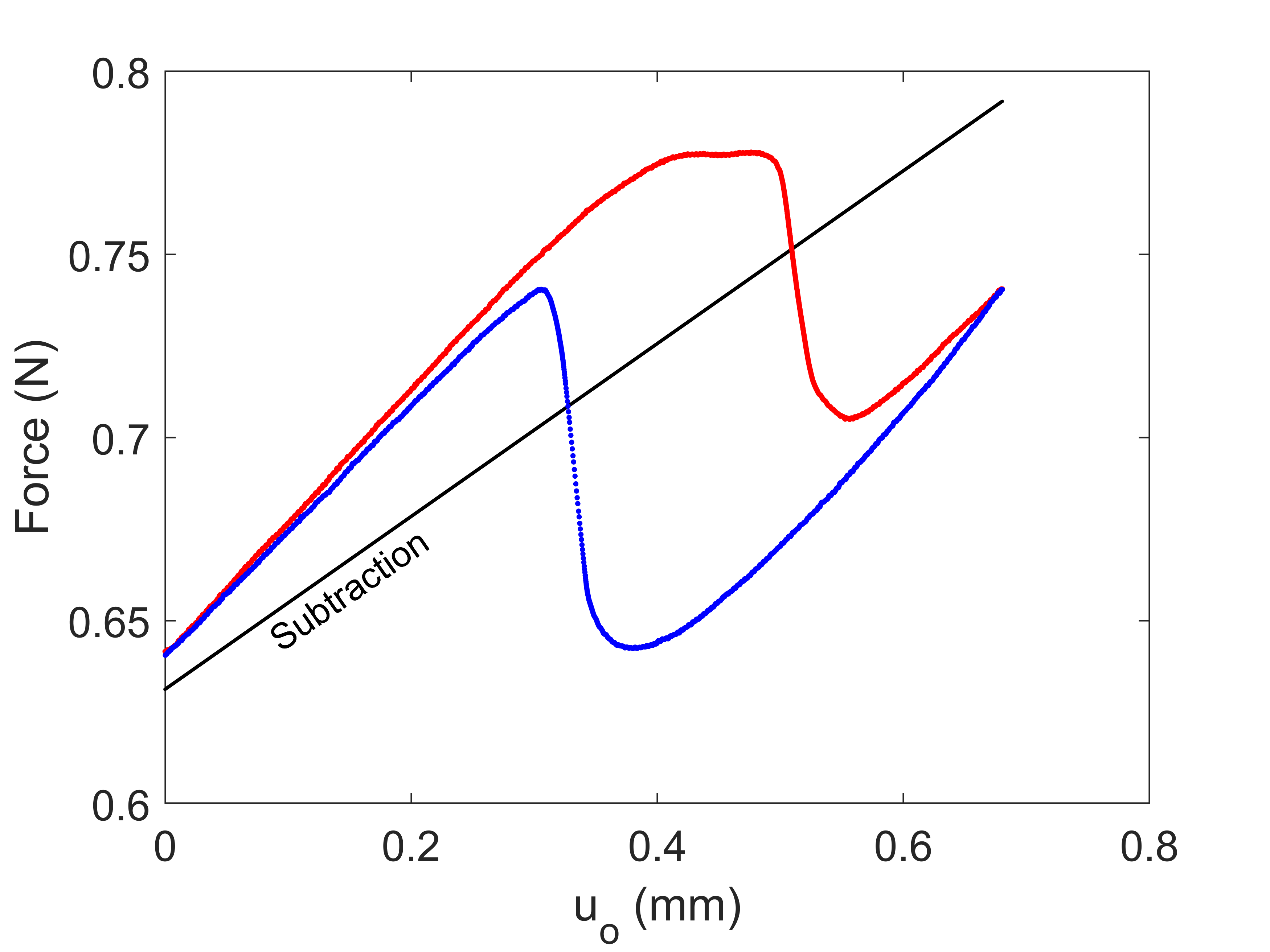}
        \includegraphics[width=0.485\linewidth]{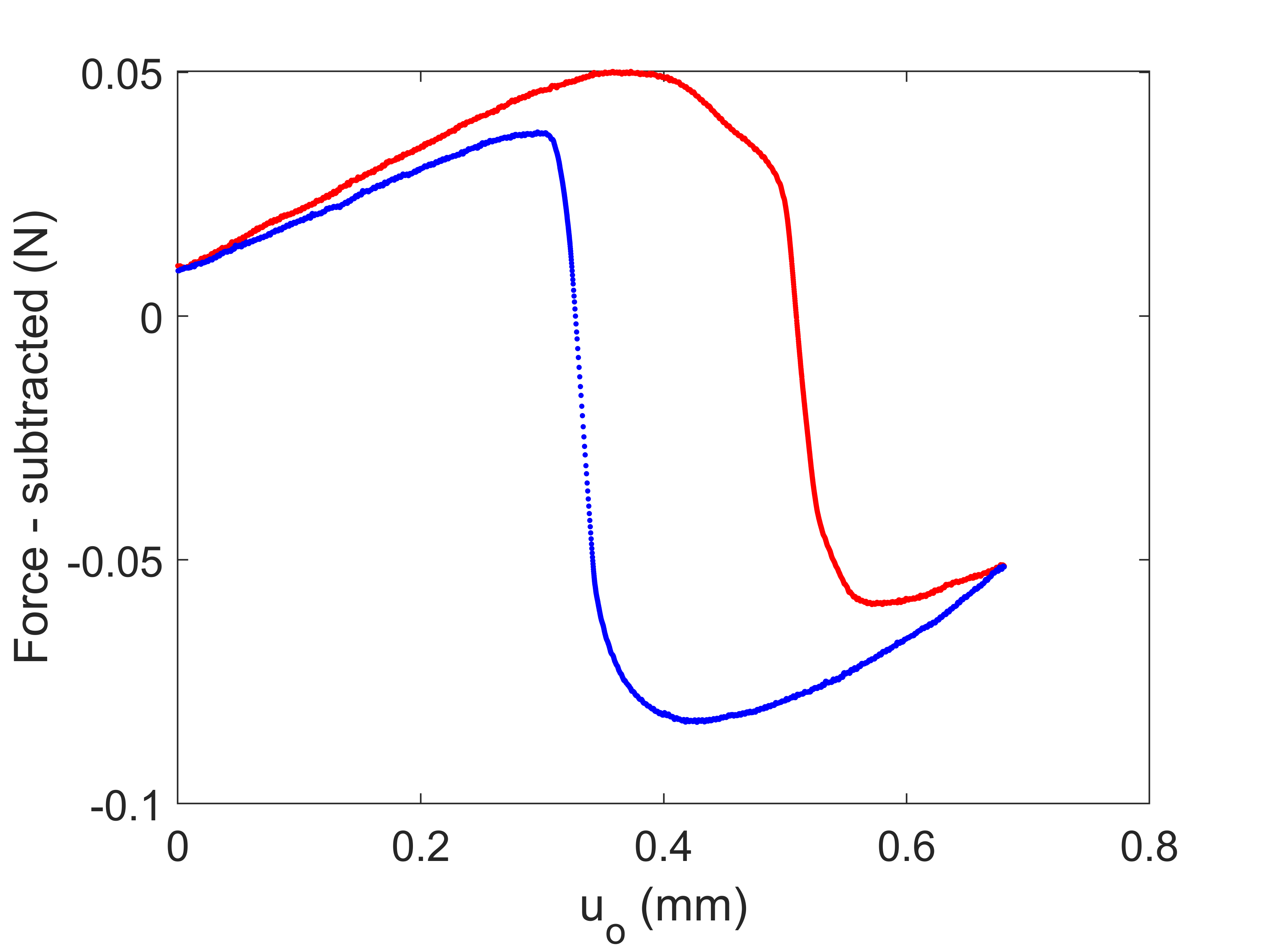}
    \caption{Alternate force subtraction for the O-valley. Subtracted line intersecting maximum crumple separation rates, and resulting curve.}
    \label{fig:subtractionaltO}
\end{figure}

Here we compare with a different method of force subtraction, in which we subtract a straight line drawn through the two points of maximum rate of crumple separation $| dD/dt |$, as shown in Figures \ref{fig:subtractionaltO} and \ref{fig:subtractionaltS}.

\begin{figure}[p]
    \centering
            \includegraphics[width=0.485\linewidth]{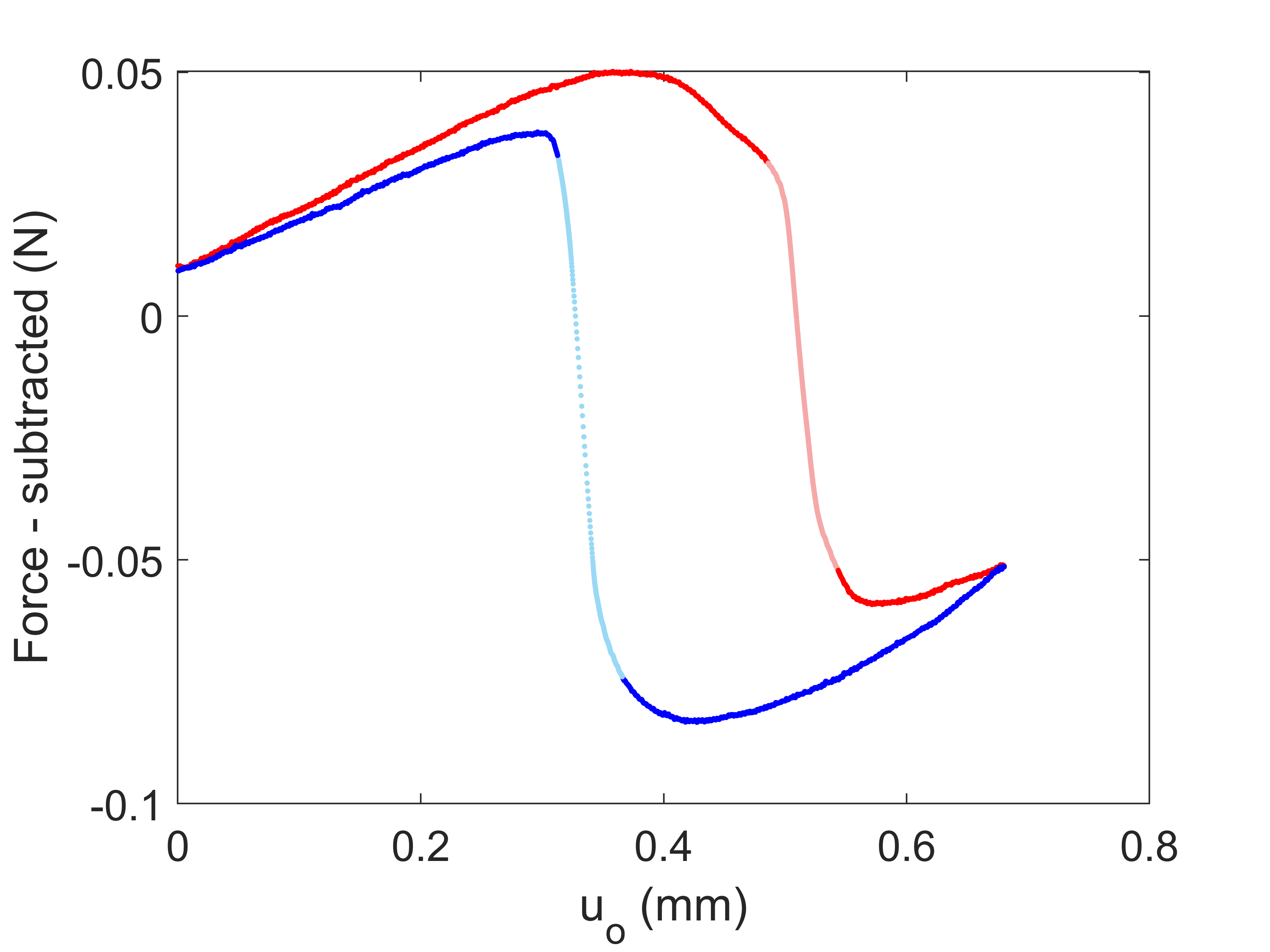}
            \includegraphics[width=0.485\linewidth]{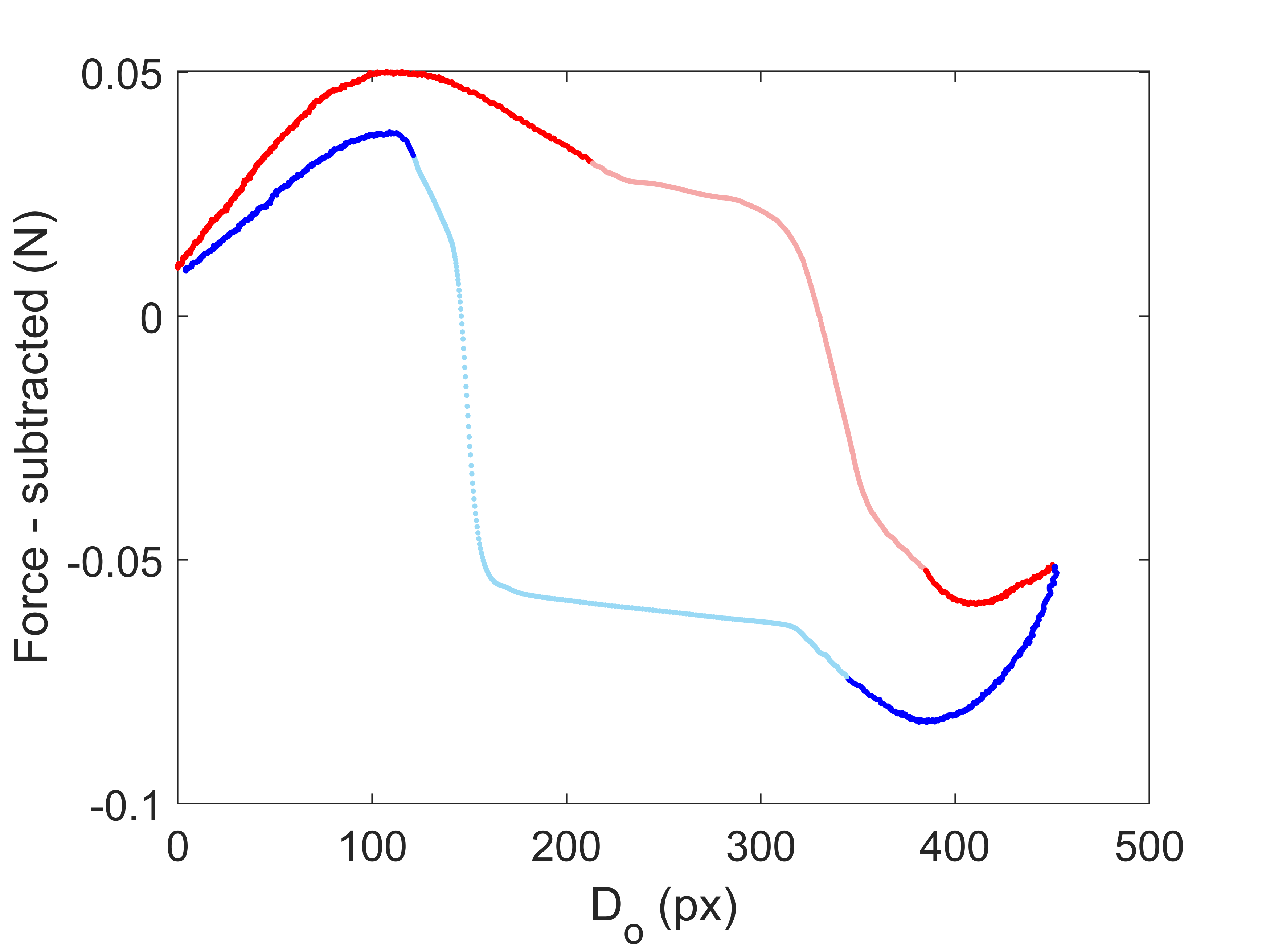}
            \includegraphics[width=0.485\linewidth]{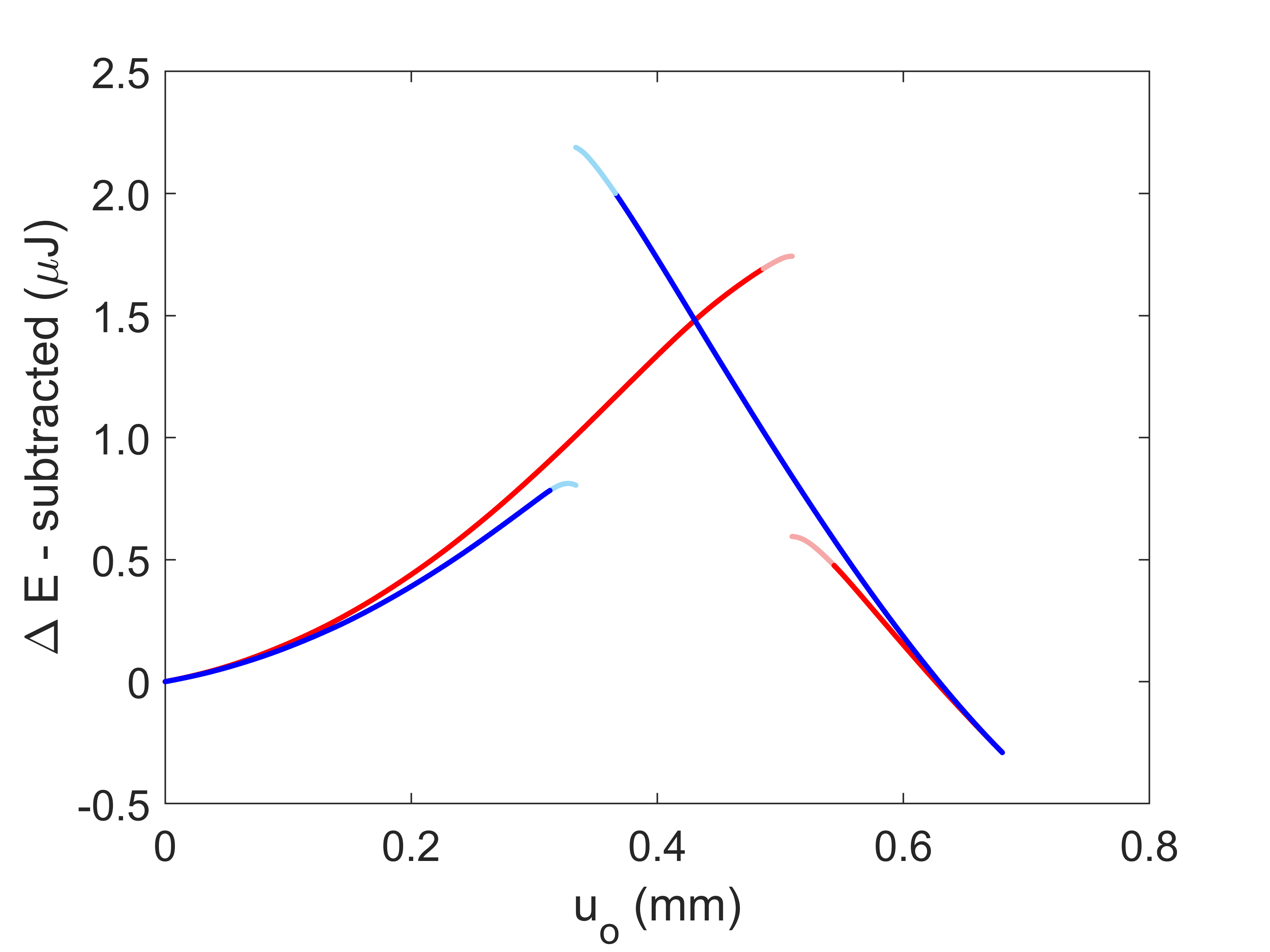}
            \includegraphics[width=0.485\linewidth]{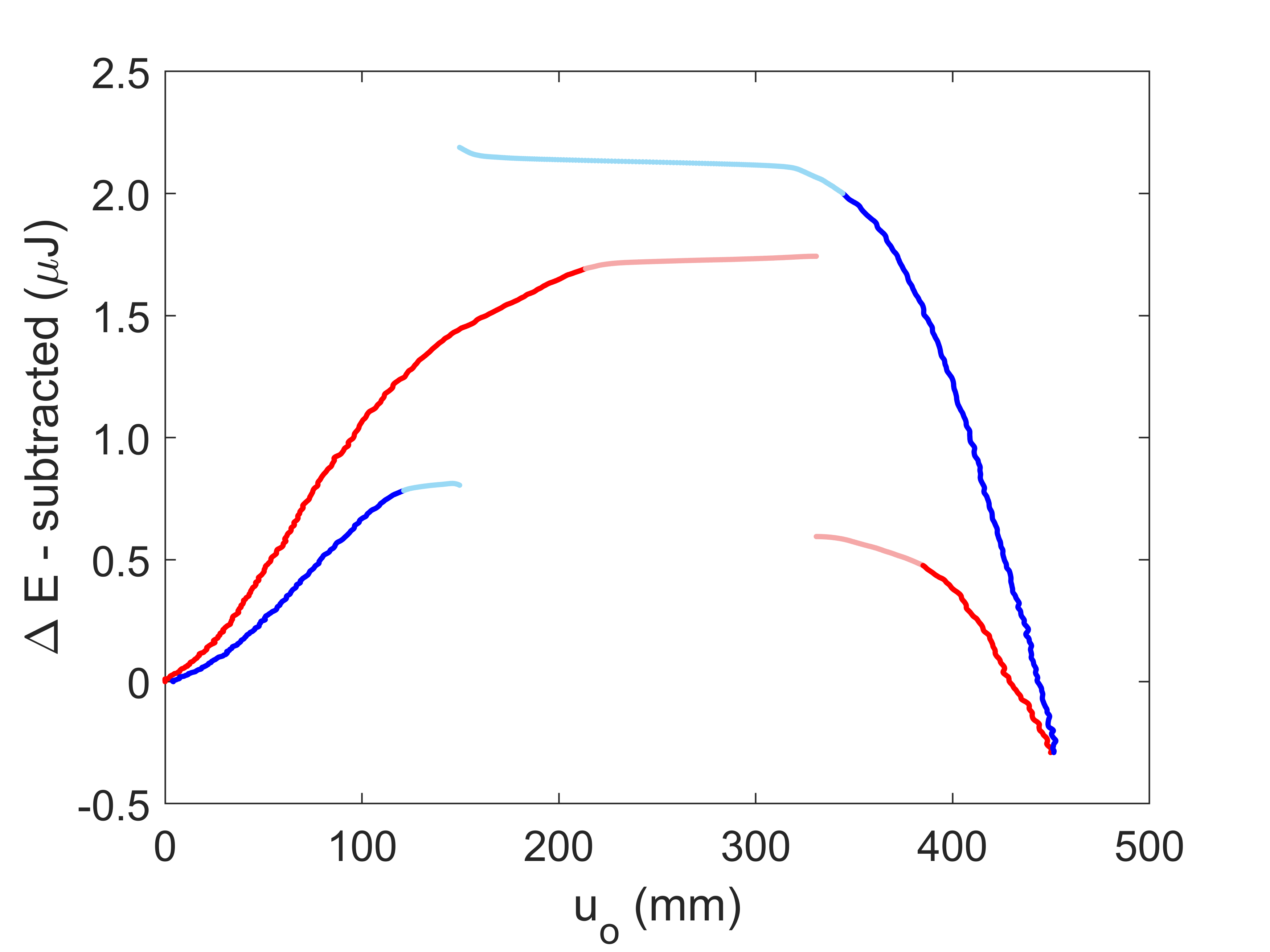}
    \caption{Processed data for O-valley using alternate method.  Compare with main text Figure \ref{compare}B.}
    \label{fig:2x2altO}
\end{figure}

\begin{figure}[p]
    \centering
        \includegraphics[width=0.485\linewidth]{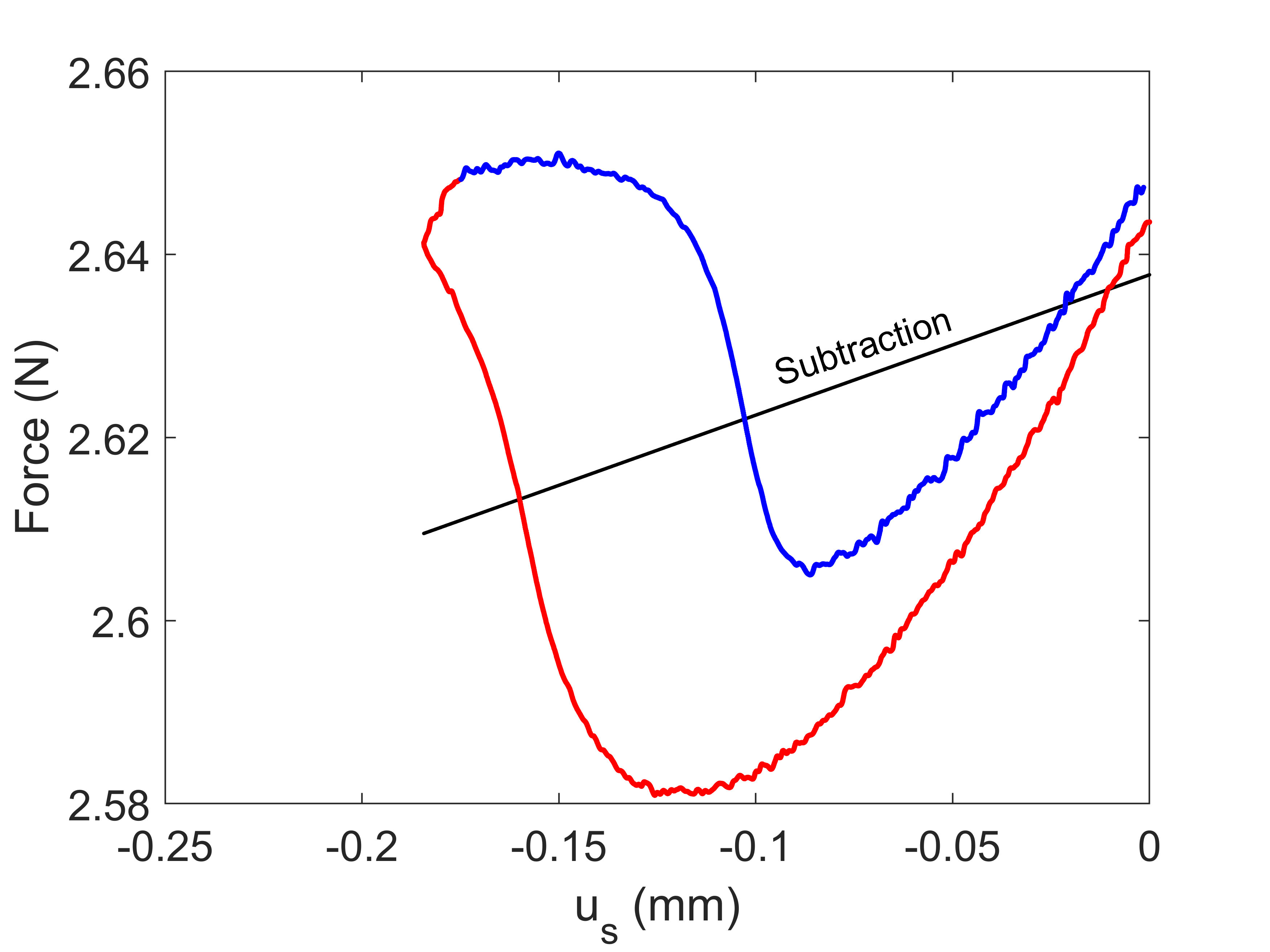}
        \includegraphics[width=0.485\linewidth]{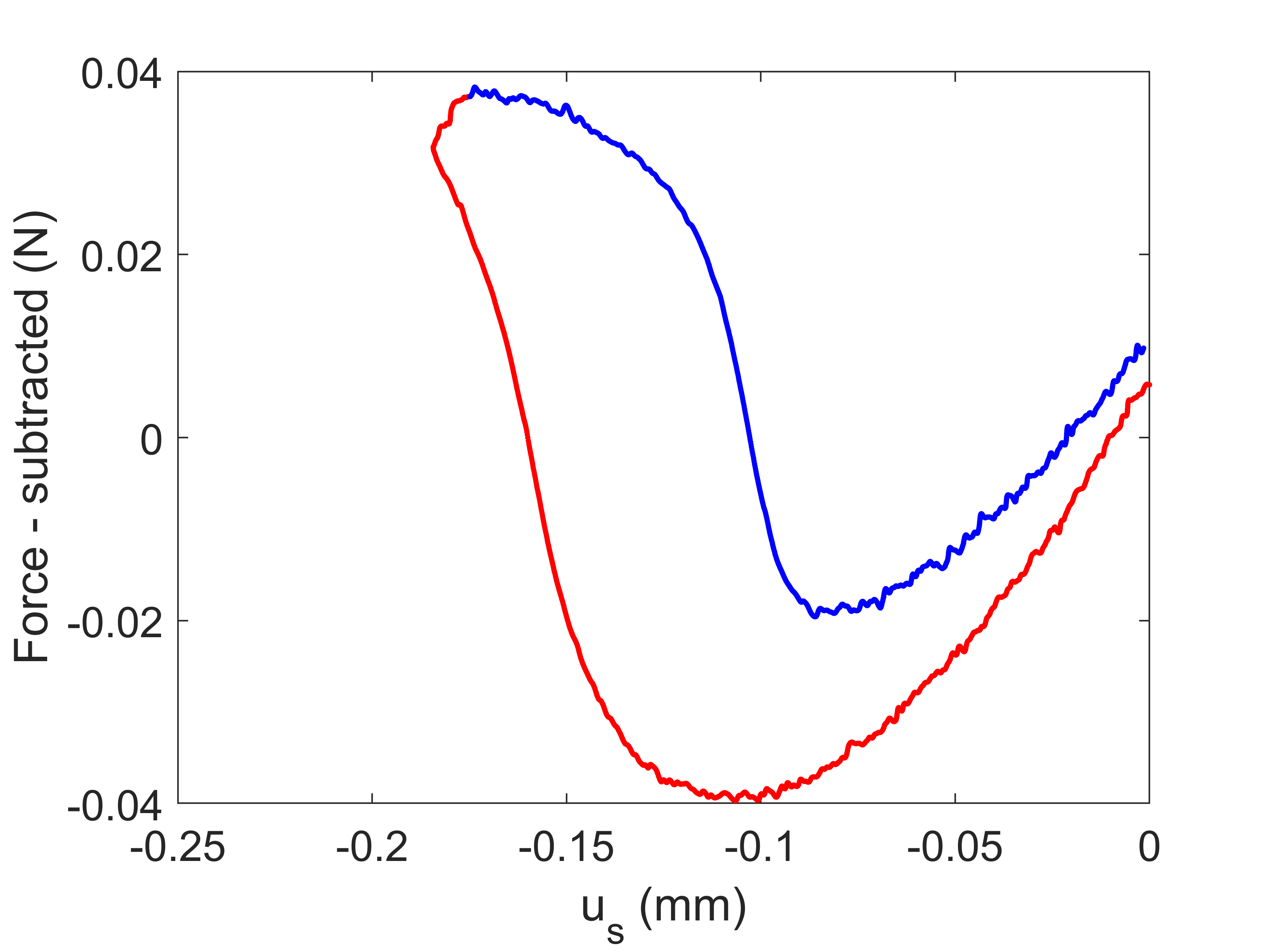}
    \caption{Alternate force subtraction for the S-ridge. Subtracted line intersecting maximum crumple separation rates, and resulting curve.}
    \label{fig:subtractionaltS}
\end{figure}

\begin{figure}[p]
    \centering
            \includegraphics[width=0.485\linewidth]{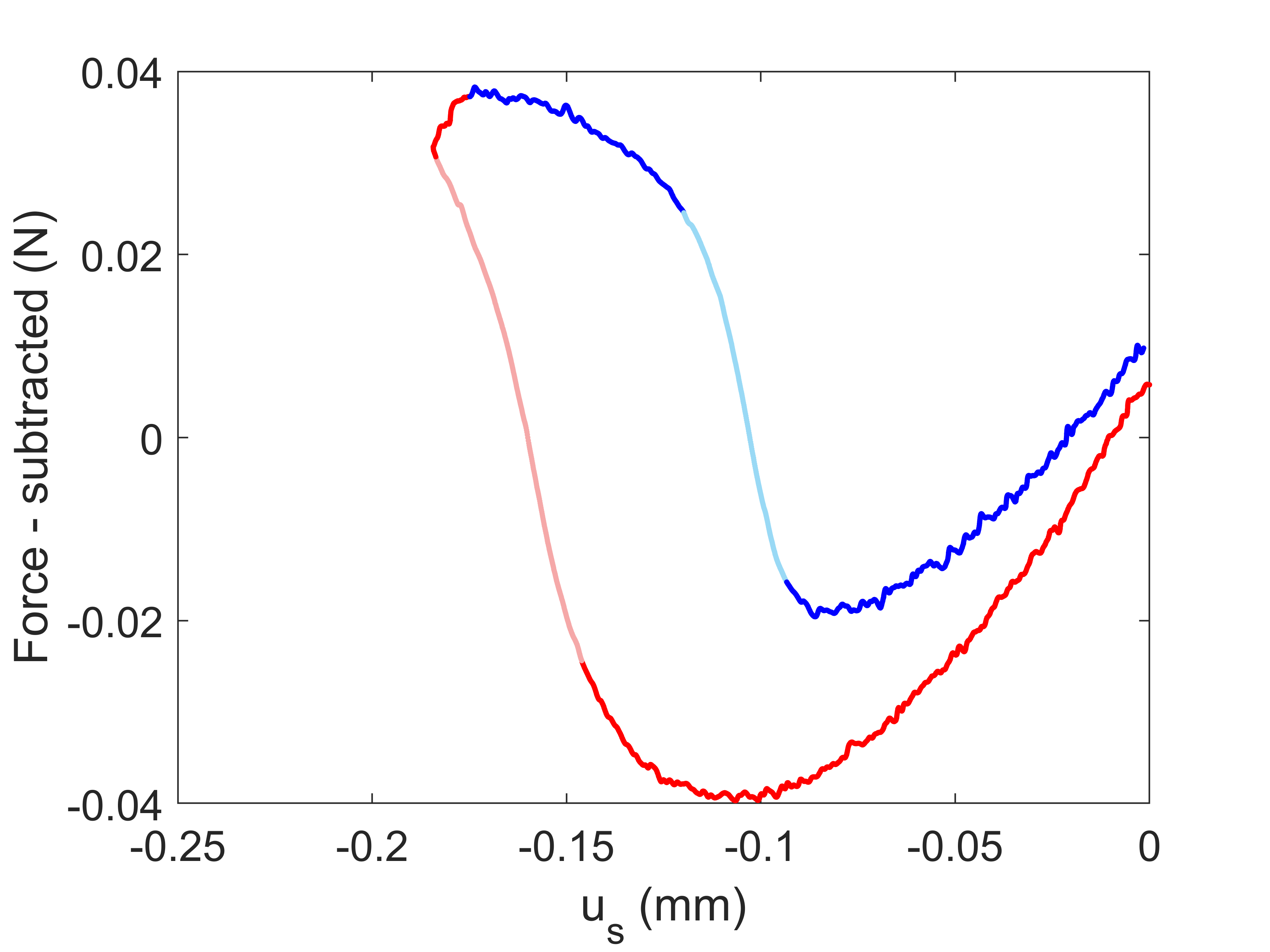}
            \includegraphics[width=0.485\linewidth]{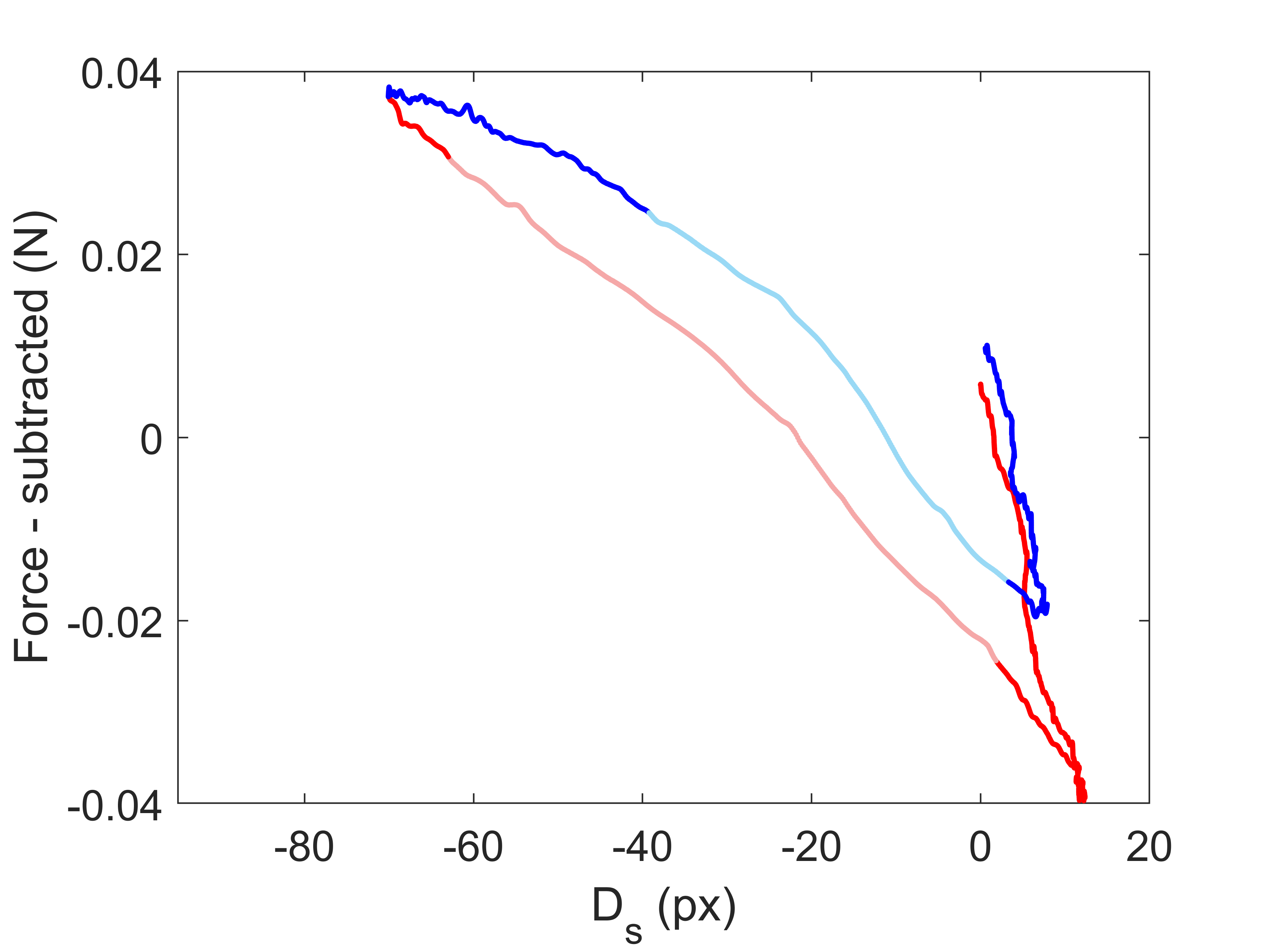}
            \includegraphics[width=0.485\linewidth]{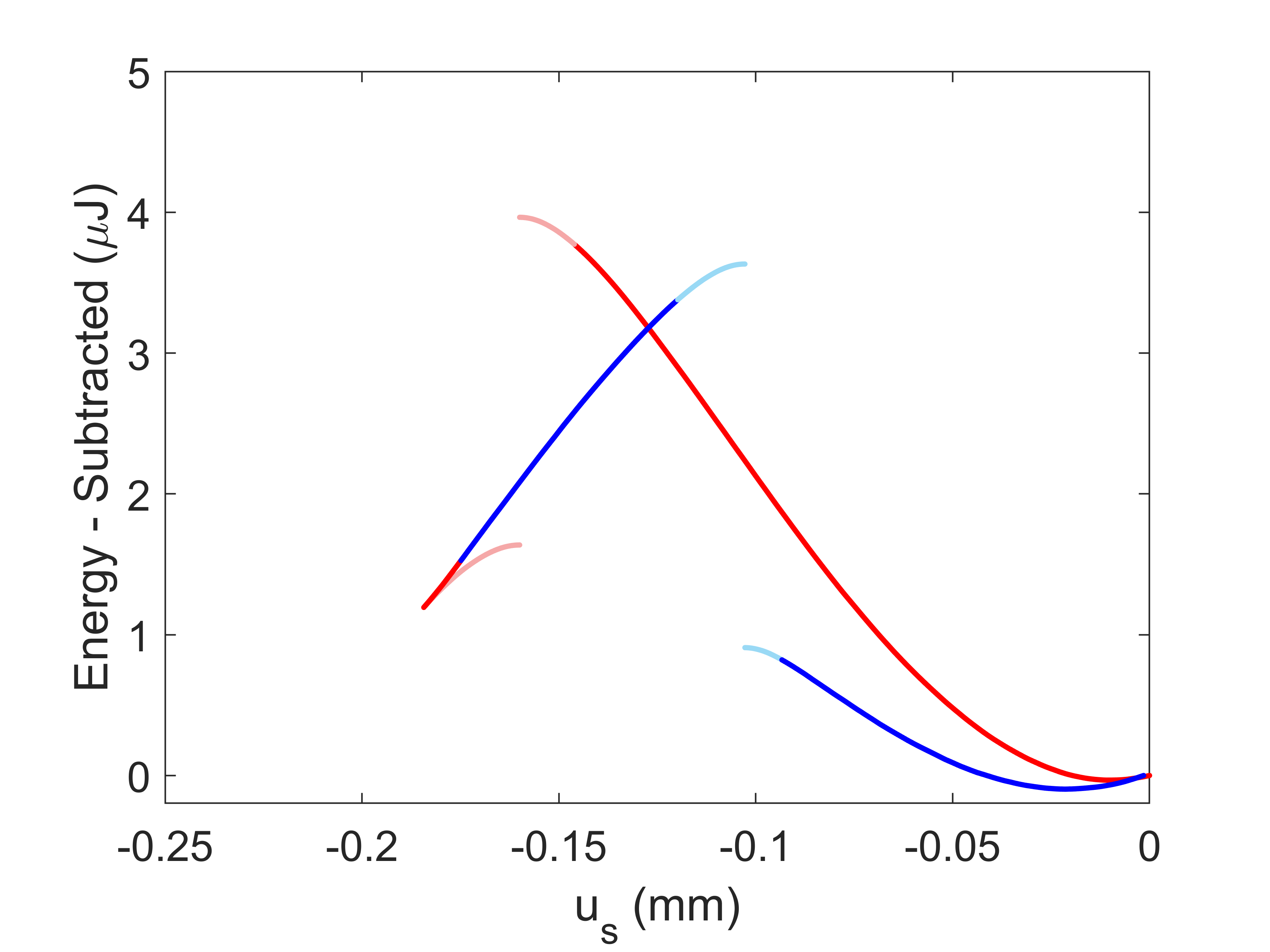}
            \includegraphics[width=0.485\linewidth]{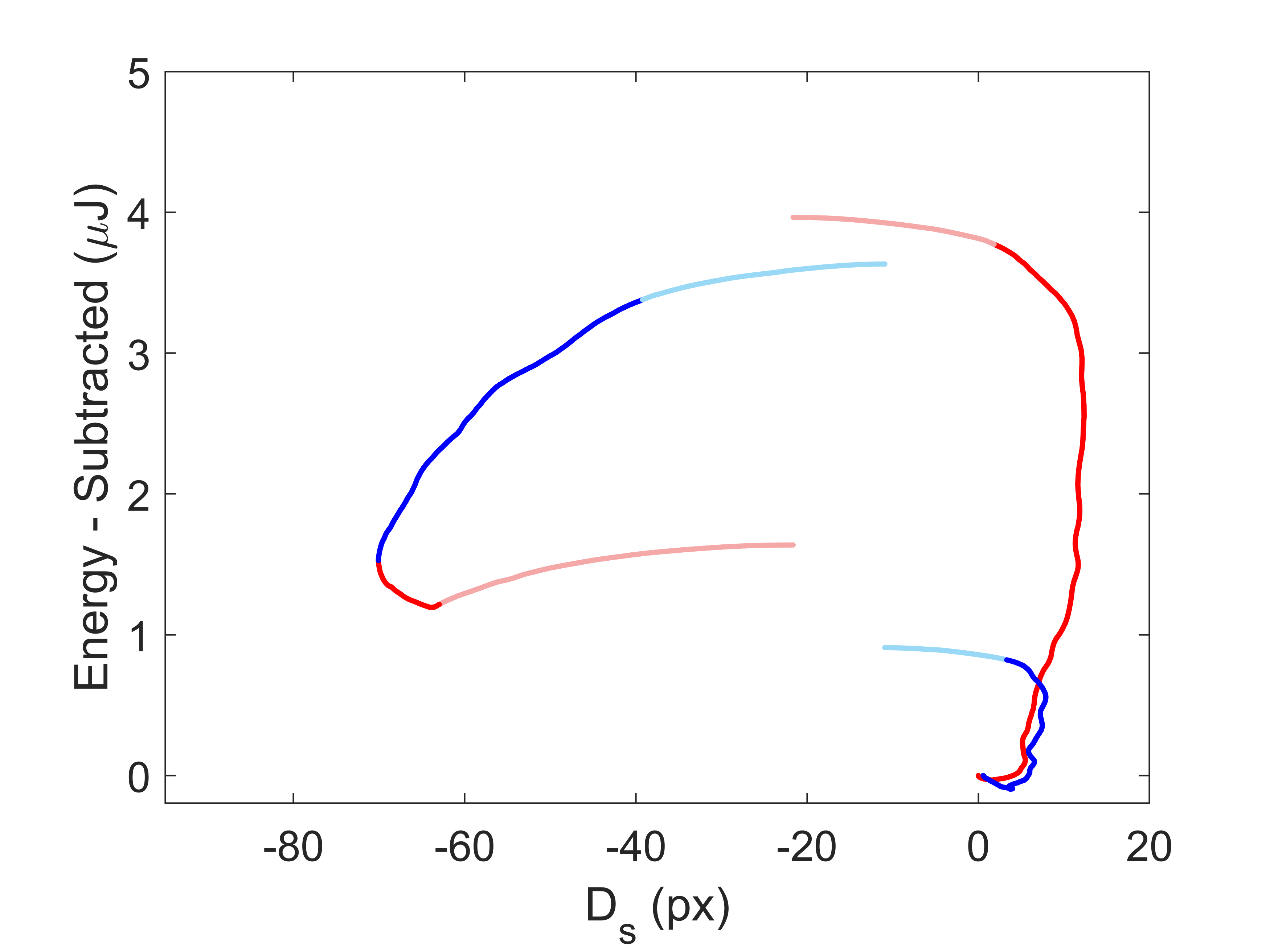}
    \caption{Processed data for S-ridge using alternate method.  Compare with main text Figure \ref{compare}C.}
    \label{fig:2x2altS}
\end{figure}

All relevant qualitative aspects of the processed data are preserved.   The resulting Figures \ref{fig:2x2altO} and \ref{fig:2x2altS} can be compared with main text Figure \ref{compare}B-C. 

\FloatBarrier

\subsubsection*{No subtraction}

\begin{figure}[h]
    \centering
            \includegraphics[width=0.485\linewidth]{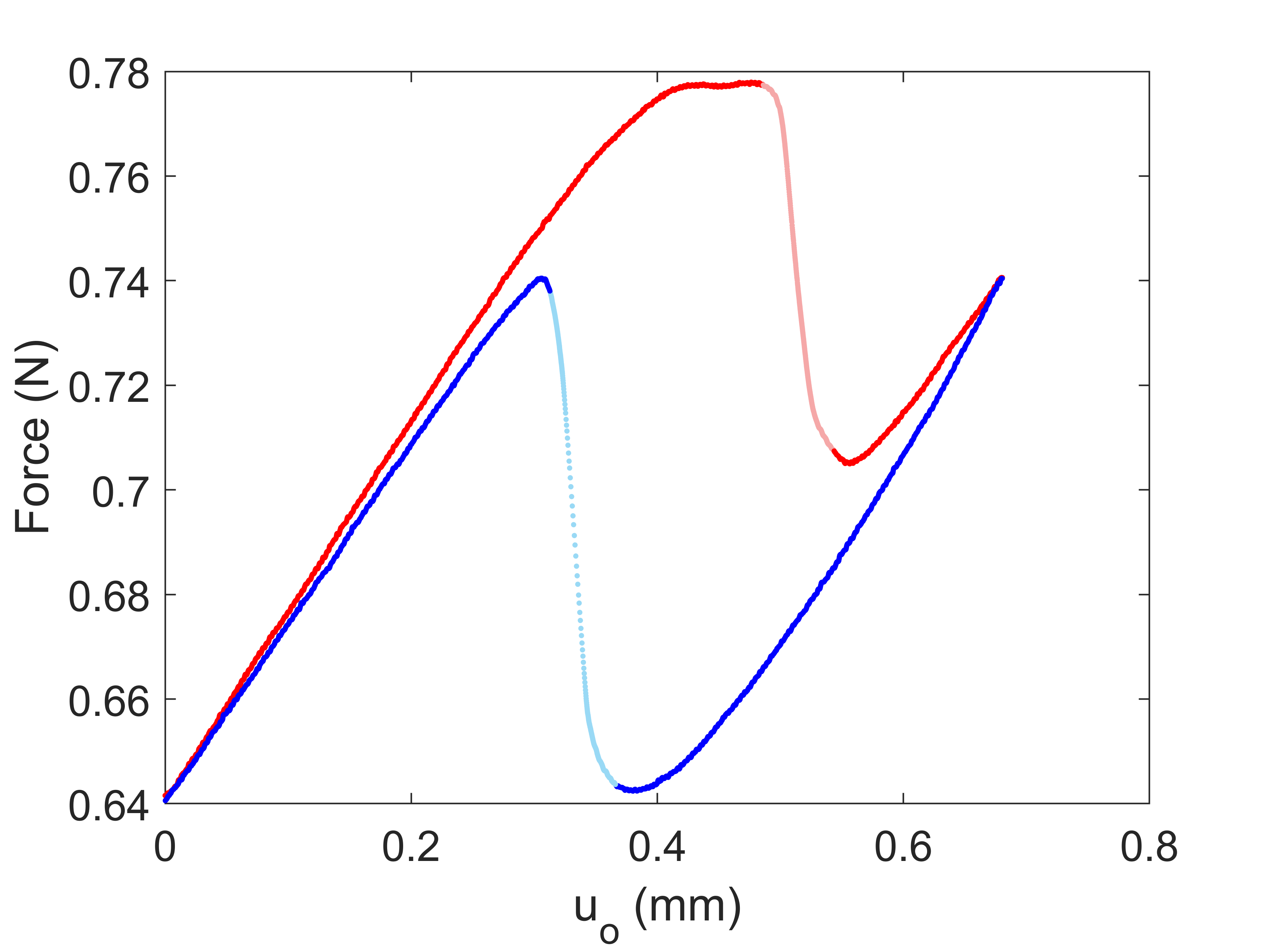}
            \includegraphics[width=0.485\linewidth]{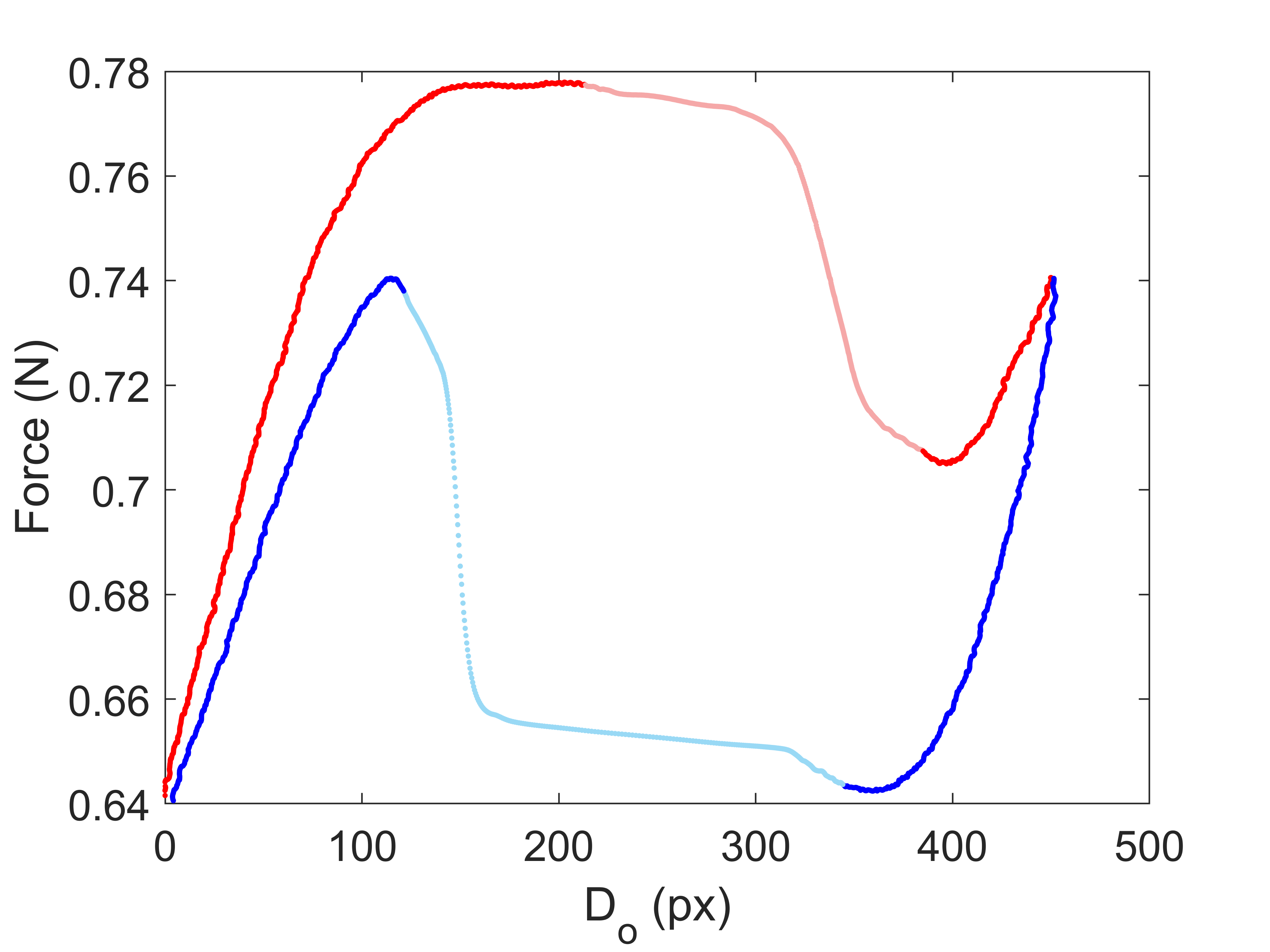}
            \includegraphics[width=0.485\linewidth]{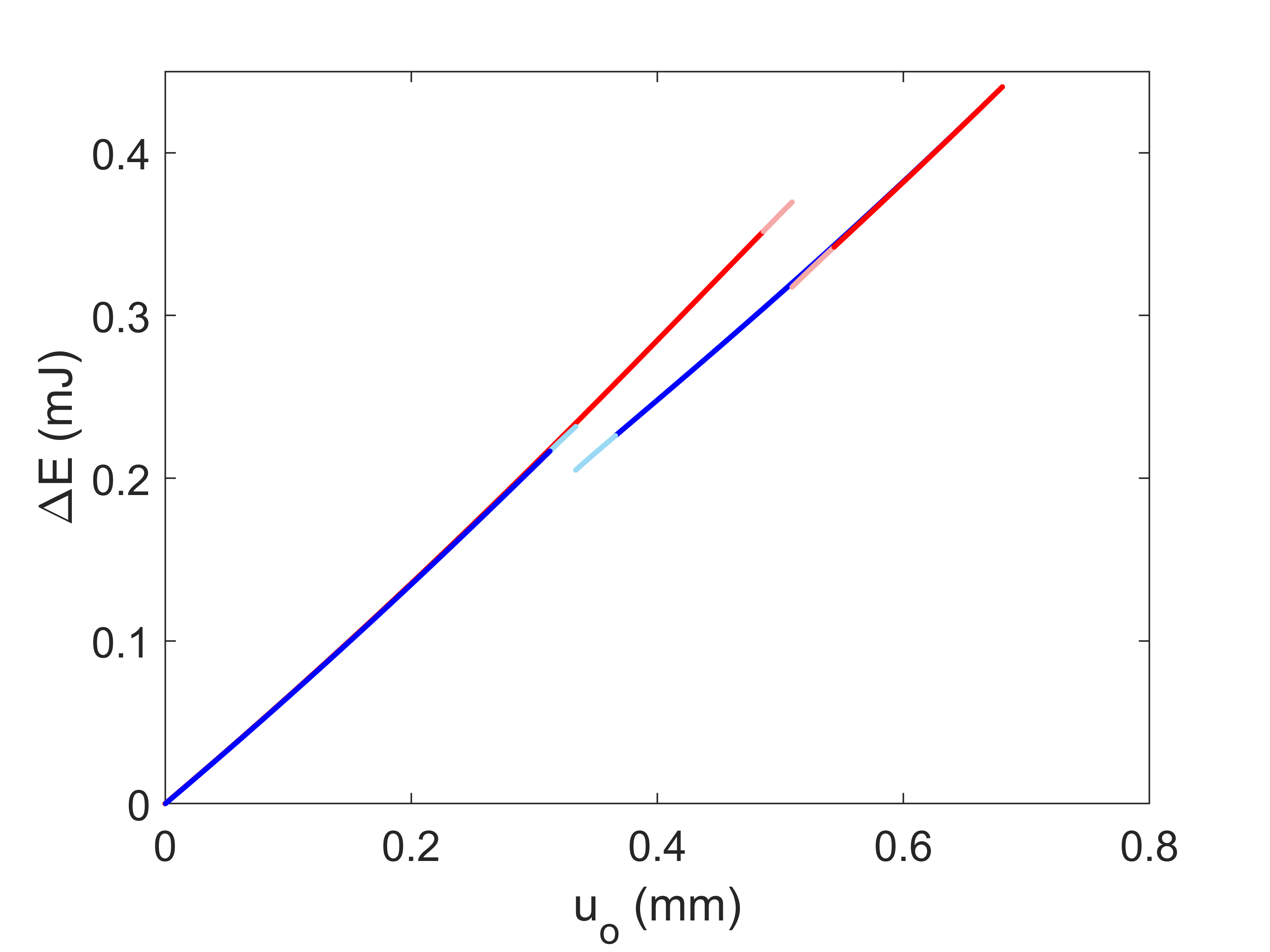}
            \includegraphics[width=0.485\linewidth]{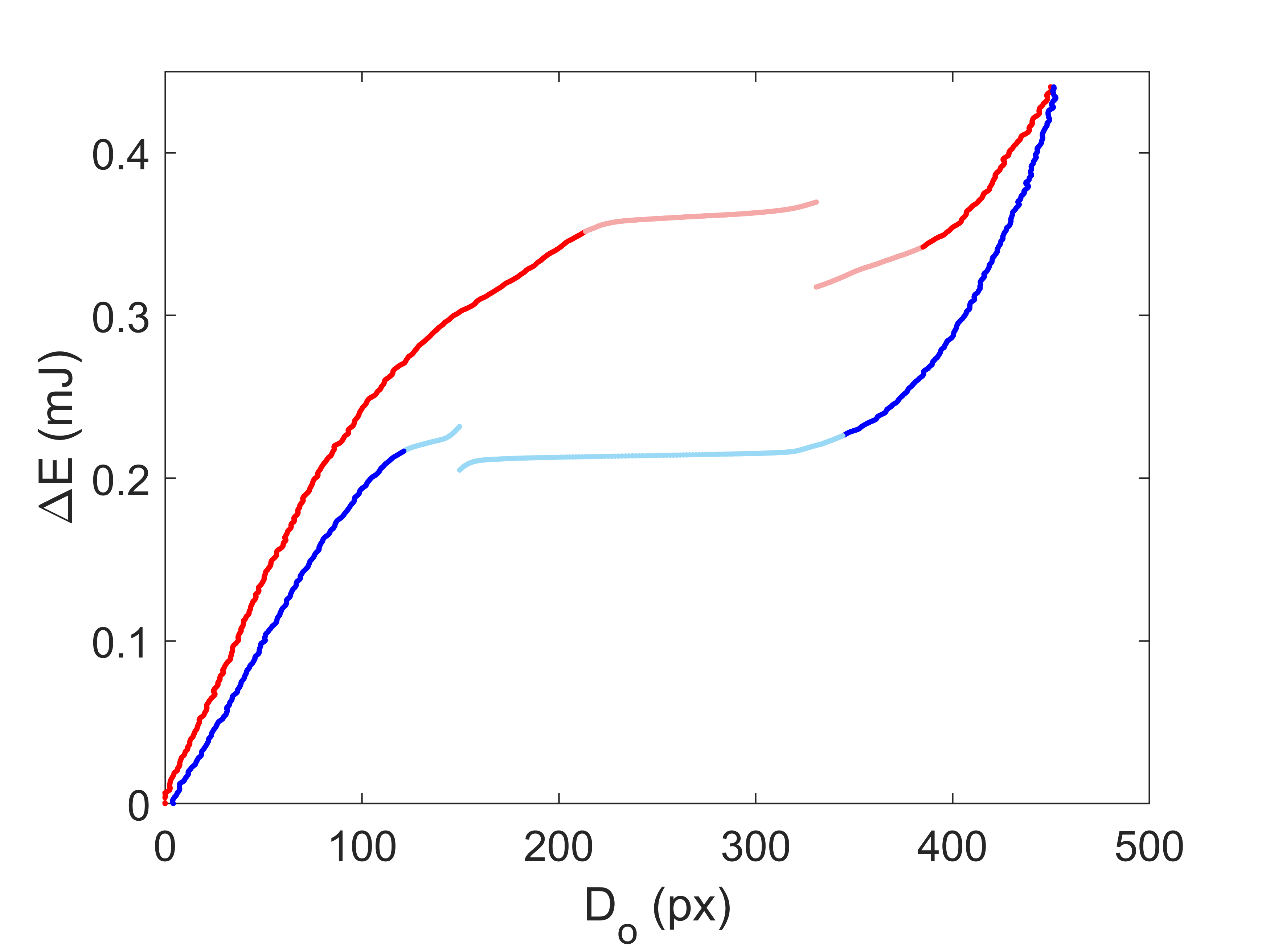}
    \caption{Processed data for O-valley using original unsubtracted force. Compare with main text Figure \ref{compare}B.}
    \label{fig:2x2nothingO}
\end{figure}

\begin{figure}[h]
    \centering
            \includegraphics[width=0.485\linewidth]{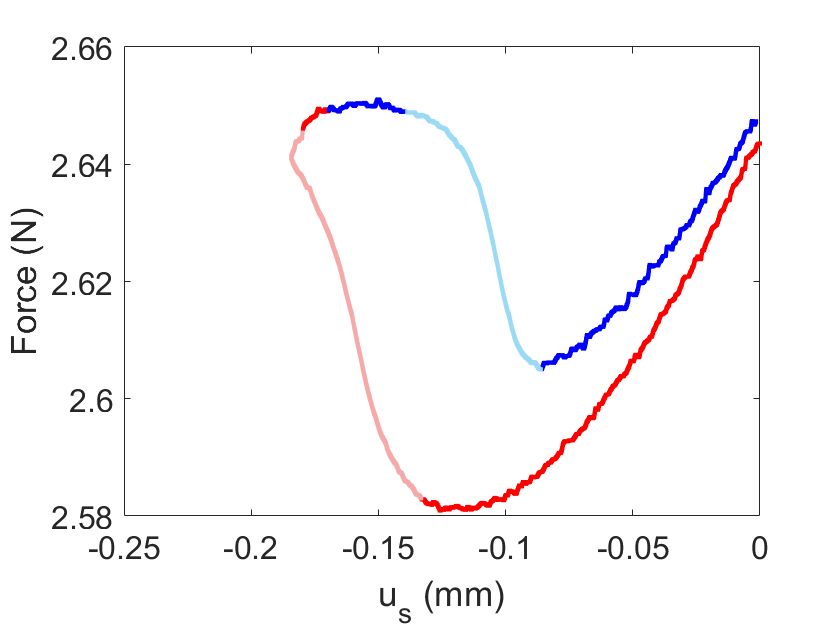}
            \includegraphics[width=0.485\linewidth]{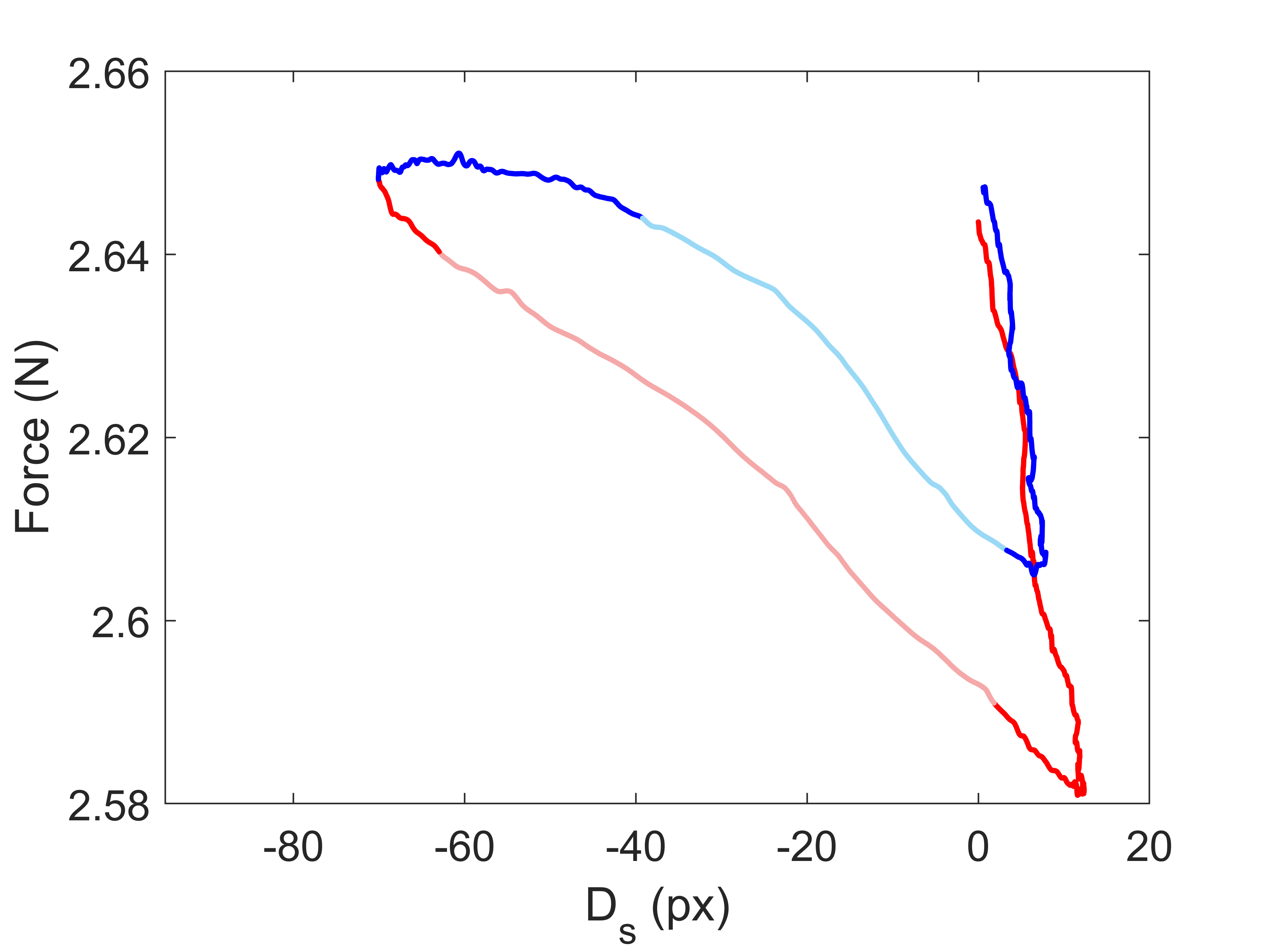}
            \includegraphics[width=0.485\linewidth]{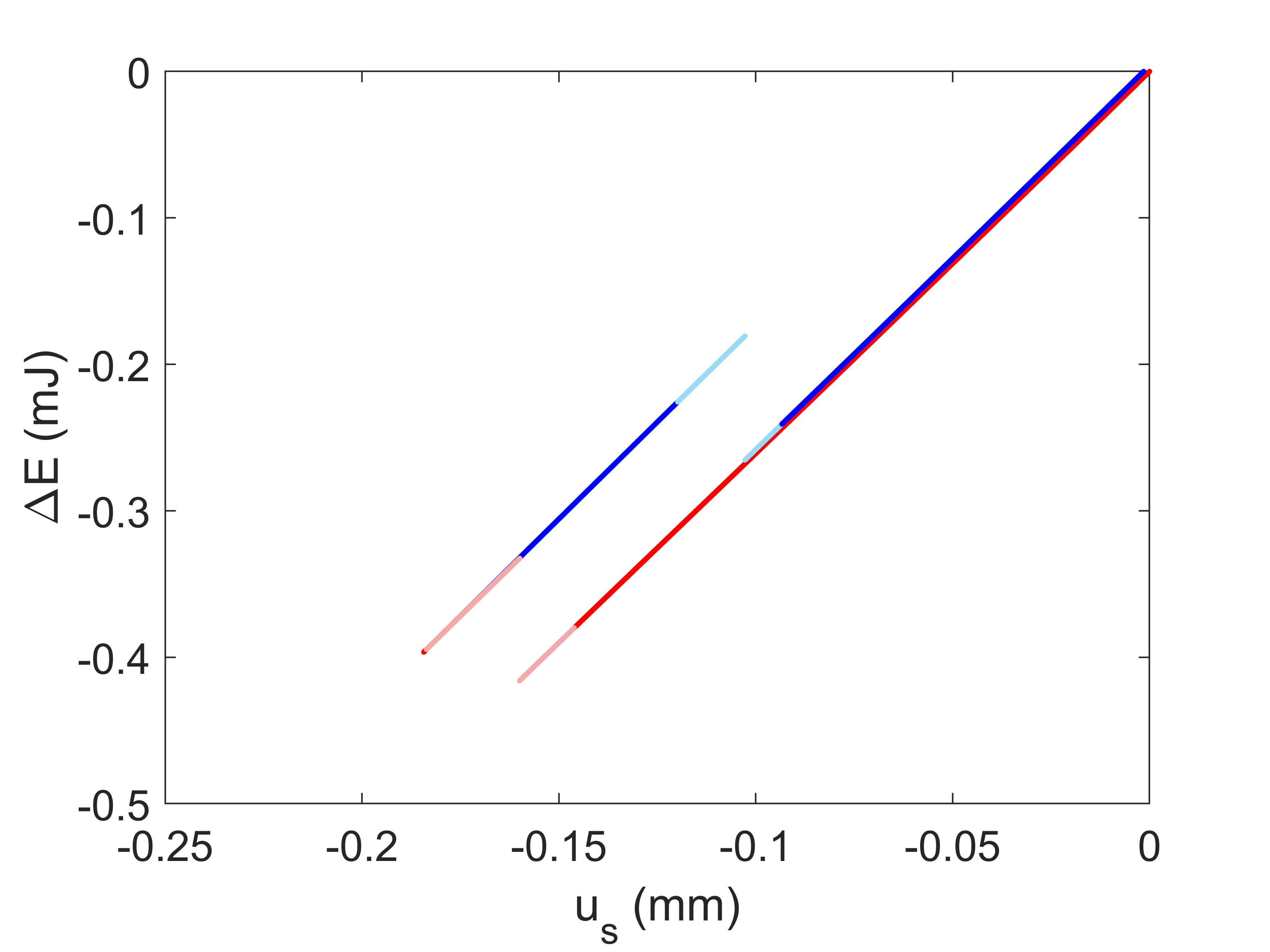}
            \includegraphics[width=0.485\linewidth]{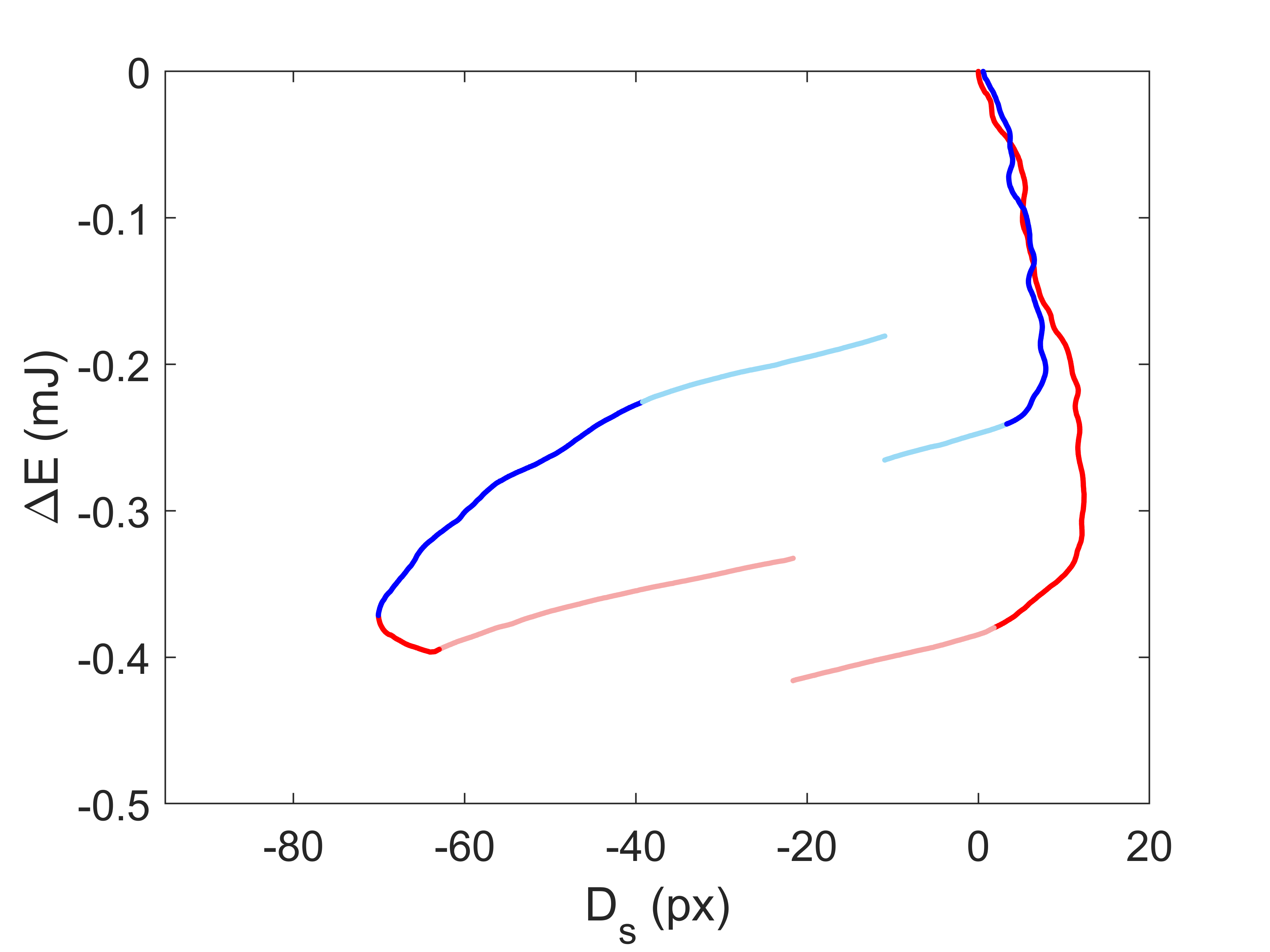}
    \caption{Processed data for S-ridge using original unsubtracted force. Compare with main text Figure \ref{compare}C.}
    \label{fig:2x2nothingS}
\end{figure}

Finally, Figures \ref{fig:2x2nothingO} and \ref{fig:2x2nothingS} show processed data using the original unsubtracted force. These can again be compared with main text Figure \ref{compare}B-C.  The background force and its associated energy dominate the energy plots and tend to obscure the behavior of the embedded crumple pair system.

\FloatBarrier

\newpage

\subsection*{Model details, and relation to experimental variables}

Our observations display behavior consistent with an underlying reduced model of similar type as that used to describe bistable structural systems \cite{BazantCedolin, Hunt20}, but further embedded in some additional background response. The quantities here are an applied force $P$, its work-conjugate variable $q_2$ related to the elastic response of the sheet background, and an internal variable $q_1$ related to the inter-crumple distance. 

There are different possible ways to treat the force-displacement curves in order to highlight the behavior of the embedded crumple pair system of interest.  One is to subtract out a constant force value, representing the average ``pre-load'' necessary to hold the structure at some chosen ``zero'' of our system.  Another is to subtract out a linear response--- we did this for our data, and also illustrated an alternate choice of line in the previous subsection. Any of these can create irrelevant zeroes of force (minima or maxima of energy). We do not know how much of the slope is background, and how much is the underlying response of the crumple pair system. 

In the experiments, we are always applying a positive force.  For the O-valley, we increase the force to separate/unbind the crumples.  For the S-ridge, we are at overall higher forces and larger displacements (greater background), and decrease the force to separate/unbind the crumples; we are not pulling them apart, but pushing them together. 

We use an energy that is quartic in the internal variable $q_1$.  The two wells of the quartic are irrelevant, only the hill between them matters. For simplicity, we place the wells at the limits of the range considered, but they could for example be pushed out to infinity (or eliminated altogether, after Dal{\'{\i}} \cite{Dali}) without changing the important aspects of the models. The symmetry of this model is also not a requirement, nor is its functional form--- in \cite{Hunt20} a different form involving square roots is used to create a double well and its hill. 
This internal variable is in turn negatively coupled to the other variable $q_2$, the imposed displacement, which is not only work-conjugate to the applied force we measure, but represents the overall elastic response of the sheet background. We note that the corresponding coupling is positive, rather than negative, in the previously mentioned canonical models \cite{BazantCedolin, Hunt20}. 

For simplicity, we present the model as symmetric about the origin, with both $q$-variables in the range between $-1$ and $1$. Interpreting these in terms of the real variables of interest in the experiment, one of these limits is going to represent a ``zero'' or bound state, and the other an unbound state. Because of the pulling/pushing distinction, the bound state zero will map onto $-1$ for the O-valley and $1$ for the S-ridge. Suitably range-normalized $u_O$ and $D_O$ will correspond, respectively, to $q_2+1$ and $q_1+1$, while normalized $u_S$ and $D_S$ will correspond to $q_2-1$ and $q_1-1$.  We do not attempt to normalize $u$ and $D$ ranges in Figure \ref{compare}. 

A simple energy with our required features has the form
\begin{align}
	\bar E = \tfrac{1}{4}\left(q_1^2-1\right)^2 + \tfrac{c}{2}\left(q_2-q_1\right)^2 + \tfrac{a}{2}(q_2-b)^2 \, ,
\end{align}
with constants $a$, $b$, and $c$. Further define a subtracted $E = \bar E - \tfrac{a}{2}(q_2-b)^2$ and $P = \bar P - a(q_2-b)$. 
The work of the applied force against its work-conjugate quantity is equal to the energy, $\int\! \bar P\mathrm{d}q_2 = \bar E$ and 
$\int\!P\mathrm{d}q_2 = E$ with appropriate zeroings.  Along equilibrium paths, virtual work considerations dictate $\bar P = \tfrac{\partial \bar E}{\partial q_2}$ and $0= \tfrac{\partial \bar E}{\partial q_1}$ ; this provides two relationships between the force and variables,  
\begin{align}
	P = c\left(q_2-q_1\right) \,, \\
	q_1\left(q_1^2-1\right) = c\left(q_2-q_1\right) \,,
\end{align}
and therefore
\begin{align}
	q_1 = q_2 - \tfrac{P}{c} \,, \\
	P = q_1\left(q_1^2-1\right) \,, \\
	P = \left(q_2 - \tfrac{P}{c}\right)\left[ \left(q_2 - \tfrac{P}{c}\right)^2-1\right] \,.
\end{align}
The last implicit relation will have two folds, in between max/min $P$ values, if the coupling constant $c$ is less than unity. The value of $c$ sets the width of the regressive branch between the two folds. For the plots in Figure \ref{compare}, we have set the value at $c = \tfrac{1}{2}$ for qualitative comparison without loss of generality. Assuming displacement control fixes $q_2$, with stability given by the second derivative with respect to $q_1$. Stability boundaries are thus at $q_1^2 = \tfrac{1-c}{3}$, which also corresponds to the turning points, so the regressive branch is unstable \cite{BazantCedolin}. 

Integrating the double-overhanging implicit cubic $P$ along $q_2$, or inserting equilibrium results into $E$, gives us a multivalued swallowtail energy $E$ as a function of $q_2$, where cusps in energy correspond to folds in force. Translating this into a function of $q_1$ transforms the swallowtail into a pair potential with attractive and repulsive regions, separated by a branch of unstable solutions.  We note a shallow minimum in the unstable region, which is perhaps a result of this particular functional form; it is not clear whether this has physical meaning.

\newpage 

\section*{Description of videos}

V1: Patterns in 0.003 inch polyester shim stock, 12 inches long, bent into a quarter-circle with 4 inches free span, clamped with lower boundary moving laterally to the right and then to the left. Boundary is moved at 0.02 mm/s, five times faster than in experiments used for data. \\

V2: Patterns in 0.002 inch Tedlar$^{\circledR}$/polyvinyl fluoride under uncontrolled loading. (Credit: C. Redman) \\

V3: Extract from a video similar to V1 showing interactions, but with boundary not properly clamped at the upper left, hence not fully controlled loading. \\

V4: Poking bound and unbound O-valley. \\

V5: Poking bound and unbound S-ridge. \\

V6: Unbinding and binding O-valley in 0.003 inch polyester shim stock, 6 inches long, bent into a quarter-circle with 4 inches free span, clamped with lower boundary moving laterally back and forth at 0.004 mm/s. Corresponds to data from sheet O7 trial 3 in Appendix~\ref{data}. \\

V7: Unbinding and binding S-ridge in 0.003 inch polyester shim stock, 12 inches long, bent into a quarter-circle with 4 inches free span, clamped with lower boundary moving laterally back and forth at 0.004 mm/s. Corresponds to data from sheet S12 trial 7 in Appendix~\ref{data}. \\

\end{document}